\documentclass[12pt]{JHEP3}

\usepackage{epsfig}
\def\eq{\begin{equation}}
\def\eeq{\end{equation}}
\def\eqa{\begin{eqnarray}}
\def\eeqa{\end{eqnarray}}

\def\bd{\begin{displaymath}}
\def\ed{\end{diplaymath}}

\def\Box{ {\,\lower 0.9pt\vbox{\hrule\hbox{\vrule height0.2cm \hskip 0.2cm
\vrule height 0.2cm }\hrule}\,}}
\def\lsim{{\
\lower-1.2pt\vbox{\hbox{\rlap{$<$}\lower5pt\vbox{\hbox{$\sim$}}}}\ }}
\def\gsim{{\
\lower-1.2pt\vbox{\hbox{\rlap{$>$}\lower5pt\vbox{\hbox{$\sim$}}}}\ }}
\newcommand\fverb{\setbox\pippobox=\hbox\bgroup\verb}
\newcommand\fverbdo{\egroup\medskip\noindent%
                        \fbox{\unhbox\pippobox}\ }
\newcommand\fverbit{\egroup\item[\fbox{\unhbox\pippobox}]}
\newbox\pippobox
\title{Lectures on String/Brane Cosmology}

\author{Fernando Quevedo
\\
 Centre for Mathematical Sciences, DAMTP\\
               University of Cambridge,\\
               Cambridge CB3 0WA UK.
\footnote{Permanent address. E-mail f.quevedo@damtp.cam.ac.uk,
fernando.quevedo@cern.ch.}\\
\\
Theory Division, CERN\\
CH-1211 Geneva 23, Switzerland .}

\abstract{An  overview is presented of some  cosmological aspects of
 string theory. Recent developments are emphasised,
  especially the  attempts to derive
 inflation or alternatives to inflation from the
dynamics of branes in string theory.
Time dependent backgrounds with potential cosmological implications,
 such as those provided by
 negative tension branes and S-branes   and
  the rolling string tachyon  are
also discussed.\footnote{Updated version of lectures delivered in
 January 2002 at the third RTN school on {\it The Quantum Structure of
 Spacetime and the Geometric Nature of Fundamental Interactions}, Utrecht.}}

\begin{document}

%\begin{document}

\section{Introduction}

This is an interesting time to think about cosmology from a string theory
 perspective.
The two subjects: cosmology and strings,
 complement each other in several ways.
 Cosmology needs an underlying
 theory
 to approach its basics questions such as
the initial singularity, if there was any,  the origin of inflation or any
alternative
 way to address the problems that
 inflation solves, such as the horizon and flatness problems and, more
 importantly, the origin of the  density
 perturbations in the cosmic microwave background (CMB).
The successes of inflation \cite{guth}\ in this regard
makes us sometimes forget that it is only a scenario in search of an
underlying theory. There are many variations of the inflationary scenario
\cite{Rev}\
 but at the moment
there is no concrete derivation of inflation (or any of its alternatives)
 from a fundamental theory such as string theory.

On the other hand string theory, although lacking a full nonperturbative
formulation,
 has been highly developed and understood in many respects and needs
a way to be confronted with physics.
One possibility is low-energy phenomenology, although it may be a
 long way before this can be tested, unless supersymmetry is discovered at
low energies and its
properties provide some hints to its high energy origin or
we get lucky and the string scale is low enough to be probed in the future
 colliders such as LHC.
Otherwise cosmology may turn out to be the main avenue to probe string theory.
This has been reinforced in view of the recent observational discoveries
that seem to indicate that the
effective cosmological constant is not exactly zero and, furthermore,
 the accuracy of the
 CMB experiments \cite{COBE, cmbexp}\,  that
have provided a great deal of information on the scalar density
 perturbations of the cosmic microwave background. This has raised the
status of cosmology to a
science subject to precision experimental tests, with a very promising
future due to the planned
 experiments for the not too far future, such as  MAP and PLANCK.

Moreover, based mostly on string theory ideas, the brane world scenario
 has emerged in the past few years
offering  dramatic changes    in our view of the universe
\cite{braneworld}.
The fact that we may be living on a
hypersurface in higher dimensions does not only imply that the scale of
 string theory could be as small as
1TeV, but also provides  completely new scenarios for the cosmological
 implications of string theory \cite{branecosmology,branecosmology2}\ .
 Actually if the brane world scenario is realized, but with a string scale
close to the Planck scale, the main
place to look at its possible implications will be cosmology,
 rather than
table top experiments or
high energy accelerators.

Finally, the study of cosmological  implications of string theory can shed
 some light into the better  understanding of
the theory itself. We already have the experience with the study of black
hole backgrounds in string theory which has
 led to some of the main successes of the theory, namely the explicit
 calculation of the black hole entropy and  the
identification of the AdS/CFT correspondence 
\cite{adscft}, 
which not only provides a
concrete realisation of  the holographic principle \cite{thooft}\
but has also  led to  important results in field and string theories.
Cosmology is the other arena where nontrivial string backgrounds can be
 explored, some of the ideas developed from other studies can be put to
 test in cosmology and probably new insights may emerge. 
In particular the recent
 realisation that our universe could be in a stage
 with a nonzero vacuum energy gives rise to an important challenge for
string theory, we need to be able to understand
string theory in such a background
\cite{desitter}. Also previous ideas about quantum
cosmology in general may find new realisations in
the context of string theory. In summary, 
we may say that cosmology presents probably the
most important challenges for string theory: the initial singularity, the
 cosmological constant, the definition of observables, the identification of
 initial conditions, 
realisation of de Sitter or quintessential backgrounds of the theory, etc.

 It is then becoming of prime importance to learn the possible
applications
 of string theory to cosmology.
These lecture notes are
 an effort to put some of these ideas together for non-experts.
Due to limitations of space, time and author's knowledge, the discussion is 
at a superficial level and incomplete. They were originally given to 
review the basic ideas on the subject, including 
brane cosmology
in static and time dependent backgrounds to conclude with 
 D-brane inflation and tachyon condensation, together with some
 details about the ekpyrotic scenario.  However, right 
after the lectures were given,
  several interesting developments have occurred related with the
 subject of the lectures that have to be briefly included for completeness
(rolling tachyon, S-branes, time dependent orbifolds).
 Fortunately there are several good reviews on the 
first part of the lectures that can be consulted for deeper insights 
\cite{Rev,
brandenberger, riotto, copeland, mcosm, carroll, easson, veneziano}.
There are hundreds of articles on brane cosmology and I cannot make justice to 
everybody working in the field.
I do apologise for omissions of important references.
The presentation tries to include only the brane cosmology ideas 
formulated in the context of string theory or that have  connections 
to it, therefore, many interesting developments in brane cosmology,
which are not clearly related to string theory are omitted.
  
I first give an overview of the standard big-bang cosmology that
introduces
 the  physical parameters, notation
and problems.
Then I  describe briefly the main ideas discussed in the past
 (before the year 2000) in string cosmology. These
 include the Brandenberger-Vafa scenario \cite{BV} where T-duality and
 winding modes could have an interesting
implication for early universe cosmology, including the possible
determination
 of the  critical dimension
 of spacetime.
Also the cosmology associated to the moduli fields, which could be
  candidates for inflaton fields \cite{moduliinflation}\ ,
but also can cause
 a serious and generic cosmological problem once they get a mass, since
they
 can either ruin
 nucleosynthesis by their decays, if they are
 unstable, or over close the universe, if they happen to be stable \cite{cmp} (see also \cite{polony}).
This has been called the cosmological moduli problem.
 Finally we mention the main ideas behind
the Gasperini-Veneziano `pre big-bang cosmology' that during the years has 
become
the string cosmology scenario subject to more detailed
 study (see \cite{veneziano} for 
a very complete review on the subject with references to the earlier work).

In the third part of the lectures I concentrate on some recent
developments.
 I will emphasise
 the role that string theory 
$p$-branes can play in cosmology. First we describe
some of the interesting results coming out of a  treatment of
brane cosmology in the simple setting of 4D brane worlds moving in
a 5D bulk \cite{branecosmology, branecosmology2, kraus, keki, ida,
 shiromizu, BCG, verlinde}. Two points are emphasised: the Einstein's equations in the 4D
brane do not have the standard behaviour in the sense that the relation
 between the Hubble parameter and the energy density is different from the
standard 4D cosmology.
We also remark the interesting possibility for understanding cosmology in
the
 brane world as just the motion of the brane in a static bulk. An observer
on
the brane feels his universe expanding while an observer in the bulk only
sees
 the brane moving in a static spacetime, this is usually known as mirage
 cosmology.

Then I discuss the  possibility that the dynamics of
D-branes may have direct impact in cosmology, in particular considering a
 pair of D-branes approaching each other
and their
subsequent collision could give rise to inflation \cite{tye}.
For a D-brane/antibrane pair it is possible to
compute the attractive potential from string theory and actually
obtain inflation with the inflaton field being the separation
 of the branes \cite{quei,dvali}.
 Furthermore, it is known from string theory that after the
 branes get to a critical distance,
an open string mode becomes tachyonic thus providing an instability
 which is precisely what is needed to end inflation \cite{quei}.
Obtaining then a realisation of the hybrid inflation scenario
\cite{hybridinflation}\  with the two
 relevant fields having well defined
stringy origin, {\it i.e.} the separation of the branes generates inflation
 and the open string tachyon finishes
 inflation and provides the mechanism to describe the process of
the brane/antibrane collision and annihilation.
Natural extensions of these ideas to include orientifold models, intersecting branes at
angles and related constructions, \cite{kali, queii, g-b, lusti}\ 
will also be discussed, which illustrates that the realisation of
 hybrid inflation from 
the inter-brane separation as the inflaton field and the open string tachyon
 as the field 
responsible to end inflation and re-heat is very generic in D-brane models.

Tachyon condensation is one of the few physical process that 
has been studied in detail
 purely from string theory techniques
\cite{banks,sen,lowerbranes,tseytlin} and can have by 
itself important 
implications to cosmology, independent
of its possible role in the brane inflation scenarios
\cite{tacos}. The rolling tachyon 
field has properties that have been 
uncovered just recently \cite{senroll,garytac},
 such as resulting in a pressure-less fluid at the end
 of its relaxation towards the minimum of 
the potential. Furthermore, its potential includes the D-branes as topological
 defects which also play 
an important role in cosmology (providing for instance 
dangerous objects such as monopoles and domain walls and less dangerous ones 
like cosmic strings). Finally it has partially motivated the introduction of 
a new type of branes known as 
space-like or S-branes 
\cite{gutperlestrom}\ 
which can roughly be thought as kinks in time, rather 
than space, of the tachyon 
potential
describing then the rolling of the tachyon. Just as for 
the case of D-branes,
 S-branes can also be obtained
 as solutions of supergravity equations, but these solutions being time 
dependent and therefore cosmological
 in nature \cite{forste,pope,andre,gqtz,gutperle,rob}.
  We illustrate in a simple example the interesting properties of 
these cosmological solutions. A  general class 
 of them have  past and future cosmological regions,
with a bounce, 
representing cosmologies with horizons and  no spacelike singularities
\cite{kl,gqtz,costa}.  
This can be interpreted as the spacetime due to the presence of negative
 tension branes with opposite charge \cite{cck,bqrtz}, similar to 
a pair of orientifold planes. These geometries are related  to 
black holes and then the mass, charge, Hawking temperature and entropy can be 
computed in a similar way. The bouncing behaviour can be interpreted 
analogous to the Schwarzschild wormhole or Einstein-Rosen bridge, but this
time connecting past and 
future cosmologies instead of the two static, asymptotically flat
regions \cite{bqrtz}.
 Stability of these solutions
may be a potential problem for their full interpretation.

We also briefly discuss the probably more ambitious proposal of  the
 ekpyrotic universe \cite{ekpyrosis,kosst,cyclic}, in the sense that
with the same idea of colliding branes, this time in the context of
 Horava-Witten compactifications rather than D-branes,
 it may not lead necessarily to inflation but could provide an alternative
to it, approaching the same questions as inflation does, especially the
inhomogeneities of the cosmic microwave background.
 This scenario has also lead to
two interesting developments. First,  resurrecting the idea of the cyclic
universe \cite{cyclicold, cyclic} and second,  the
realisation of cosmological string backgrounds by just orbifolding flat
spacetime \cite{hs, kosst, TDbackgrounds, hp}. 
This process guarantees an exact
solution of string theory and has opened the possibility to approach
issues concerning a big bang-like singularity
performing explicit string calculations. Possible problems with this
approach to time dependent backgrounds are
briefly mentioned.

\section{Cosmology Overview}

\subsection{Standard FRW Cosmology}

The standard cosmological model has been extremely successful given its
simplicity. The starting point
is classic Einstein equations in the presence of matter. The requirements
of
 homogeneity and isotropy of the
4D spacetime determines the metric up to an arbitrary function of time
$a(t)$,
  known as the scale factor,
which measures the
time evolution of the Universe and a discrete parameter $k=-1, 0, 1$ which
 determines if the Universe is
open, flat or closed,  respectively. The Friedmann-Robertson-Walker (FRW)
metric describing the evolution
of the
Universe can then be written as:

\eq
\label{frw}
ds^2\ = \ -dt^2\ + \ a^2(t)\ \left[\frac{dr^2}{1-kr^2} +
r^2\left(d\theta^2+\sin^2\theta d\phi^2\right)\right]\ .
\eeq
The  scale factor $a(t)$ is given by solving
Einstein's equations

\eq
G_{\mu\nu}\ \equiv R_{\mu\nu}-\frac{1}{2} g_{\mu\nu} R= -\ 8\pi G\ T_{\mu\nu}\ .
\eeq

In natural unites $h=c=1$, Newton's constant $G$ can be written in terms of the Planck mass
$8\pi G = 1/M_{Planck}^2$.
The stress-energy tensor $T_{\mu\nu}$ is usually taken to correspond to a
 perfect fluid (latin indices are 3-dimensional):

\eq
T_{00}\ =\  \ \rho\ ,  \qquad\qquad T_{ij}\ =\ p\ g_{ij}\ ,
\eeq
with the energy density $\rho$ and the pressure $p$ satisfying an equation
 of state of the form $p\ =\ w\rho $. Here $w$ is a parameter which, for many
interesting cases is just a constant describing the kind of matter
dominating in the stress-energy tensor.
We present in the table the values of $w$ for
common cases corresponding to matter, radiation and vacuum domination.

\TABLE{\renewcommand{\arraystretch}{1.7}
%\begin{center}
\begin{tabular}{|c|c|c|c|}
\hline
Stress Energy & $w$ & Energy Density & Scale Factor $a(t)$ \\
\hline
\hline
 Matter  &  $w=0$  & $\rho\sim a^{-3} $ & $ a(t)\sim t^{2/3} $ \\
\hline
 Radiation & $ w=\frac{1}{3}$ & $ \rho\sim a^{-4} $ & $a(t)\sim t^{1/2}$\\
\hline
Vacuum ($\Lambda$)  & $ w=-1$ & $ \rho\sim \frac{\Lambda}{8\pi G}$ & $a(t)\sim
\exp(\sqrt{\Lambda/3} t)$\\
\hline
\end{tabular}
\caption{\small Behaviour of scale factor and energy density for matter,
radiation and vacuum dominated universes. The solution for the scale
factor is written for the case $k=0$.
\label{Scaleu}}}
%\end{center}}
%\end{table}
%

%
%%%%%%%%%%%%%%%%%%%%%%%%

Einstein's equations for the ansatz (\ref{frw}) above reduce to the
 Friedmann's equations\footnote{The second equation is sometimes referred to as the Raychaudhuri equation.}:
\eqa
H^2 & = & \frac{8\pi G}{3}\ \rho\ - \ \frac{k}{a^2} \\
\frac{\ddot a}{a} & = & -\frac{4\pi G}{3}\ \left(\rho+ 3p\right)\ , 
\eeqa
 with $H$ the Hubble function $H\equiv \frac{\dot a}{a}$.
These two equations imply the energy conservation equation $\nabla_\mu
T^{\mu\nu}=0$ or:

\eq
\label{conservation}
\dot\rho\ = \ -3 H\ \left(\rho + p \right)\ . 
\eeq
 Therefore, after using the equation of state $p=w\rho$ we are left with
two equations which we can take as the
first Friedmann equation and the energy conservation, for $\rho $ and $
a(t)$,  which can be easily solved.
Equation (\ref{conservation}) gives immediately $\rho \sim a^{-3(1+w)}$
introducing this in Friedmann's equation gives
the solution for $a(t)$.
We show in the table the behaviour of $\rho$ and $a(t)$  for typical
equations of state.
Notice that for the expressions for  $a(t)$ we have neglected the
curvature term
$k/ a^2$ from Friedmann's equation  and therefore we write the solution
for the flat universe case. The exact solutions for the other cases can
also be found.

For the more general case we may say several things without solving the
equations explicitly. First of all, under very general assumptions, based
mainly on the
positivity of the energy density $\rho$, it can be
shown that the FRW ansatz necessarily
implies an initial singularity, the big-bang, from which the universe
starts expanding. For instance
if $\rho+ 3p>0$, which is satisfied for many physical cases, the
acceleration
 of the universe
measured by $\ddot a$ is negative, as seen from the second Friedmann's
equation.
 For $k=-1,0$ the first equation tells
us that, for positive energy density, the universe naturally expands forever
whereas for $k=1$ there will be a value of
$a$ for which the curvature term  compensates the energy density term and
 $\dot a=0$, after this time $a$ decreases
and the universe re-collapses. A word of caution is needed at this point,
 which is usually a source of confusion.
Because of the previous argument, it was often claimed that, for instance, a
 closed universe ($k=1$) will re-collapse.
However we can see from the second Friedmann equation, which is
independent
of $k$, that if $\rho +3p<0$ the universe
 will always accelerate. This happens for instance for vacuum domination
($w=-1$) where for $k=1$ the solution is
$a(t)\sim \cosh(\sqrt{\Lambda/3} t)$, which is clearly accelerating.

We can illustrate the structure of the big-bang model in terms of a
 spacetime diagram, known as Penrose or conformal diagram,
see for instance \cite{hawking}. This diagram
not only pictures the relevant parts of the spacetime, in this case the
initial singularity, but it is such that by a 
conformal transformation, it  represents
the points at infinity in a compact region and furthermore, even though
the
spacetime is highly curved, especially close to the singularity, light rays
follow lines at $45$ degrees just as in standard Minkowski space. These 
diagrams are usually, but not always (depending on the symmetries of the metric)
two-dimensional. In the FRW case it includes the $t-r$ plane, so each 
point in the diagram represents a 2-sphere for $k=1$ or the 2D flat and hyperbolic
 spaces for
$k=0,-1$ respectively. The wiggled line of figure~1 is the spacelike surface at $t=0$ 
representing the singularity. At this point the scale factor $a=0$ and the
 radius of the sphere (for $k=1$) is zero and $\rho\rightarrow\infty$. The Penrose diagrams extract in 
a simple way the causal structure of the spacetime. In this case we can see 
that 
extrapolating to the past from any point in the diagram necessarily hits the 
big-bang singularity.

\EPSFIGURE[r]{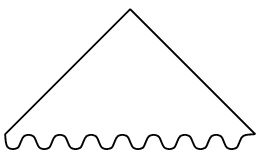,width=7cm}{Penrose diagram for a FRW spacetime,
 the initial spacelike singularity
is represented by the wiggled lines, and future infinity is mapped to finite 
values.}
{\label{figure1}}

A useful concept to introduce is the critical density
\eq
\rho_{critical}\ \equiv \ \frac{3 H^2}{ 8\pi G}\ .
\eeq
Which, for the present time $H=H_0\sim 65 \rm{Km/s/Mpc^{-1}}$ gives
$\rho_c\sim 1.7\times 10^{-29} \rm{g/cm^3}$ ({$1 Mpc$ (mega-parsec)
 $=3\times 10^{22}$ meters}).
This allows us to define a dimensionless parameter $\Omega$ which
corresponds
 to the ratio of the energy density of a given system to the critical
density:
$\Omega_i\equiv \rho_i/ \rho_{critical}$, with the index $i$ labelling
 the different contributions to the energy density, and the total ratio is
$\Omega=\sum_i \Omega_i$.
With these definitions we can write the first  Friedmann equation as:
\eq
\Omega\ = \ 1\ + \ \frac{k}{H^2 a^2}\ .
\eeq
From this we can see the clear connection between the curvature of the
spatial sections given by $k$
and the departure from critical density  given by
$\Omega$. A flat universe ($k=0$) corresponds to a critical density
($\Omega=1$) whereas open ($k=-1)$ and closed ($k=1$) universes correspond
to
$\Omega < 1$ and $\Omega>1 $ respectively.

With all this information in mind,  we just assume that the early universe
 corresponds to an expanding gas of particles and, with the input of the
 standard model of particle physics, and some
thermodynamics \footnote{As long as gravity is weak we can safely define the
concept of
thermal equilibrium and therefore a temperature.}
we can trace the evolution of the system. We present in  table 1 some of
the
important points through the evolution and refer to the standard
literature for
 details \cite{Rev}. There are few things to keep in mind: the gas is considered
to be in equilibrium. The main two reasons for a particle to leave
equilibrium is
that its mass threshold is reached by the effective temperature of the
universe
and so it is easier for this particle to annihilate with its antiparticle
than being produced again,  since, as the universe cools down, 
there is not enough energy to produce such
a heavy object. Also, if the expansion rate of the relevant reactions
$\Gamma$  is smaller than the expansion rate of the universe, measured by
$H$,
some particles also get out of equilibrium.
 For instance, at temperatures above
$1$ MeV the reactions that keep neutrinos in equilibrium are faster than
the
expansion rate but at this temperature  $H\geq \Gamma$ and they
decouple from the plasma, leaving then an observable, in principle,
trace of the very early universe. Unfortunately we are very far from
being able to detect such radiation.

At the atomic physics scale, the universe is cold enough for atoms to be
 formed and  the photons are out of equilibrium, giving rise to the famous
cosmic microwave background. At approximately the same
 time also the universe changes from
being
 radiation dominated ($w=1/3$) to  matter dominated ($w=0$).
After this, the formation of structures such as clusters and galaxies can start,
 probably due to the quantum fluctuations of the early universe,
leading to our present time.

\TABLE{\renewcommand{\arraystretch}{1.7}
%\begin{center}
\begin{tabular}{|c|c|c|c|}
\hline
Temperature & Time & Particle Physics & Cosmological Event \\
\hline
\hline
 $10^{19}$ GeV  &  $10^{-43}$ s  &  String Theory? &  Gravitons decouple ?
\\
\hline
$10^2$ GeV \  -   & $10^{-43}$ s\  -  & Grand Unification?
 & Topological defects? \\ $10^{19}$ GeV & $10^{-12}$ s
& Desert? String Theory? &  Baryogenesis? Inflation? \\
& & Extra dimensions? & \\
\hline
 $10^2$ GeV & $ 10^{-12}$ s & Electroweak Breaking & Baryogenesis?\\
\hline
$0.3$ GeV  & $10^{-5}$ s  & QCD scale & Quark-Hadron transition \\
\hline
$10- 0.1$ MeV & $10^{-2}-10^2$ s & Nuclear Physics scale &
Nucleosynthesis,  \\
& & & Neutrinos decouple\\
\hline
$10$ eV & $10^{11}$ s & Atomic Physics scale & Atoms formed, CMB\\
& & & Matter domination\\
\hline
\end{tabular}
\caption{\small A brief history of the universe (or time in a nutshell).
The temperature units can be translated to $K$ by using
 $1$ GeV $= 1.16\times 10^{13} K$.
\label{Scale}}}
%\end{center}}
%\end{table}
%

The standard cosmological model has strong experimental evidence which can
be summarised as follows:

\begin{itemize}

\item
The original observation of Hubble and Slipher at the beginning of the 20th
century, that the
galaxies are all separating from each other, at a rate that is roughly
proportional to  the
separation, is clearly realised for $H$ approximately constant at present,
$H=H_0>0$. This has been overwhelmingly verified during the past few decades.

\item
The relative abundance of the elements with  approximately
$75 \%$ Hydrogen almost $24\%
$ Helium, and other light elements such as Deuterium $D$ and helium-4
$^4 He$, with
small fractions of a percent,
 is a big success of nucleosynthesis, and at present is
the farther away in the past that we have been able to compare theory and
observation.

\item
The discovery of the cosmic microwave background, signalling the time of
last photon scattering,
 by Penzias and Wilson in 1964
was perhaps the most spectacular test of the model. Starting in the 1990's
with the discoveries of the COBE satellite
 and more recent balloon experiments such as BOOMERANG, Maxima and DASI, 
cosmology has been brought to the status of
precision science. In particular, the confirmation of the black body
nature of the CMB is known with excellent
precision, but, more importantly, the fluctuations in the temperature
$\frac{\delta T}{T}$
signalling density fluctuations $\frac{\delta\rho}{\rho}$
in the early universe
provide a great piece of information about the possible 
microscopic origin of the
large scale structure formation. The temperature fluctuations are analysed
 in terms of their spherical harmonics decomposition $\frac{\delta T}{T}
=\sum_{lm} a_{lm} Y_{lm}(\theta,\phi)$, with the power spectrum
$C_l=\langle |a_{lm}|^2\rangle$ showing a peak structure (see figure~2). 
The higher the multi-pole moment the smaller angular separation in the sky. 
The location
and height of the peaks provides precise 
information about the fundamental parameters
of FRW cosmology, such as $\Omega$, $\Omega_{baryon}$, the cosmological
constant, etc.
See for instance \cite{kamkos,triangle}. 
In particular the first peak being at approximately $l=200$ provides a 
very strong evidence in favour of  a flat universe.

\cleardoublepage
\EPSFIGURE[r]{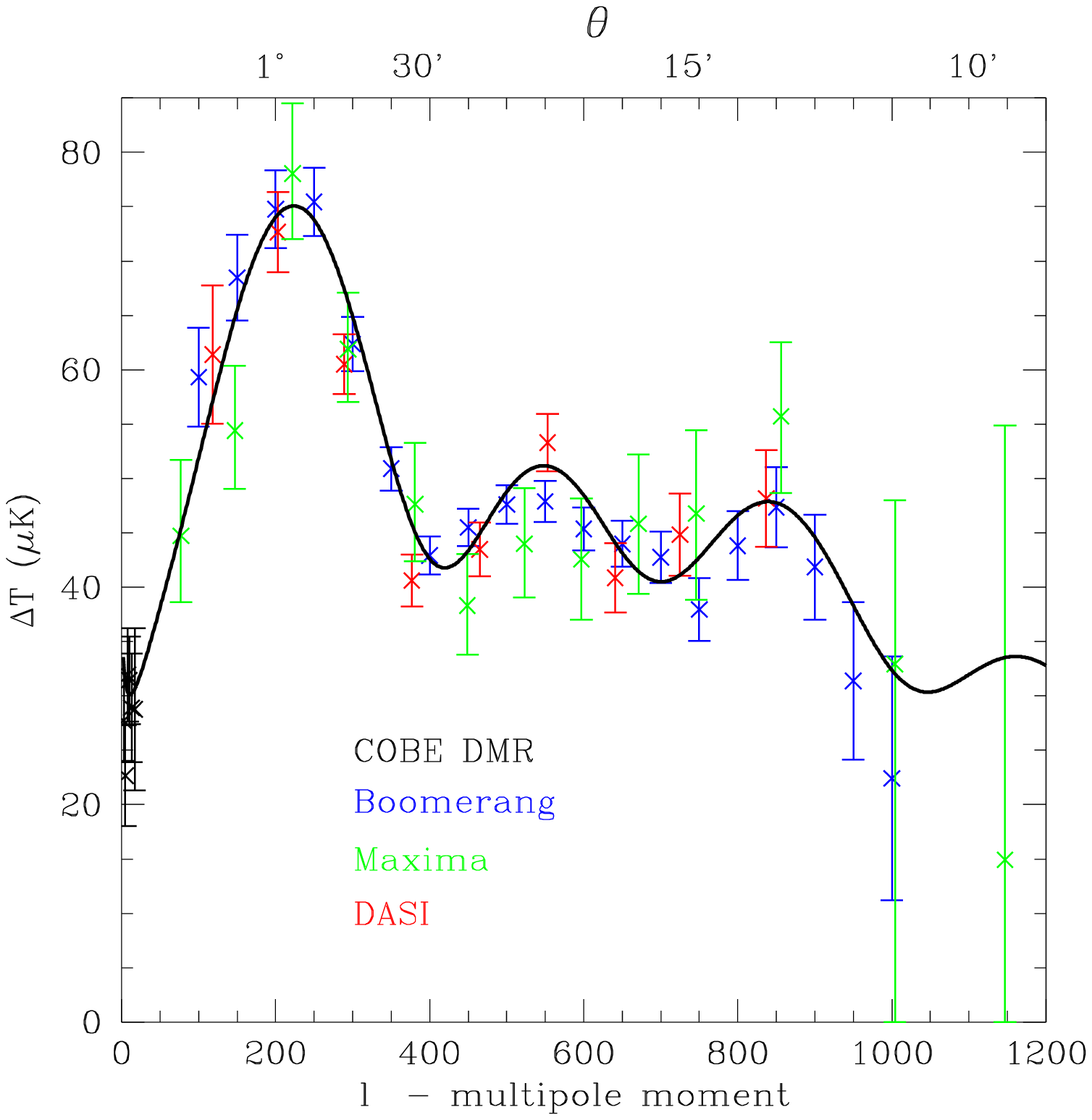,width=16cm}{The cosmic microwave background spectrum of temperature fluctuations.
The different location and height of the peaks as a function of multi pole moments put strong constraints
 on the parameters of the standard model of cosmology. The solid line
represents the best fit of the combined observations. Courtesy
of Rob Crittenden.}
{\label{CMB}}
\clearpage
\end{itemize}

Further observations and theoretical developments
 point also to some problems of this model which 
we can summarise as follows

\begin{itemize}

\item
The breakdown of classical relativity near the initial singularity is 
certainly the main conceptual problem in cosmology. 

\item
The horizon problem. The isotropy of the universe reflected by the CMB is actually the source of a problem.
 Assuming the standard expansion of the universe
we receive the same  information from points in the space that do not appear
 to be in causal contact with each  other. Therefore it is actually a puzzle 
why the radiation is so uniform. 

\item 
Origin of CMB anisotropies. The definitely observed anisotropies in the
 CMB are expected to be produced from physics of the early universe which is 
not explained in the standard model.

\item
Flatness. The universe is almost flat in the sense that $0.2<\Omega <2 $. 
This evidence is being strengthened by the more recent results on the CMB,
 essentially the position and height of the first acoustic peak on the 
spectrum of the CMB precisely provides evidence for $\Omega\sim 1$ at
present.
See for instance \cite{kamkos,triangle}.
The flatness problems refers to the fact that for $\Omega$ to be so close to
 one at present it had to be essentially one in the early universe with a
 precision of many significative figures. There is no explanation for
 this.

\item
Baryogenesis. Combining the standard models of particle physics and cosmology 
we cannot explain why there seems to be an excess of matter over antimatter.
The requirements for this to happen, {\it ie} process out of equilibrium,
baryon number violation and CP violation need to be combined in a model 
beyond the standard models but at the moment has no explanation.

\item
Dark matter. The survey and study of the behaviour of matter, 
such as rotation curves for galaxies, at many different scales, has given
 evidence that there should be a new kind of matter not present in the 
standard model of particle physics. This should play an important role in the 
explanation for
the large scale structure formation.

\item
Dark energy. Recent results form the study of high redshifted supernovae,
 combined with the CMB, has provided strong evidence
for the fact that the universe is actually accelerating at present. This,
as mentioned before, indicates that there should be a form of `dark energy'
which provides $\rho+3p<0$ and causes the universe to accelerate. 
An effective cosmological constant or a time varying scalar field are 
the main proposals for this dark energy. In any case this stresses the 
cosmological constant problem (why is the cosmological constant almost zero?)
and makes it more interesting to explain why it has the value it seems to
 have at present $\Lambda = 10^{-120} M_{Planck}^4 = (10^{-3} \rm{eV})^4$.
Present observations point towards $\Omega=\Omega_\Lambda + \Omega_B +
\Omega_{DM}=1$ with the contribution from dark energy $\Omega_\Lambda\sim 0.7$
whereas the dark matter and baryonic contributions together only make
$\Omega_B +
\Omega_{DM}\sim  0.3$. Another way to rephrase this challenge is by calling
 it the coincidence problem which essentially states: why each of these contributions happen to be of the
 same order by the time of galaxy formation. For a recent review see \cite{straumann}

\end{itemize}

All of these problems are strong motivations to guide us into the 
possible ways to modify both the standard models of particle physics and
 cosmology. Furthermore,
 their extensions could not only offer solutions to these problems but generate new ones also. For instance grand 
unified models typically imply the existence of topological defects such as domain walls, cosmic strings and monopoles 
which could have an important impact 
in cosmology. In particular the existence of monopoles and domain walls would
over-close the universe and therefore cause new problems, named the domain wall and monopole problems.
 Cosmic strings on the other hand were thought to
be useful for galaxy formation, although by themselves would predict an 
spectrum of density perturbations which do not fit the CMB results. 
  
\subsection{Inflation}

More than 20 years ago, the inflationary universe was proposed
\cite{guth}, offering 
a possible solution to the flatness, horizon  and monopole problems.
It was soon realised that, more importantly, it could also 
provide an explanation for the possible CMB
 anisotropies and therefore for structure formation.

The main idea behind inflation is that in the early universe there is a short 
time when the universe expanded very fast, usually an exponential expansion.
If the inflationary period is long enough, it would flatten the universe 
quickly (solving the flatness problem), it would also explain why some regions
could be in causal contact with each other, solving the horizon problem. 
Finally the fast expansion would dilute many objects, such as monopoles and 
other 
unwanted massive particles in such a way as to make them harmless for the 
over-closure of the universe. 

The simplest realization of inflation is to introduce a scalar field $\psi$ with 
a potential $V(\psi )$, the value of the potential provides an effective cosmological constant. 
If it is flat enough then we would be in a situation similar to the case 
$w=-1$ for which we already saw that the scale factor $a(t)$ increases 
exponentially.

The Friedmann's equation for this system becomes:
\eq
\label{frinfl}
H^2\ = \ \frac{8\pi G}{3} \ \left( V+ \frac{\dot\psi^2}{2} \right)
- \frac{k}{a^2}\ , {\label{finfl}}
\eeq
whereas the scalar field equation is
\eq
\label{inflaton}
\ddot\psi\ - \ 3 H\dot\psi\ = \ -V'\ . 
\eeq
Where $V(\psi)$ is the scalar field potential and $V'\equiv dV/d\psi$.
Notice that the second term in this equation acts like a friction term for 
 a harmonic oscillator (for a quadratic potential) with the friction determined by the Hubble parameter $H$.

The right hand side of equation (\ref{finfl}) is the energy density due to 
the scalar field $\psi$. We can easily see that if the potential energy 
dominates over the kinetic energy and $V\sim \Lambda>0$ we have the 
$w=-1$ case with exponential expansion
 $a\sim e^{Ht}\sim exp(\sqrt{\Lambda/3}) $ in units of the Planck mass, for
 $k=0$, and similar
 expressions for other values of $k$. The important point is that the scale
 factor increases exponentially, therefore solving the horizon, flatness and 
monopole problems.

The  conditions for inflation to be realised can be summarised in two 
useful equations, known as the slow roll conditions:
\eqa
\epsilon & \equiv & \frac{M_{Planck}^2}{2}\ \left( \frac{V'}{V}\right)^2\  
\ll  \ 1\ , \\
\eta & \equiv & M_{Planck}^2\ \frac{V''}{V}\  \ll  \ 1\ . 
 \eeqa

The parameters $\epsilon $ and $\eta$ have become the standard way to 
parametrise the physics of inflation. If the first condition is satisfied 
the potential is flat enough as to guarantee an exponential expansion.
If the second condition is satisfied the friction term in 
equation (\ref{inflaton}) dominates and therefore implies the 
slow rolling of the field on the potential, guaranteeing the inflationary 
period lasts for some time.

If the potential were a constant we would be in a de Sitter universe
expansion and the amount of inflation would be given by the size of $H$
(since $a(t)\sim e^{Ht}$). More generally the number of e-foldings is 
given by:

\eq
N(t)\equiv \int_{t_{init}}^{t_{end}} H(t') dt'\ = \int_{\psi_{init}}^{\psi_{end}}
\frac{H}{\dot\psi} d\psi\ =\ \frac{1}{M_{Planck}^2}
\int_{\psi_{end}}^{\psi_{init}}
\frac{V}{V'} d\psi\ . 
\eeq
 
A successful period of inflation required to
solve the horizon problem needs at least $N\geq 60$.
The recipe to a successful model of inflation is then to find a 
scalar field potential $V$ satisfying the slow roll conditions in such a way 
that the number of e-foldings exceeds $60$ (slightly smaller values are
 sometimes allowed depending on the scale that inflation occurs). 
It is of no surprise that many potentials have been
 proposed that achieve this.
For a collection of models see for instance the book of Liddle and Lyth in 
\cite{Rev}. Usually getting a 
potential flat enough requires
certain amount of fine tuning unless there is a theoretical motivation for
 the potential. It is fair to say that 
at the moment there are no compelling candidates.

\EPSFIGURE[r]{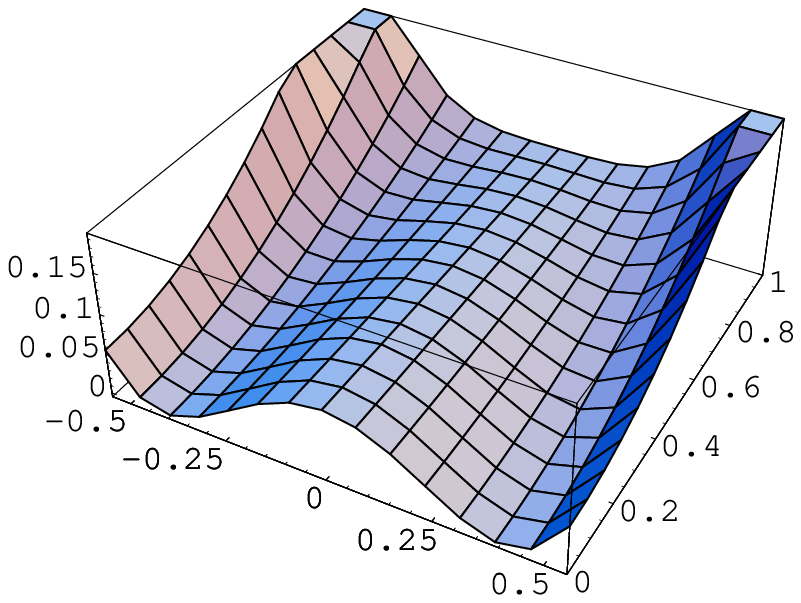,width=8cm}{A typical potential for hybrid
 inflation, an inflaton field rolls
 slowly, 
for a critical value the other field becomes massless and then
 tachyonic constituting the
 direction of larger
 gradient and ending inflation.}
{\label{figure3}}

The scenario that has been used recently as a concrete paradigm is  hybrid 
inflation, introduced by Linde in 1991 \cite{hybridinflation}. The idea here is to
 separate the inflaton with the ending of inflation. This means that there
 are at least two fields. The inflaton having a flat potential that satisfies
 the slow roll conditions and then a second field for which its mass 
depends on the inflaton field in  such a way that before and during inflation 
the squared mass is positive but then after inflation the mass squared 
becomes negative and the field becomes tachyonic, 
signalling an instability in that direction. This means that the stationary
 point for this field is now a maximum instead of a minimum and the field 
wants to roll fast towards its true vacuum, ending  inflation. 
Solving in this way the `graceful exit' of inflation problem. This scenario permits that 
the inflaton field does not have to take values larger than the Planck scale as
usually happens in single field potentials. Furthermore it is easier to realize
in concrete examples either in supersymmetric theories and, as we will see,
in string theory.

A typical potential for these fields takes the form:
\eq
V(X,Y)\ = \ a\left(Y^2-1\right) X^2 + b X^4 +c
\eeq
with $a,b,c$ suitable positive constants. 
We can easily see that for $Y^2>1$ the field $X$ has a positive mass$^2$, at
$Y^2=1$, $X$ is massless and for $Y^2<1$ the field $X$ is tachyonic,
 with a potential similar to the Higgs field. Therefore the potential in the 
$Y$ direction can be  very flat, see  figure~3, and $Y$ can be identified
 with the inflaton field $\psi$. The tachyon field $X$ is responsible for 
finishing inflation since it provides the direction of maximum gradient 
after it becomes tachyonic. Notice that adding more fields usually does not help
into improving the conditions for inflation, since the direction of maximum
 gradient is at the end the dominant. What helps in this case is that the 
field $X$ changes from being massive to tachyonic and then allows 
$Y$ to roll slowly and induce inflation.

Probably the most relevant property of inflation is that it can provide an 
explanation for the density perturbations of the CMB and therefore indirectly 
account for the large scale structure formation. Quantum fluctuations of
 the scalar field give rise to fluctuations in the energy density that at the 
end provide the fluctuations in the temperature observed at COBE \cite{Rev,riotto,Mukhanov}. 
Furthermore most of the models of inflation imply a scale invariant,
 Gaussian and adiabatic spectrum which is consistent with
 observations. This has made inflation becoming the standard cosmological 
paradigm to test  present and future observations.  

The typical situation is that any scale, including the perturbations,
will increase substantially during inflation whereas the Hubble scale
remains essentially constant. Therefore the scale will leave the
horizon
(determined essentially by $H^{-1}$) and the fluctuations get frozen. After inflation, the Hubble scale will
increase faster and then the scales will re-enter the horizon.
The amplitude of the density perturbation ($\delta\rho/\rho$)
 when it re-enters the
horizon, as observed by Cosmic Microwave Background (CMB) experiments
 is given by:
\eq\label{deltaH}
\delta_H = \frac{2}{5} {\mathcal P}_{\mathcal R}^{1/2} =
   \frac{1}{5 \pi \sqrt{3}}\,{V^{3/2}\over M_p^3\,V'}= 1.91\times 10^{-5}\,,
\eeq
where ${\mathcal P}_{\mathcal R}$ is the power spectrum computed 
in terms of the two-point correlators of the perturbations.
Here the value of $\delta_H$ is implied by the COBE
 results~\cite{COBE}.

In order to study the
 scale dependence of the spectrum, whatever its form is, one can
 define an effective {\it spectral index} $n(k)$ as $n(k)-1 \equiv
 \frac{d\,\ln{{\mathcal P}_{\mathcal R}}}{d\,\ln{k}}$. This is
 equivalent to the power-law behaviour that one assumes when defining
 the spectral index as ${\mathcal P}_{\mathcal R}(k)\propto k^{n-1}$
 over an interval of $k$ where $n(k)$ is constant. One can then work
 $n(k)$ and its derivative by using the slow roll conditions defined
 above \cite{riotto,Rev}, and they are given by
\eq\label{sindex}
n-1 = \frac{\partial\ln{\mathcal P}_{\mathcal R}}{\partial\ln k }
       \simeq 2\eta - 6\epsilon  \,,\qquad\qquad
{dn\over d\ln k} \simeq 24 \epsilon^2 - 16\epsilon\eta +2\xi^2 \,.
\eeq
%where $k$ is the length scale%
where $\xi^2\equiv M^2_P \frac{V'\,V'''}{V^2}$. Showing that for slow rolling
($\eta,\epsilon \ll 1$) the spectrum is almost scale invariant ($n\sim 1$).

The gravitational wave spectrum 
can be calculated in a similar way. The
gravitational spectral index $n_{grav}$ is  given by
\eq\label{tensorpert}
n_{grav}= {d\ln{\mathcal P}_{grav}(k)\over d\ln k} =
         -2 \epsilon \,.
%\ll 1
\eeq
Therefore we have a simple recipe to check if any potential can give
rise to inflation: compute the parameters $\epsilon,\eta$, check the
slow roll conditions, if they are satisfied we can right away find the
spectral indices and the COBE normalisation (\ref{deltaH}) puts a constraint on
the parameters and scales of the potential.

Finally, quantum fluctuations can move the field up the potential,
providing more inflation. This gives rise to eternal inflation since
parts of the universe will keep expanding forever, each of them
releasing the field to a lower value of the potential, which will lead
to standard inflation, in a process that
induces a self-reproducing universe (see for instance \cite{lindeekp}
and references therein).

\section{Pre D-branes String Cosmology}

Since the mid 1980's some effort has been dedicated to the cosmological 
implications of string theory. One possible approach was to look at
cosmological solutions of 
the theory starting from a 10D effective action. 
A collection of many of these solutions can be seen in \cite{wittenbh}.
They correspond to solutions of Einstein's equations in the presence of
dilaton and antisymmetric fields with a
 time dependent metric.
Some solutions were found for which some dimensions expand and others contract.
Furthermore, starting in 1991, Witten found an exact conformal field theory 
corresponding to a coset $SL(2,R)/U(1)$ that written in terms of the
WZW action gave rise to the metric of a 2D black hole
\cite{wittenbh}. This opened the way towards looking for non trivial spacetimes as exact CFT's in 2D by investigating 
different cosets \cite{tv,gq,nw}. One interesting observation was made in 
\cite{kl} for which changing the sign of the Kac-Moody level provides a 
spacetime for which time and space were interchanged, and therefore a
black hole geometry turned into a cosmological
 one. A similar structure has been found   recently and we will mention it in section 4.6.
  
Besides looking for time-dependent solutions \cite{copeland}, there were several 
interesting issues discovered at that time. 
We can summarise the main results of those investigations as follows.

\subsection{Brandenberger-Vafa Scenario}

In 1987, $T$-duality was discovered in string theory (for a review see
\cite{tduality}). This refers to the 
now well known
$R\rightarrow 1/R$ symmetry of the partition functions and mass spectrum 
of string theories compactified in a circle
of radius $R$. 
In particular the bosonic and heterotic strings are known to be self-dual 
under 
this transformation. The mass formula takes the form
\eq
M^2\ = \  \frac{n^2}{4R^2} + m^2 R^2 + N_L + N_R -2\ . 
\eeq
In units of the inverse string tension $\alpha'=1/2$. The integers $n$ and
 $m$ give the quantised momentum in the 
circle and the winding number of the string in the circle, $N_{L,R}$ are the 
left and right oscillator numbers.
We can easily see that the spectrum is invariant under the simultaneous
 exchange $R\leftrightarrow 1/2R $
and winding and momenta $n\leftrightarrow m$. This symmetry has had many
 important implications in the development of string theory. Regarding
 cosmology, 
 Brandenberger and Vafa soon realised that it could have interesting
 applications. First they emphasised that
the concept of distance has different interpretation in the two dual 
regimes. There is a minimum distance in string theory.
 \footnote{This
 statement has been modified in the
last few years due to the fact that for instance D0 branes can probe
 distances smaller than the string scale.
 This has no direct implications for the argument we are presenting
 here.} 
 At large radius the
position coordinate is the conjugate variable to momentum $p= n/R$, 
as usual. But 
 at distances smaller than the self-dual radius
we have to use the dual coordinate which is the conjugate variable to
 winding $W= mR$. 
There is no sense to talk
 about distances smaller than the string scale, 
since they will be equivalent to large distances. Brandenberger and Vafa 
claimed that if the universe is though to be a 
product of circles this may be a way to eliminate the initial singularity
 also \footnote{In their original article it 
was claimed that Einstein's equations were not symmetric under the $T$
 duality symmetry. However at that time it was not 
realized that
the equations are rendered invariant by the appropriate transformation of the 
dilaton.}.

The second interesting observation of BV was that this could provide a 
dynamical explanation of the
reason why our universe looks four-dimensional. The argument goes like this:
 imagine that the universe starts with all 
spatial dimensions of the string size, then the existence of winding modes
 will prevent the corresponding dimension
from expanding (imagine a rope wrapping a cylinder). However winding modes
 naturally annihilate with anti-winding modes.
In the total ten dimensions a winding string will naturally miss to meet the
 anti-winding string just because their 
world-sheets can have many different trajectories in ten dimensions. However 
these world-sheets naturally overlap in a 
four-dimensional hypersurface of the total space in which winding and 
anti winding strings can annihilate and allow the expansion of the 
three spatial 
dimensions. This could then mean that
 the winding strings will prevent six of the
 dimensions from expanding and will leave three spatial dimensions to expand.
 This intuitive argument has been
 recently revived to include the presence of higher dimensional objects such
 as 
D-branes \cite{BV}. The claim is that even in the configuration of a gas of D-branes of
 different dimensionalities, the
original BV argument still holds since the string winding modes are still the
 most relevant for the
expansion argument \cite{brandenberger}.

Even though this a very rough argument, that needs much refinement before it
 can be taken too seriously,
 it is essentially 
the only concrete proposal so far to approach the question of why 
we feel only  three large  dimensions. Only because of this reason it 
deserves further investigation. One 
criticism to this is the assumption that the dimensions are toroidal, 
something that generically is not
 considered very realistic for both our spatial dimensions and the extra ones.
 Regarding the assumptions
 in the extra dimensions, recently   intersecting D6
 brane models with realistic
 properties in toroidal compactifications were constructed. The chirality comes from the
 intersection even though the 
background space, being a torus, was not expected to give realistic
 models.
 Furthermore, there has been an
 attempt to extend the result of BV to more realistic compactifications such
 as
orbifolds \cite{greene}. 

\subsection{Moduli and Inflation}

\EPSFIGURE[r]{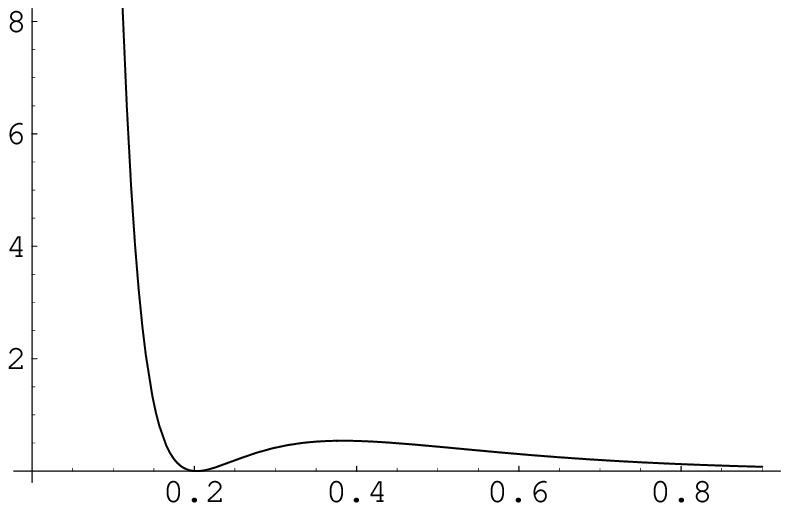,width=6cm}{A typical potential for the dilaton
 field. It has a runaway minimum on which the theory is free and a nontrivial
 minimum at finite values. The potential is too steep to produce inflation 
and prevents any other field from becoming the inflaton. It is hard to stop the 
field from passing through the finite vacuum towards the one at infinity.}
{\label{figure2}}

One of the few things that can be called a prediction of string theory is the
 existence of 
light scalar particles with gravitational strength interactions such as the
 complex dilaton
field $S$  and
the moduli fields that describe the size and shape of the extra dimensions,
 which we can refer generically as 
$T$. In supersymmetric string models these fields are completely undetermined 
reflecting the 
 vacuum degeneracy problem. In the effective field theory this is realised by
 the fact that those fields have 
vanishing potential to all orders in perturbation theory due to the 
non-renormalisation   theorems of supersymmetric
field theories. This is usually taken to be an artifact of perturbation 
theory and the potentials are
 expected to be lifted hopefully breaking the continuous vacuum degeneracy
 and fixing the value of the
 moduli to the preferred
phenomenologically (the dilaton leading to weak coupling and the
 radius larger than the string scale).
However nonperturbative potentials are not well understood and the fixing of
 the moduli remains as one of the main open 
question in string models.

The fact that the potential is flat to all orders in perturbation theory
 may be an indication that after the breaking of 
supersymmetry the potential may be flat enough as to satisfy the slow roll 
conditions and generate inflation. 
Therefore string theory naturally provides good candidates for inflaton 
fields. However this has not been realised
 in practice.

 A detailed study of the general properties expected for the 
dilaton potential was performed in
 \cite{brustein} with very negative conclusions. The main problem is that
 nonperturbative superpotentials in string
 theory are expected to depend on $e^{-aS}$ which give rise to runaway
 potentials, in specific scenarios such as the 
racetrack case on which the superpotential is  a sum of those exponentials,  
similar to the one in  figure~4,
  not only are too steep for inflation but also does not allow any other 
field to be the inflaton field since that will 
be the fastest rolling direction 
towards the minimum. Furthermore the nontrivial
 minimum creates a problem since if the field 
configuration starts to the left of the minimum it would naturally roll 
through the minimum towards the runaway 
vacuum which 
corresponds to zero string coupling. Instead of helping to solve cosmological
 problems, the dilaton potential 
actually creates a new one. A proper treatment of the situation in a
 cosmological background can cause enough
 friction as to stop the field in the nontrivial minimum as suggested
 in \cite{beatriz}, see also \cite{steinhardt}.

\subsection{The Cosmological Moduli Problem}

A more generic and probably more serious problem was pointed out in
 \cite{cmp} (see also \cite{polony}). This is the so-called
`cosmological moduli problem'. This refers in general to any scalar 
field that has gravitational strength interactions
and acquires a nonzero mass after supersymmetry breaking. It was shown 
in general \cite{cmp},  that independent of the 
mechanism for supersymmetry breaking, as long as it is mediated by gravity, 
the masses of the dilaton and moduli fields
are expected to be of the same order as the gravitino mass. This causes a 
serious cosmological problem because if the 
scalar field happens to be stable, it will over-close the  universe. Otherwise
if the 
field decays, it will do it very late in the
history of the universe due to the weakness of its couplings. This will then
 ruin the results of nucleosynthesis, like destroying $^4 He$ and $D$
 nuclei and therefore changing their relative abundance.

Notice that we may say the same about the gravitino and the fermionic 
partners of the moduli, since they are also expected to get a mass 
of order $m\sim 1$ TeV. If they are stable, the could over-close the universe
 unless their mass is of order $m\sim 1$ keV. If unstable, 
their decay rate would then be 
$\Gamma\sim m^3/M_{Planck}^2$, 
imposing that they decay after nucleosynthesis, otherwise their decay
 products give unacceptable alteration of the primordial $^4$ He and D 
abundances, imposes a lower bound on their masses of 
$m> 10$ TeV, which is already a bit high.
 However for fermionic fields the problem can 
be easily cured
 by diluting the fields by inflation in a similar way that inflation solves
 the monopole problem. This can be done as long as the reheating temperature
after inflation
satisfies 
\eq
T_{RH}\ \lessapprox 10^8\left(\frac{100 {\rm GeV}}{m_{3/2}}\right)\ {\rm GeV}\ . 
\eeq

The problem is more serious for scalar fields because
 even after inflation the scalar field  $\phi$ 
will naturally be displaced from its
equilibrium position by an amount $\delta\phi$ and oscillations around the 
minimum of its potential rather than thermal production are the
main source for their energy. After inflation the field behaves like
 non-relativistic matter and therefore its energy density decreases with
 temperature
$\rho\sim T^{-3}$ whereas radiation decreases faster $\rho_{rad}\sim T^{-4}$
 so these fields dominate the energy density of the universe $\rho/\rho_{rad}
\sim 1/T$ as the universe cools down. From their many couplings to the 
standard model fields it is expected that these moduli fields will decay,
 but again since their interactions are of gravitational strength this occurs
 very late, again their decay modes destroying the $^4$ He and D nuclei and 
the successful
 predictions of nucleosynthesis unless their mass satisfies $m> 10$ TeV.
This again does not look very strong constraint, however now inflation does not
 help and furthermore, the decay of the scalar field
  leads to an entropy increase of the order
\eq
\Delta\ \sim \ \frac{\delta\phi^2}{m M_{Planck}}\ .
\eeq
If $\Delta$ is very large this would erase any pre-existing baryon asymmetry,
 therefore this condition requires $\delta\phi \ll M_{Planck}$ which is not
 natural given that the moduli fields are expected to have 
thermal and/or quantum fluctuations that can be as large as 
 the Planck scale.

Even though this problem has been discussed at length during the past 8
 years, it is fair to say that there is not  
a completely satisfactory solution. Inflation at low energies and low energy
 baryogenesis ameliorate the problem considerably \cite{cmp, moduliinflation}.
Probably the best proposal for the 
solution is thermal inflation \cite{edetal}. 
A simple way out would be if the moduli
 fields are fixed at the string scale and then supersymmetry is broken at
 low energies, in that case these fields do not survive at low energies and 
the problem does not exist. Otherwise this is an important
 challenge for any realistic effort in string theory cosmology.

\subsection{Pre Big-Bang Scenario}

A natural next step from the BV proposal is to consider the possibility of
 $T$ duality in backgrounds closer to  the FRW type. For this let us 
first consider 
a low energy string effective action including the metric $g_{MN}$, 
the dilaton $\varphi$ and the NS-NS antisymmetric tensor of the bosonic
 and heterotic strings $B_{MN}$. In an arbitrary number of dimensions $D=d+1$
the bosonic action takes the form:
\eq
S\ = \ \int d^{D}x\sqrt{-g}\ e^{-\varphi}\left( R + \partial_M\varphi
\partial^M\varphi -\frac{1}{12} H_{MNP} H^{MNP}+\cdots \right).
\eeq
Where  $H=dB$. 
For an ansatz  of the type:
\eq
ds^2= -dt^2 + \sum_i^d a_i^2(t)\ dx_i^2
\eeq
we can easily see that $T$ duality is a symmetry of the 
equations of motion acting as:
\eq
a_i(t)\rightarrow \frac{1}{a_i(t)}
 \qquad \varphi\rightarrow \varphi - 2 \sum_i\log a_i
\eeq
Since $a_i(t)$ represent in this case the scale factors, like in FRW, this 
has been named scale factor duality. Thus we can see that expanding and
contracting universes are related by this symmetry.

Furthermore,  Veneziano and collaborators \cite{pbb}, realised that this symmetry can be
 combined with the standard symmetry for this kind of backgrounds
 corresponding to the
exchange:
\eq
a(t)\leftrightarrow a(-t)
\eeq
This simple observation opens up the possibility of considering a period 
before $t=0$ for which the Hubble parameter increase instead of decrease.
That is without duality the symmetry under $t\rightarrow -t$ would send
$H(t)\rightarrow -H(-t)$ but combining
 this with duality provides four different sign 
combinations for $ H(t)$. 
If the universe at late times is decelerating $H$ would be a decreasing
 monotonic function of time for `positive' $t$, then a combination of duality
 and the 
$t\rightarrow -t$ transformation can give rise to an $H(-t) = H(t)$ so 
that this
 function can be  even, see figure~5. So we can see that there is  a
 possible scenario in which the universe accelerates from negative times
 towards the big bang and then decelerates after the big-bang. The 
acceleration would 
indicate a period of inflation before the big-bang without the need of an
 scalar
 potential. 

A concrete solution for this system corresponds to the isotropic case
$a_i=a_j\equiv a(t)$ for which:
\eq
a(t)= t^{1/\sqrt{d}}\qquad t>0\ , 
\eeq
with a constant dilaton. For this $H(t)\sim 1/t$ decreases monotonically 
with time. By applying the transformation $t\rightarrow -t$ and duality we
 can generate the four different branches of solutions:

\eqa
a(t) & = & t^{\pm1/\sqrt{d}}\qquad t>0\nonumber\\
 & = & \left(-t\right)^{\pm1/\sqrt{d}}\qquad t<0\ .
\label{ven6}
\eeqa
With 
\eq
\varphi_\pm (\pm t)\ = \ \left(\pm\sqrt{d} - 1\right) \log(\pm t)\ .
\label{ven7}
\eeq

\EPSFIGURE[r]{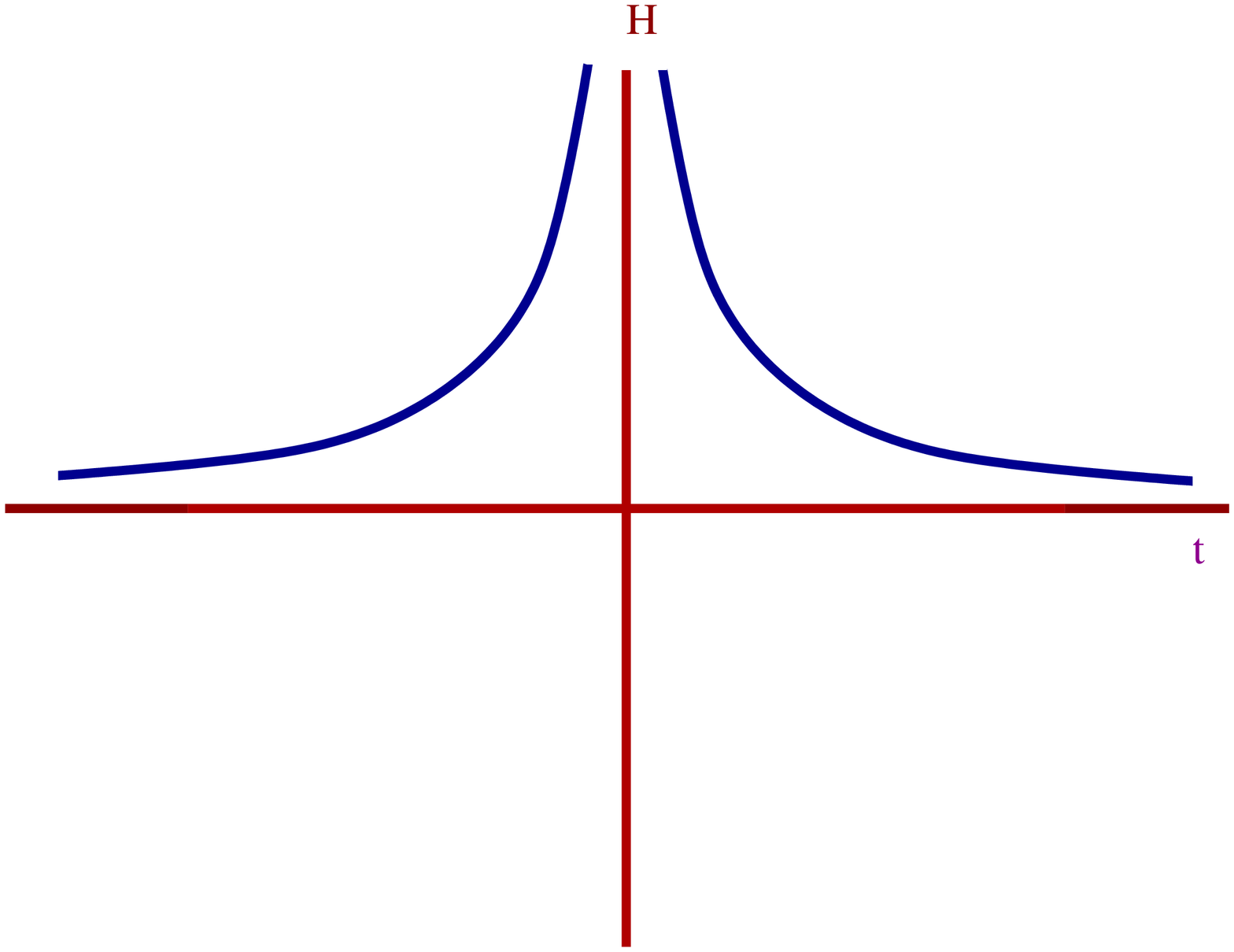,width=8cm}{Possible realisation of the pre big bang scenario with the past and future regions expected to match at the singularity with strong coupling and large curvature}
{\label{figure4}}

The two branches for which the universe expands $H>0$ provide an 
interesting realization of the pre big-bang scenario, see  figure~5.
 Notice that 
the solutions are such that have a singularity at $t=0$ but also in this
 region the dilaton blows up implying strong coupling. 
It is expected that nonperturbative string effects would provide a smooth 
matching between these two branches. The weak 
coupling perturbative string vacuum appears as a natural initial condition 
in the pre big-bang era. Therefore the scenario consists of an empty cold
 universe in the infinite past that expands in an accelerated way 
towards a region of higher curvature until it approaches the region of 
strong coupling and large curvature which is assumed will match smoothly
 to the post big bang branch in which the universe continues expanding but
 decelerates.

This cosmological 
string scenario is probably the one that has been  subject to more 
detailed     
investigation during the past decade. It has several attractive features
 such as the 
possibility of having a period before the big-bang helping to provide the 
initial conditions and giving rise to an alternative to scalar field inflation, with the advantage of being motivated by string theory.
Furthermore a study of the density perturbations for this scenario 
has attracted alternatives to inflation. The spectrum  of density 
perturbations has been estimated and claimed not to contradict the recent 
observations.
Also it provides testable differences with respect to the tensor 
perturbations that could be put to test in the future.

The scenario has been also subject to criticism for several reasons. First, as
 the authors point out,
 the main problem to understand is the graceful exit question,
 that is how to pass smoothly from pre to post big-bang period. 
The argument is that close to the big-bang the perturbative treatment of 
string theory does not hold since the dilaton and the curvature increase,
 implying strong string coupling. Therefore there is no concrete way to
 address this issue in the framework that the theory is treated. 
Another important
 problem is the fact that the moduli are neglected from this analysis and 
there has to be a mechanism that stabilises the extra dimension. This is a
 standard problem in string theory so it is not particular of this scenario.
 Also 
the scale factor duality symmetry that motivated the scenario is not
 clearly realised
in a more realistic setting with nontrivial matter content. The fact that the 
dilaton will eventually be fixed by nonperturbative effects may change the 
setting of the scenario. 

Furthermore, 
issues of fine-tuning have been pointed out in the literature \cite{turner,
linde} as well as not reproducing the  CMB spectrum.
For this, the density perturbations coming from the dilaton, 
are not scale invariant with a blue power spectrum
 with  too small to account for the COBE data. Considering an axion
 field, dual to the NS-NS antisymmetric tensor  of string theory, 
does not work in principle since even though the spectrum is scale
 invariant, the perturbations are isocurvature instead of adiabatic,
 implying a pattern of acoustic peaks different from what has been observed,
 especially at BOOMERANG and DASI. A possible way out has been
 proposed for which a nonperturbative potential for the axion is
 expected to be generated after the pre-big bang era. In this case,
 under the assumptions that the axion field is away from its minimum
 after the big-bang and it dominates the energy density before
 decaying, the perturbations change from isocurvature to adiabatic
 after the decay of the axion. This general mechanism has been named `curvaton'\cite{bozza,copeland,veneziano}. On the other hand tensor perturbations have a
 blue spectrum and could be eventually  detected at antennas or
 interferometers. This could be a way to differentiate, observationally,
 this mechanism from standard scalar field inflation. Finally,
 electromagnetic perturbations are generically amplified in this
 scenario due to their coupling to the dilaton, something that may
 eventually be tested. This enhacement is interesting since it could be present
 in other string theory scenarios where the dilaton plays a
 cosmological role.

It is not clear if this scenario can be promoted to a fully realistic
early cosmological framework.  Nevertheless this is an interesting effort that 
deserves further investigation, it has kept the field of string cosmology 
active for several years, it has resurrected the old ideas of 
having cosmology before the big-bang \cite{cyclicold} (remember that 
even eternal inflation needed a beginning of time), in a string setting
 and has influenced
in one way or another the recent developments in this area.

%\cleardoublepage
\EPSFIGURE[r]{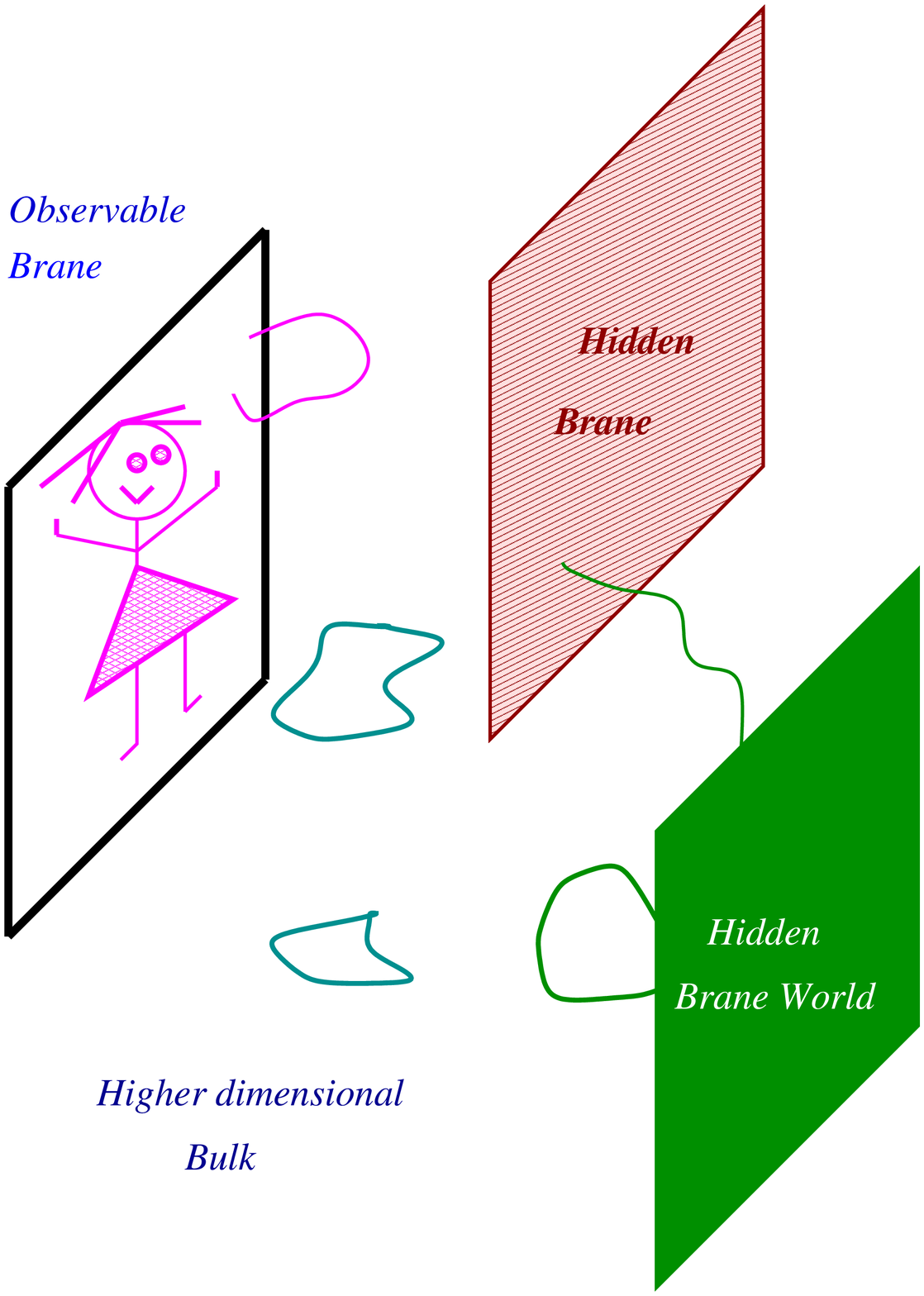,width=8cm}{The Brane World. A picture of the brane-world scenario.
 Open strings (including matter and gauge fields) are attached to 
branes, whereas closed strings (including gravity and other
 fields, like the dilaton) can propagate in the bulk.}
{\label{figure6}}
%\clearpage

\section{Brane Cosmology in String Theory}

The discovery of D branes and the Horava-Witten scenario, have opened the
 way to the realisation 
 of the brane-world scenario in string theory. The typical situation in type IIA, IIB 
and I strings is that there
may be many D branes sustaining gauge and matter fields, at least one of 
them should include the Standard
 Model of particle physics where we would be living. All this is due to the 
open string sector of the theories 
for which the end points  are
constrained to move on the brane. The closed string sector including gravity 
and the dilaton probes all the
 extra dimensions. For spacetimes resulting from a product geometry, the
 effective Planck mass in 4D is given by 
\eq
M_{Planck}^2\ \sim  \ M_s^8 R^6\ ,
\eeq
with $R$ the size of the extra dimension and $M_s$ the string scale.
This is at the source of the claim that large extra dimensions allow the
 possibility of small
$M_s$ as long as the Planck mass is fixed to the experimentally known value.

In five dimensions, Randall and Sundrum generalised this adding a warp 
factor to the metric depending on the extra 5th dimension:
\eq
ds^2\ = \ W(y)\ g_{\mu\nu} dx^\mu dx^\nu + dy^2,
\eeq
where $y$ is the  fifth dimension and $W(y)$ is the warp factor.
This metric shares the same symmetries as the direct product case $W=1$
but allows now the possibility of $W(y)$ playing a role. In particular if
 there are branes at locations fixed in $y$, they will feel different scale in their metrics due to different values of $W(y)$, Randall and Sundrum found an exponential dependence in $W$ that made the scales change very fast,
 allowing for the
 possibility of small fundamental scale
even for not that large radii. Furthermore they even discovered that starting
 in 5D anti de Sitter space and 
fine tuning the cosmological constant it is possible to have infinitely large
 extra dimensions and having 
gravity  localised in the brane. These discoveries triggered a large amount
 of interest, beyond the string 
community, on the physical implications of the extra dimensions and rapidly
 included cosmology. 

In string theory it was found that there are explicit realisations of the brane world in at least three different ways.
\begin{itemize}
\item
Having branes trapped in singularities supports 
chiral fermions and nonabelian gauge symmetries
with $N=1,0$ supersymmetry 
and then allows for the possibility of having the standard model in one 
stack of branes at singularities \cite{ads}. 
Realistic models with $N=1,0$ supersymmetry
 have been found \cite{aiq}\ with the small string scale $M_s\sim 10^{12}$ GeV or 
even $M_s\sim 1$ TeV.

\item
D-branes intersecting at nontrivial angles
\cite{doug}. The intersection of the branes can have chiral fermions and therefore there can be found explicit 
realistic models in many possible bulk backgrounds, including torii.
Again the standard model lives on D-branes and gravity in the 
bulk realising the brane world scenario. Nonsupersymmetric models require the string scale close to $1$ TeV, whereas supersymmetric models may have a larger scale \cite{intersec}.

\item
Horava-Witten scenario in which 11D M-theory compactified in one interval
\cite{braneworld}.
The two 10D surfaces at the endpoints have $E_8$ gauge theories 
and when compactified to 4D give rise to a brane world also \cite{ovrut}. 
Compactifications of M-theory in terms of $G_2$ holonomy manifolds, although 
not constructed explicitly, are known that in order to have chiral
fermions, 
the matter needs
to be at singular points, therefore implying that if constructed they 
may have also a brane world structure \cite{g2}.

\end{itemize}
Therefore the brane world appears naturally in string constructions and
its cosmology can be studied bearing in mind each of the particular string realisations.

\subsection{Brane Cosmology in 5D}

Binetruy et al \cite{branecosmology},
 made the first concrete description of a 5D cosmology with branes and found interesting results.
 We will briefly describe here some of the developments in this directions but only as an introduction to the 
real topic of these lectures which is brane cosmology in string theory. Therefore we will limit to mention the 
developments in that direction that will be used in the next sections.

Let us start with the simple action:
\eq
S \ = \ S_{Bulk}\ + \ S_{Brane}\ = \ \int d^5x\sqrt{-g_5}\left[R \ - \ \Lambda\right]-\int d^4x\sqrt{-g_4}
\left(
{\cal K} +{\cal L}_{matter}\right)\ ,
\eeq
with ${\cal K}$ the trace of the extrinsic curvature and the rest in a self explanatory notation.
Looking for cosmological solutions from the bulk action we look at the most general solution which is homogeneous and
 isotropic in 4D. If we choose the most general metric depending on $t$ and the 5th dimension $y$,
 we can take the brane to be at  particular point in the space, say $y=0$. We can work 
the bulk metric in 
a conformal frame (where the $t,y$ part of the metric is conformally flat). In this case
the metric can be written as \cite{BCG}:
\eq
ds^2\ =  e^{2\nu(t',y)}B^{-2/3}(t',y)\left( -dt'^2 + dy^2\right) +
 B^{2/3}\left[\frac{d\chi^2}{1-k\chi^2}+\chi^2 d\Omega^2\right]\ .
\eeq
Where the functions $\nu$ and $B$ are  completely arbitrary. Now we state 
Birkhoff's theorem which essentially says that 
the most general homogeneous and isotropic metric depends on only one variable. This can be seen as follows.
Define the light cone coordinates:
\eq
u\ = \ \frac{t'-y}{2}\ , \qquad v\ = \ \frac{t'+y}{2}
\eeq
Einstein's equations reduce in these variables to 
\eqa
B_{uv} & = & \left(2\Lambda B^{1/3}\ - \ 6 k B^{-1/3}\right) e^{2\nu}\nonumber \\
\nu_{uv} & = & \left(\frac{\Lambda}{3}\ B^{-2/3}\ + \ k B^{-4/3}\right) e^{2\nu}\nonumber\\
B_u\left[\log B_u\right]_u & = & 2\nu_u B_u\nonumber\\
B_v\left[\log B_b\right]_v & = & 2 \nu_v B_v
\eeqa
This implies that $B=B\left[ U(u)+ V(v)\right]$ and $ e^{2\nu}= B'U'V'$ where primes refer to derivatives
 with respect to the variable that the function depends on. Without loss of generality 
we can fix $V(v)=v$ and setting 
$r=B^{1/3}$, $t=3(v-U)$ the first equation above can be integrated to give:
\eq
ds^2\   =   -\ h(r)\ dt^2\ + \ \frac{dr^2}{h(r)}\ + \ r^2 d\Omega^2
\eeq
with $h(r)= k-\frac{\Lambda}{6} r^2 - \frac{\mu}{r^2}$. We can identify this 
metric clearly as Schwarzschild AdS! This shows that even though we 
started with a time dependent  metric, the most general solution can be 
reduced to a static
solution.
As we said above this is just a statement of 
the  general Birkhoff's theorem.
 So far we have not yet used the brane.
However without performing any calculation we can see already the implications
 of this result for the brane cosmology. 
In our starting point the brane was assumed to be fixed at $y=0$ and then the 
metric was a function of time and $y$.
The time dependence providing for the cosmological nature of the metric.
 However we have found that the metric at the end is static. Where is the 
time dependence?. It so happens that with the change of variables we have made
 we can no longer claim that the brane is at a fixed value of $r$. The 
location of the brane is determined by some function describing its 
trajectory in this static background $r=r(t)$. Therefore the brane is 
moving in time and it is this motion that will appear as 
a cosmological evolution for an observer on the brane. This is quite 
remarkable, since we can imagine that  our present  feeling that the universe expands, 
could only be an illusion and actually our universe would be moving with 
some non vanishing 
 velocity in a higher dimensional, static bulk spacetime. This effect has
 been named mirage cosmology \cite{kraus,ida,BCG,keki}.

Notice that we could have worked all the time with the original coordinates
 $t,y$ and look for cosmological solutions. The end result is the same.
 In both procedures we need to take into account the presence of the brane by
 considering the Israel matching conditions, which can be derived in a
 straightforward way from standard techniques in general relativity, 
including the Gauss-Codazzi equations
\cite{wald}. Or simply we can use standard Green 
function techniques to match the solutions in the
brane, taken as a delta function source. It is usually simpler to assume a
 ${\bf Z}_2$ symmetry across the brane making the a cut and paste of the
 spacetime to make it symmetric around the brane. The Israel conditions 
then can be written as:
\eq
\left[{\cal K}_{ij}\right]^+_-\ = \ -\left( T_{ij} - 
\frac{1}{3} g_{ij} T^k_k\right)
\eeq
Where ${\cal K}_{ij}$ is the extrinsic
 curvature and $T_{ij}$ the energy momentum
 tensor. The $\pm$ refer to the two regions separated by the brane.
Imposing these conditions we arrive at the Friedmann's equations for the 4D
 brane. They give  the standard energy conservation
\eq
\dot\rho\ + \ 3\left(p+\rho\right)\frac{\dot a}{a}\ = \ 0
\eeq
with $a_0$ the scale factor on the brane. The first Friedman's equation 
takes an interesting form:
\eq
H^2\ \sim \ \rho^2 +...
\eeq
This instead of the standard behaviour $H^2\sim \rho$ in FRW. It was then 
suggested that brane worlds would give rise to 
non-standard cosmologies in 4D. This has been addressed in several ways. 
The most accepted argument at the moment is that
 we have to have a mechanism to fix the value of the 5th dimension, before
 making the comparison. Once this is achieved,
 the standard behaviour is recovered. It is also claimed that in general $\rho$
 has to be substituted by the contribution 
to the energy density of the brane and then instead of just $\rho^2$ we get
 $(\rho +\Lambda_4)^2$ which when expanding
  the squares recovers the linear term in $\rho$ which then, at late times, 
 will be dominant
 over $\rho^2$ since $\rho$ decreases in time.

Here we will follow Verlinde \cite{verlinde}\  who, in a very elegant way, recovered the Friedmann
 equation (see also \cite{renjie}). The idea is to use the AdS/CFT correspondence where the
 brane takes the place of
 the boundary (therefore the CFT becomes interacting).
The point is as follows. Start with the AdS$_5$ metric above and specify the
 location of the brane 
in parametric form $r=r(\tau)$, $t= t(\tau )$. We can choose the $\tau$
 parameter such that the following equation is satisfied:

\eq
\frac{1}{h(r)}\left(\frac{dr}{d\tau}\right)^2-
 h(r)\left(\frac{dt}{d\tau}\right)^2\ = \ -1
\label{constrain}
\eeq
this guarantees that the induced 4D metric in the brane takes the FRW form:
\eq
ds_4^2\ = \ -d\tau^2 \ + \ a^2(\tau)\ d\Omega_3^2
\eeq
with the scale factor $a(\tau)= r(\tau)$. This is already an interesting
 piece of information that the
scale factor of the brane is just the radial distance from the centre of the
 black hole (which is identified with the renormalisation group parameter in the field theory dual). Assuming that the 
matter Lagrangian in the brane is just a constant tension term with tension
 $\kappa$, the equation of motion for the
brane action gives simply
\eq
{\cal K}_{ij}\ = \ \frac{\kappa}{3}\ g_{ij}
\eeq
with $g_{ij}$ the induced metric on the brane. This then gives us for the 
metric above:
$\frac{dt}{d\tau}= \kappa r/h(r)$. Combining this with
(\ref{constrain})
and tunning the cosmological constant $\kappa^2=\Lambda/6$,  we
 get the Friedman equation:
\eq
H^2\ = \ -\frac{1}{a^2}\ + \ \frac{\mu}{a^4}
\eeq
which corresponds to the Friedmann  equation for  radiation
 ($\rho\sim 1/a^4$).
In reference \cite{verlinde}  this interpretation goes further by using the
 AdS/CFT correspondence identifying the 
radiation with the finite temperature CFT dual to the AdS solution. They also
 find general expressions for entropy and temperature. Finding in particular
 a general expression between the entropy and the energy density that 
generalises a result 
on 2D CFT by Cardy to the general dimensional case. This is known as the 
Cardy-Verlinde formula. This will not be used in what follows and we refer 
the reader to the literature for the details of this result.

Before finishing this section we may wonder if the general result used here
 about the Birkhoff's theorem holds in general in string theory. 
Unfortunately this is not the case once we introduce the dilaton field. 
It can be shown
 explicitly that Birkohffs theorem does not hold in this case and 
therefore the cosmological evolution of a brane will have two 
sources, one the motion of the brane and two the time dependence of the
 bulk background \cite{gqtz,unpublished}.

\subsection{D-Brane Inflation}

So far we have discussed string cosmology in a way  that does not make 
contact with the scalar field inflation which has been the dominant
 topic from the cosmology point of view. The difficulty with this is that 
as we mentioned before, there is not very much known about scalar 
potentials from string theory.
In most cases under control the potentials are just zero and the lifting by
 nonperturbative effects usually leads to runaway potentials or in general 
potentials which are too steep to inflate. It is then an open question as 
how to
derive inflating potentials from string theory.

\EPSFIGURE[r]{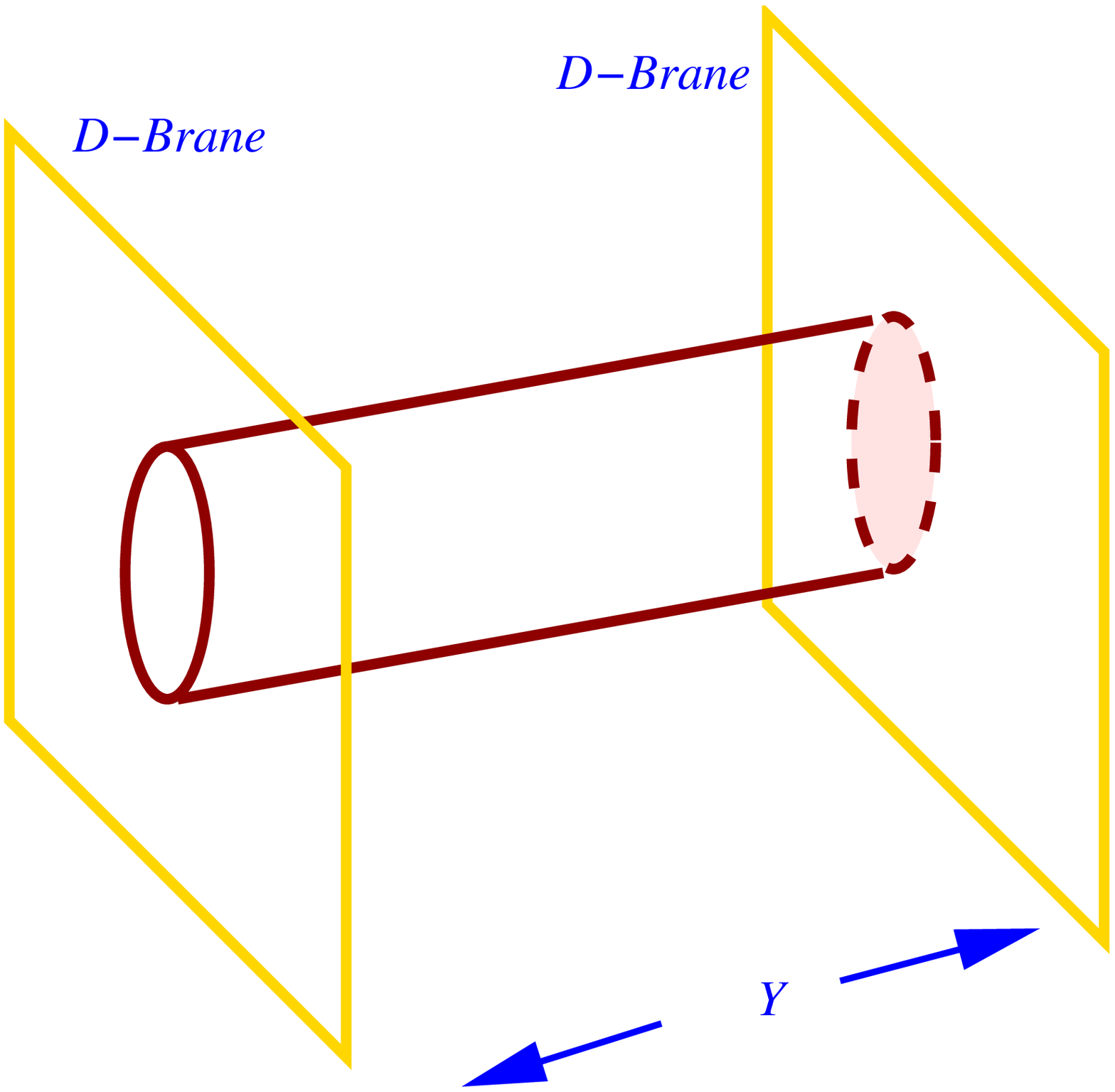,width=6cm}{Cylinder interaction between
 two branes. It could be in two dual ways, as 
a tree-level exchange of closed strings, valid at large distances or as
 a one-loop exchange of open strings, dominant at small distances.}
{\label{figure7}}

The interactions between D-branes offer a new avenue to investigate these
 issues and has led to  the first concrete examples of 
scalar field inflation from string theory, providing also a nice
geometrical and stringy picture of the inflationary process, as well as the 
ending of inflation. 
Furthermore, these ideas lead to interesting new cosmological scenarios
 for which inflation is only a part.

In 1998, Dvali and Tye came up with a very interesting proposal to 
derive inflation from D-branes. They argued that two D-branes could generate 
inflation as follows. If both branes are BPS, meaning that they preserve part of the 
original supersymmetry of the system, and satisfy a
 Bogomolnyi-Prasad-Sommerfeld bound, the net force between them
 vanishes. The reason for this is that both have a positive tension and,
 therefore, are naturally attracted to each other by gravitational 
interactions. Also the exchange of the dilaton field naturally leads to
 an attractive interaction.
 However, both branes are also charged under antisymmetric Ramond-Ramond 
fields for which the interaction is repulsive, given that both branes have
 the same charge. Therefore the combined action of the three interactions
 cancels exactly if the branes are BPS.

This calculation can be done explicitly, the interaction amplitude corresponds
 to the exchange of closed strings between the two branes. The amplitude can
 be computed by calculating the one-loop open string amplitude corresponding 
to a 
cylinder \cite{poli}:
\eq
{\cal A} = 2\int{\frac{dt}{2t} {\rm Tr} e^{-t H}}= 
2 T_p\int \frac{dt}{2t}\left(8\pi^2\alpha't\right)^{-(p+1)/2}
e^{-\frac{Y^2t}{2\phi^2\alpha'}}\left[Z_{NS}-Z_R\right]
\equiv {\cal }A_{NS} - {\cal }A_{R}
\eeq
With
\eqa
Z_{NS} & = & 
\frac{-16\prod_n\left(1+q^{2n}\right)^8+q^{-1}\prod_n\left(1+q^{2n-1}\right)^8}{\prod_n\left(1-q^{2n}\right)^8}
\nonumber \\
Z_R & = & \frac{q^{-1}\prod_n
\left(1-q^{2n-1}\right)^8}{\prod_n\left(1-q^{2n}\right)^8}
\eeqa
Here $q\equiv e^{-t/4\alpha'}$ and $t$ is the 
proper time parameter for the cylinder.
 $H$ is  the Hamiltonian for each sector of
 the theory. It is easy to see that 
$Z_{NS}=Z_R$. Therefore the interaction potential just vanished.
The cancellation between R-R and NS-NS sectors is a reflection of 
the BPS condition for the D-branes.
(This is the reason that D-branes can
 generally be stacked together increasing the gauge symmetry.)

\EPSFIGURE[r]{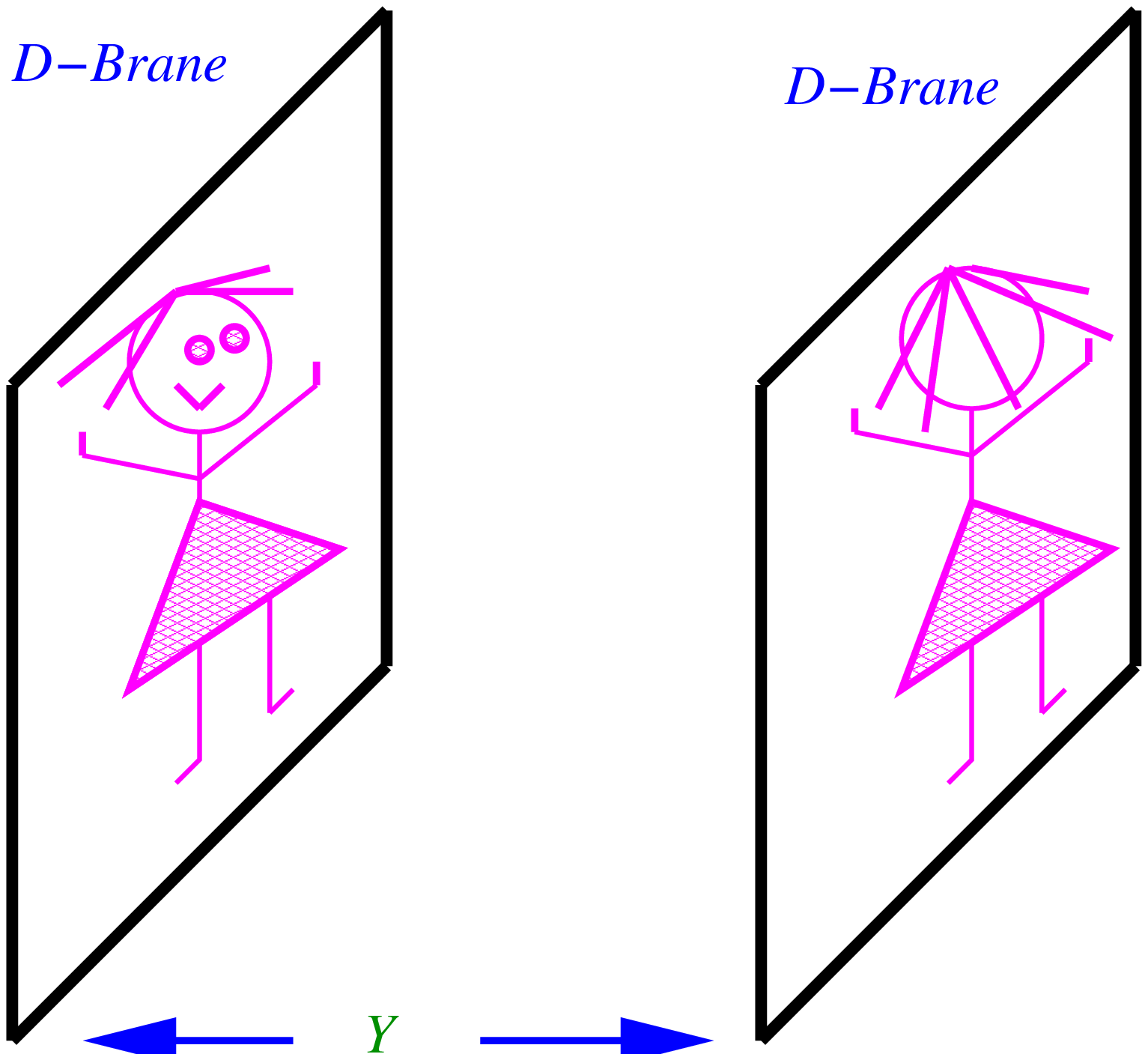,width=8cm}{D brane- D brane interaction is zero as long as supersymmetry is unbroken.}
{\label{figure8}}

The proposal of Dvali and Tye was that, after supersymmetry gets broken, 
it is expected that the RR field and the dilaton may get a mass, whereas the 
graviton stays massless. Therefore the cancellation of the inter-brane force
 no longer holds and a nonzero potential develops which is generally 
attractive, given that gravity is the dominant force whereas the other 
interactions will have a
Yukawa suppression due to the mass of the carrier modes.
Therefore they propose that the effective potential would be a sum of two
 terms, the first one coming from the sum of the two tensions of the branes 
is like a cosmological constant term and the second is the uncancelled 
interaction potential. At distances large compared to the string length
it takes  the form:
\eq
V\approx 2T + \frac{a}{Y^{d-2}}\left( 1+ \sum_{NS} e^{-m_{NS}Y}
-2\sum_{RR} e^{-m_{R} Y}\right)
\eeq
Where $Y$ is the separation between the branes
and $a$ a dimension-full constant. In the
limit of zero RR and NS-NS masses $m_{R},m_{NS}$
the interaction potential vanishes.
But when they are massive we can see the potential takes a form that has the
properties to lead to inflation since it is very flat, due to the exponential 
terms and has a positive value at infinity due to the tension
term. The authors argued that this kind of potential satisfies the slow roll
 conditions and can give rise to inflation in a natural way. A  minor problem of
 their scenario is that they assumed that the string scale was
 $1$ TeV, therefore the density perturbations were of order 
$\delta_H\sim \frac{H}{\epsilon M_{Planck}}$ for an undetermined parameter
 $\epsilon$ which for $H\sim M_s^2/M_{Planck}$ and $M_s\sim 1$ TeV
 would imply an extremely small value of $\epsilon$ to get the COBE
normalisation $\delta_H\sim 10^{-5}$. This problem can be easily solved by 
just assuming the string scale to be closer to the Planck scale or even
an intermediate scale.

\EPSFIGURE[r]{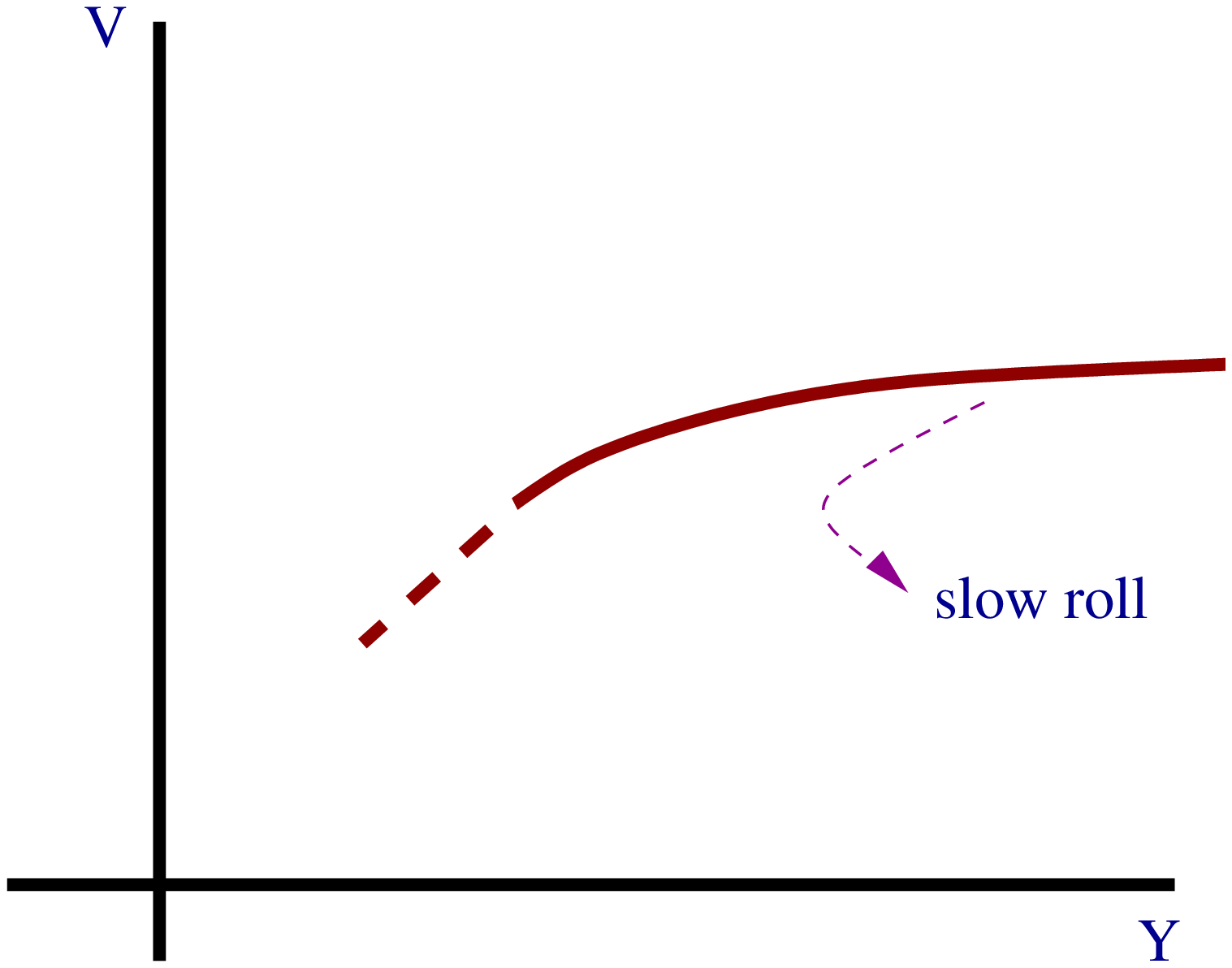,width=8cm}{An expected shape of the inflation potential for D-branes, after
 supersymmetry breaking}
{\label{figure9}}

A more serious drawback of this scenario is the lack  of computability.
After all, the proposed potential is only motivated by physical intuition
 but does not correspond to an honest-to-God string calculation. This
 made
 it difficult to make explicit progress. Furthermore, the authors did not
 address the issue of what happens after the branes collide and how to 
finish inflation. Nevertheless, this scenario provided an interesting
possibility of realising inflation from brane interactions, the 
shape of the potential looks naively correct and opened up the 
idea that the attraction and further collision of branes could have interesting
cosmological implications.

\EPSFIGURE[r]{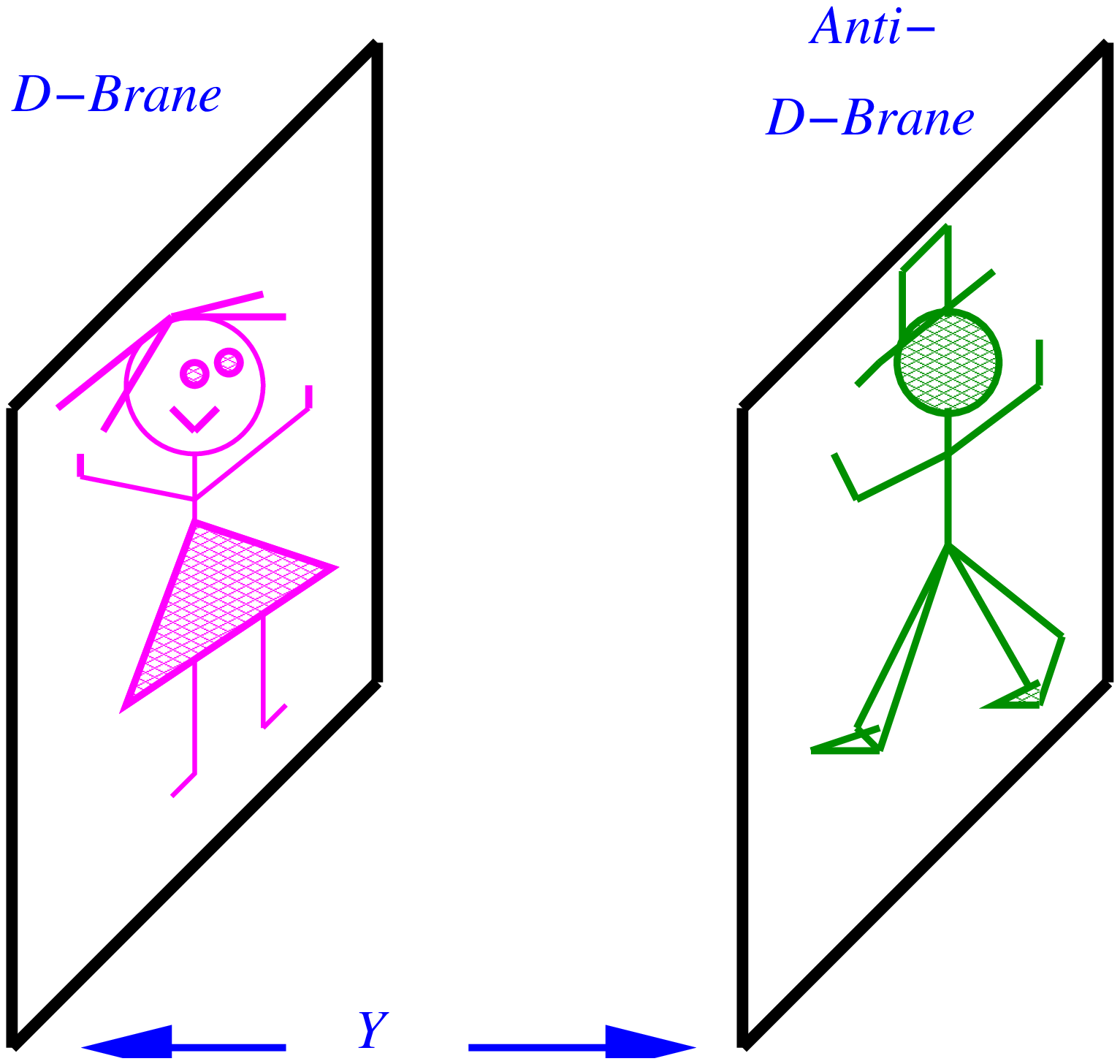,width=8cm}{D branes and anti D branes attract each other.}
{\label{figure5}}

\subsection{D-Brane/Antibrane Inflation and Tachyon Condensation}

Let us consider now a  brane/antibrane pair, that means a pair of branes with 
opposite RR charge. We know that their interaction does not cancel since now,
 the cylinder diagram will give an amplitude for which the RR contribution 
changes sign and therefore we have: 
\eq
{\cal A}={\cal }A_{NS} + {\cal }A_{R}= 2{\cal }A_{NS}\neq 0.
\label{amplitude}
\eeq
This gives rise naturally to an attractive force. Contrary to the case
of the  brane/brane
  potential, that required uncomputable nonperturbative corrections,
this case 
 is computable in an explicit way (from the cylinder 
diagram, taking the limit of $t\rightarrow \infty$ gives rise to the Newtonian
 potential at distances larger than the string scale) in perturbation
theory.

 Let us see in some 
detail 
how to compute the corresponding potential. We will start with one 
D$p$-brane and an $\bar {\rm D}p$-brane in a large four-dimensional
bulk with extra dimensions compactified in tori. The 4D effective action
can be written as the sum of bulk and branes contributions:

\eq
S\ = \ S_{Bulk}+ S_{{\rm D}}+ S_{\bar {\rm D}},
\eeq
with the bulk action

\eq 
S_{Bulk}\ = \ 
\frac{1}{2}\int d^4x d^6z\sqrt{-g}\left\{M_s^{8} e^{-2\varphi} R+
\cdots \right\},
\eeq
where we are denoting by $x^\mu$ the four spacetime coordinates and $z^m$ the 
coordinates in the extra dimensions. The  branes actions can be 
obtained expanding the Born-Infeld action as:

\eq 
S_{{\rm D}}\ = \ -\int d^4x d^{p-3}z\sqrt{-\gamma}\left\{ T_p\ +
 \ \cdots \right\}, 
\eeq
where $\gamma_{ab}=g_{\mu\nu}\partial_a x_i^\mu \partial_bx_i^\nu$ is the
induced metric on the brane and $T_p =M_s^{p+1} e^{-\varphi}$ is the brane
 tension.
We will assume the branes to be parallel and the separation is given by
$Y^m\equiv \left( x_1-x_2\right)^m$ where the sub-indices $1,2$ 
refer to the brane
 and antibrane respectively. Expanding in powers of $\partial_aY^m$ we get:
\eq
\label{braneaction}
S_{{\rm D}}+ S_{\bar {\rm D}}\ = -\int d^4x\  d^{p-3}Y\sqrt{-\gamma} T_p
\left[2+\frac{1}{4}g_{mn}\gamma^{ab}\partial_aY^m\partial_bY^n \  \ \cdots
 \right].
\eeq

The interaction part can be directly obtained by the calculation of the
 cylinder amplitude mentioned above. For large separations
 $M_s^{-1}\ll  Y$ it
 simply takes the Newtonian form $V_{interaccion}\sim Y^{d_\perp -2}$
where $d_\perp \equiv 9-p$ is the number of dimensions transverse to the
 branes. Combining this with the derivative dependent part of the branes
 action give us a potential of the form:
\eq
V(Y)\ = \ A\ - \ \frac{B}{Y^{d_\perp -2}},
\eeq
where 
\eqa
A & \equiv & 2 T_p V_{||}=\frac{2 e^{\varphi}}{(M_s r_\perp)^{d\perp}} M_s^2
 M_{Planck}^2,\\
B& \equiv & \frac{\beta e^{2\varphi}}{M_s^8} T_p^2 V_{||}= \frac{\beta
 e^{\varphi} M_{Planck}^2}{M_s^{2(d_\perp -2)}r_\perp^{d_\perp}}.
\eeqa
Here the symbols $||$ and $\perp$ refer to parallel and perpendicular to the
 branes, then $V_{||}$ is the volume parallel to the brane and $r_\perp$ is the
radius of the space perpendicular to the brane, which is assumed constant.
Also the constant $\beta$ is given by 
$\beta=\pi^{-d\perp/2} \Gamma\left(\frac{d\perp -2}{2}\right)$.

Notice also that equation (\ref{braneaction}) provides the normalisation of 
the kinetic energy and therefore we can work with the canonically normalised
 field
$\Phi \equiv \sqrt{\frac{T_p V_{||}}{2}} Y$ when analysing the consequences of
 the potential.

Now that we have the full information for the scalar potential we can ask if 
this potential gives rise to inflation. The fact that in the limit
$Y\rightarrow \infty$ it goes to a positive constant $A$ is encouraging. 
To check the slow roll conditions we compute the constants $\epsilon$ and
 $\eta$
and find that $\epsilon<\eta$ as usual, and:
\eq
\eta\ = \ -\beta \left(d_\perp -1\right) 
\left(d_\perp -2\right)\left(\frac{r_\perp}{Y}\right).
\eeq

\EPSFIGURE[r]{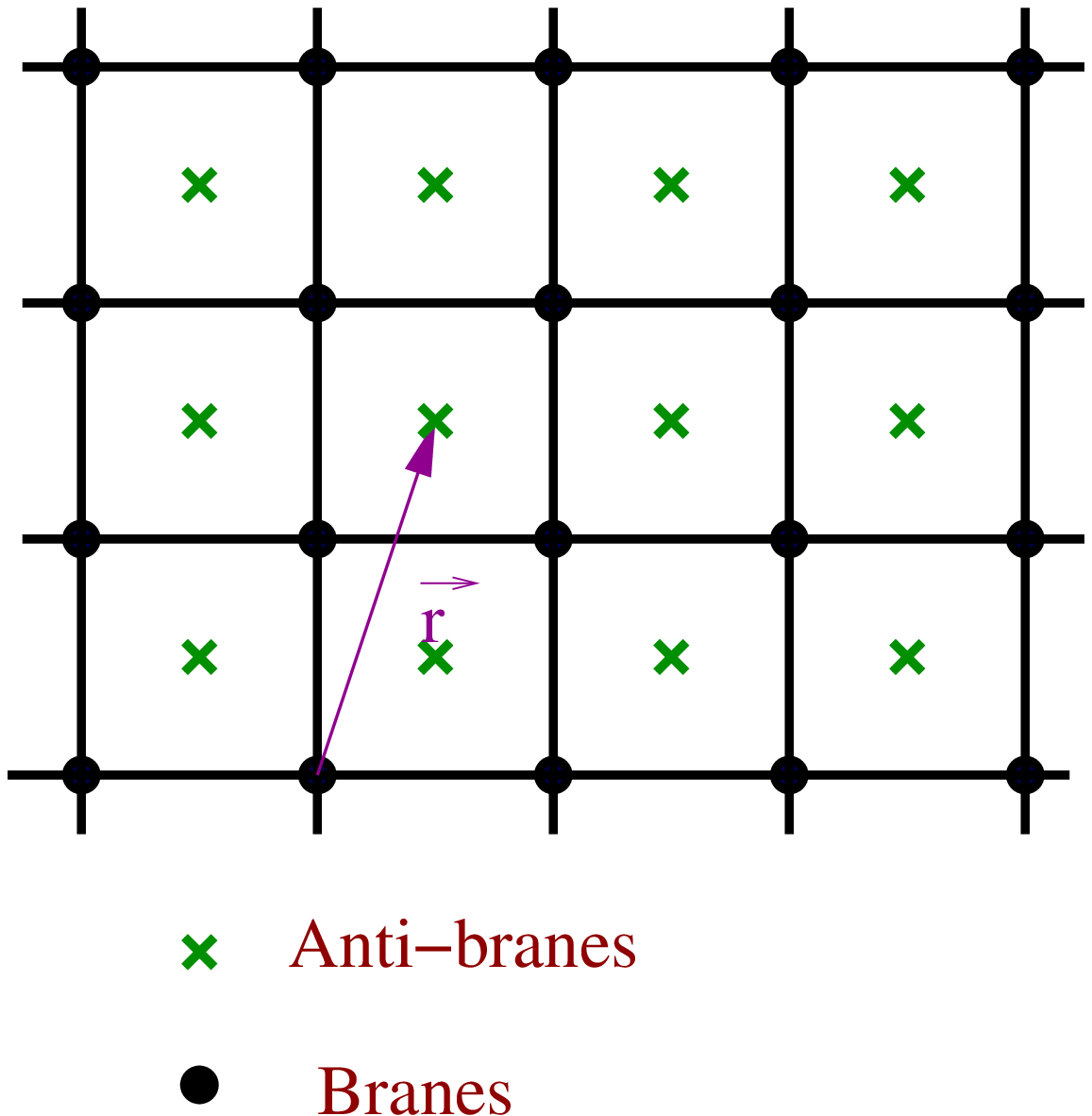,width=6cm}{A lattice describing the
equilibrium  configuration of branes and antibranes.}
{\label{figure11}}

We can see that since $\beta$ has a fixed value in string theory and the
 parameter $\eta$ is proportional to the ratio 
$\left(d_\perp -2\right)\left(\frac{r_\perp}{Y}\right)\gg 1$ (since the
 separation of
 the branes is assumed to be much smaller compared to the size of the extra 
dimension for the approximation of the potential to be valid). Therefore, we 
conclude that slow roll requires $\eta\ll 1$, implying 
that $Y\gg r_\perp$ which is inconsistent. This means that  this 
potential does not give rise to inflation. The cases of D7 and D8 branes ($d_\perp=1,2$)
have to be treated separately but the same conclusions hold.

We may wonder if this situation can be improved. We have realised that if the
distance between the branes is $M_s^{-1}\ll  Y \ll r_\perp$ then generically it
does not give rise to inflation. Now we ask the following question.
 Relaxing the condition $Y\ll r_\perp$, is it possible to get inflation?
For this we have to be able to compute the potential in a torus \cite{quei}.
 Let us consider the simplest case of a square torus (see figure~11).
 We assume the compactified transverse manifold to be
a $d_\perp$-dimensional square torus with a uniform circumference
$r_\perp$.  When the brane-antibrane separation is comparable to $r_\perp$, we
have to include contributions to the potential from $p$-brane images,
{\it i.e.}, we have to study the potential in the covering space of
the torus, which is a $d_\perp$-dimensional lattice, $({\bf
{R/Z}})^d_\perp$.

The potential at the position of the antibrane is 
\begin{equation}
V({\vec r})\,=\,A\,-\,\sum_{i}\,{B\over|{\vec r}-{\vec r}_i|^{d_\perp-2}}\,,
\end{equation}
where 
$\vec r$ and ${\vec r}_i$ are the vectors denoting the positions of
the $p$-branes and antibranes in the $d_\perp$-dimensional coordinate
space, and the summation is over all the lattice sites occupied by the
brane images, labelled by $i$.  We schematically show our set-up in
Figs.~11, 12.
 It looks that the value of the potential diverges by simply summing
over the infinite number of lattice sites, 
however this is just an artifact of working with the method of images
and the value of the potential can be
 unambiguously computed taking into account the finite size of the torus (see appendix of \cite{queii} for a detailed
calculation.).

\EPSFIGURE[r]{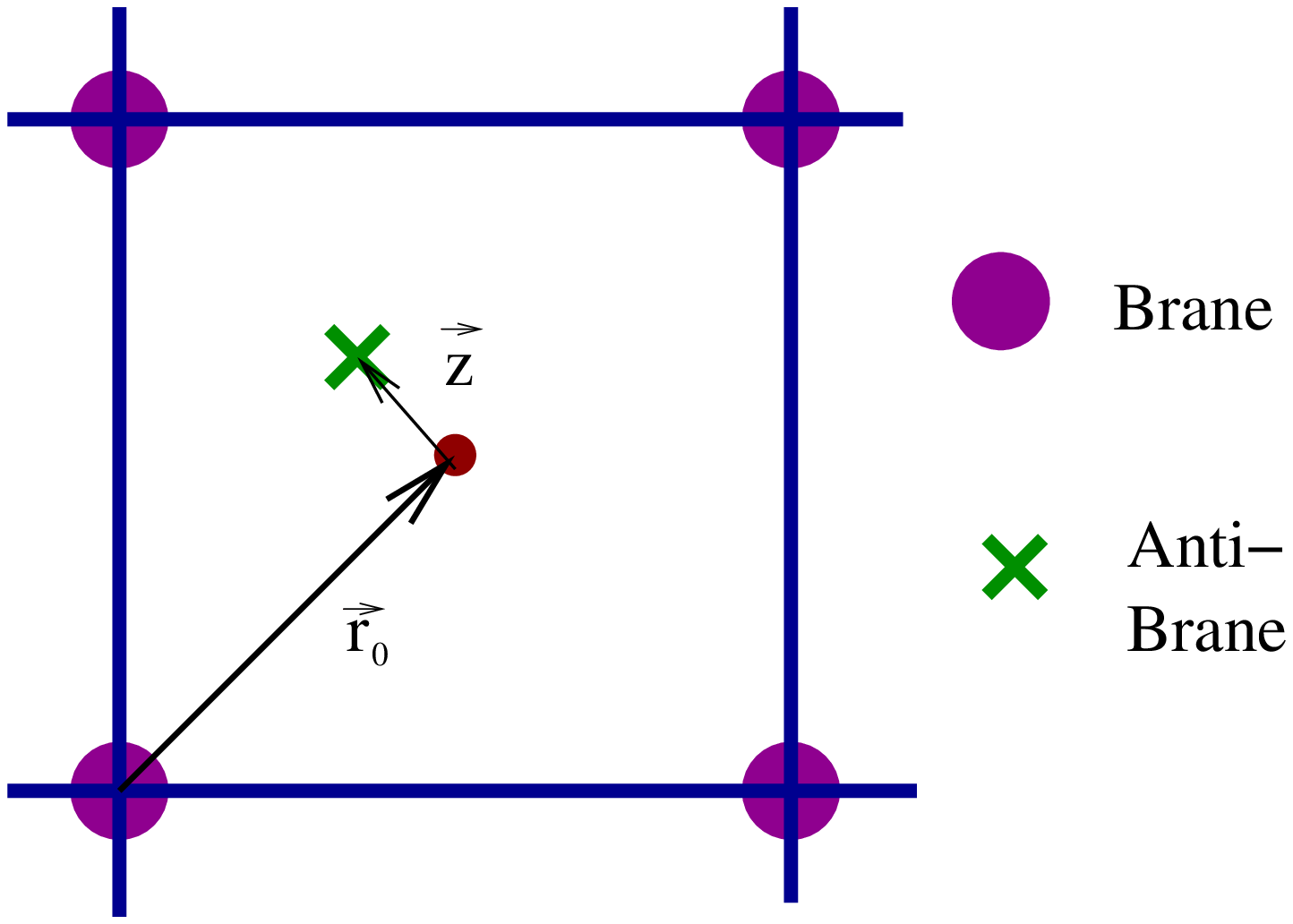,width=6cm}{The location of the antibrane in a
configuration that gives 
rise to inflation if the separation from the critical point in the middle is small enough.}
{\label{figure12}}

Consider now the  antibrane motion when the antibrane is near the centre of
the  hyper-cubic cell. In this case we may expand the potential in terms
of power series of this small displacement $\vec z$ from
the centre. From a simple symmetry argument, one can easily see that
all the odd powers in $\vec z$ vanish.  Furthermore, 
 the quadratic terms in $\vec z$ also vanish, so that the
leading contribution to the potential is the quartic term in $\vec z$.
We can therefore model the relative motion of the branes in a
quartic potential
\begin{equation}
V(z)\,=\,A-{1\over4}\,C\,z^4\,,
\end{equation}
where $A$ is defined as before and
$C=\gamma\,M_s^{-8}\,e^{2\varphi}\,T^2_p\,V_\parallel\,
r_\perp^{-(2+d_\perp)}$ with $\gamma$ a constant of order ${\cal
O}(1)$.  It is straightforward to derive for this potential the slow
roll parameter $\eta$ and the density perturbation $\delta_H$ 
\eq 
\eta \approx -3\,\gamma \left({z\over r_\perp}\right)^2 \,, \qquad\qquad
\delta_H \approx {2\over5\,\pi}\,\sqrt{\,\gamma\over3}\, {N^{3/2}
\over M_{Planck}\, r_\perp} \,, 
\eeq 
where again we have used the standard slow-roll equations. We see now that
slow-roll is guaranteed for sufficiently small $z$.
Therefore we have succeeded in obtaining inflation from a completely computable
string potential. This was the first example of such a string theory
 derived inflationary potential \cite{quei}. Furthermore, to obtain
the minimum number of efoldings $N\geq 60$ and  the COBE normalised
value of the density fluctuations we can easily see that 
$\delta_H\approx 10^{-5}$ is obtained for a compactification radius
 $r_\perp^{-1}\approx 10^{12}$ GeV
corresponding to  an intermediate string scale $M_s\approx 10^{13}$ GeV.
The spectral index $n\approx 1-3/N$ is in the favoured range. 
Therefore this is a string theory derived inflation that has all the
 properties of successful inflationary models, with the advantage of having a
 fundamental origin with a geometrical interpretation for the inflaton field 
as the distance, in the extra dimensions, of the colliding worlds, described by the
 brane and antibrane respectively.

Furthermore string theory also provides a way to end  inflation
\cite{quei}.
This is probably the most interesting part of this scenario.
We first recall that the string potential we have been discussing is valid
 for distances larger than the string scale. However the potential is 
attractive
 and at some point the branes get closer to each other and this approximation
 also will not be valid. We expect something different to happen at those 
separations and fortunately it happens to be understood. The point is that 
the amplitude (\ref{amplitude})
has a divergence appearing at a critical distance $Y_c=\sqrt{2\alpha'}\pi$
 \cite{banks}. What happens at this distance is that an
open string mode
that was massive at large separations becomes massless and, at separations
 smaller than this, it becomes tachyonic. The corresponding tachyon potential
 has been proposed to take an approximate form:
\eq
V(Y,T)\ =\ \frac{1}{4\alpha'}\left(\frac{Y^2}{2\pi^2\alpha'} -1\right) T^2 
+ C T^4+\cdots
\eeq
With $C$ a constant. Notice that this reproduces the change of the mass for
 the field $T$ as a function of the separation $Y$.
We can immediately see, that taking the effective potential as a
 function of both $T$ and $Y$, gives us a potential precisely of the form 
proposed for hybrid inflation! Therefore string theory provides with a 
natural way to end inflation.

Moreover, the tachyon potential has been studied in some detail during the
past few years and its  general  structure has been extracted.
In particular, Sen conjectured that at the overlap point ($Y=0$) the potential
 should be of the Mexican hat form with the height of the maximum equal to 
the sum of the brane tensions $2T_p$. The minimum would correspond to 
the closed string vacuum being  supersymmetric where the
potential vanishes.
These conjectures have been verified with more than $90\% $ accuracy
using string field theory techniques \cite{tseytlin,sen}.
This allows us to estimate the reheating temperature after inflation which is
 essentially the energy difference between minimum and maximum, giving:
\eq
T_{RH} = 
\left(\frac{2e^{\varphi}}{(M_s r_\perp)^{d_\perp}}\right)^{1/4}
\sqrt{M_sM_{Planck}},
\eeq
which for instance for $p=5$ gives $T_{RH}\sim 10^{13}$ GeV.

 Finally the tachyon potential also has topological defects which correspond to
D $(p-2)$ branes (and antibranes) \cite{sen}. In fact all BPS D-branes are expected to
 appear as topological defects of a tachyon potential. This has implied an 
elegant 
classification of D branes from the mathematics of $K$-theory \cite{k}. For the 
cosmological purposes that interests us here this can have very interesting
 implications in several ways. We have seen that inflation is possible but
 not generic. That means  that only for distances between the branes close to the 
equilibrium position, the potential is flat enough as to give rise to  inflation,
 otherwise the slow roll conditions are not satisfied and there is no
inflation \cite{quei}.

\EPSFIGURE[r]{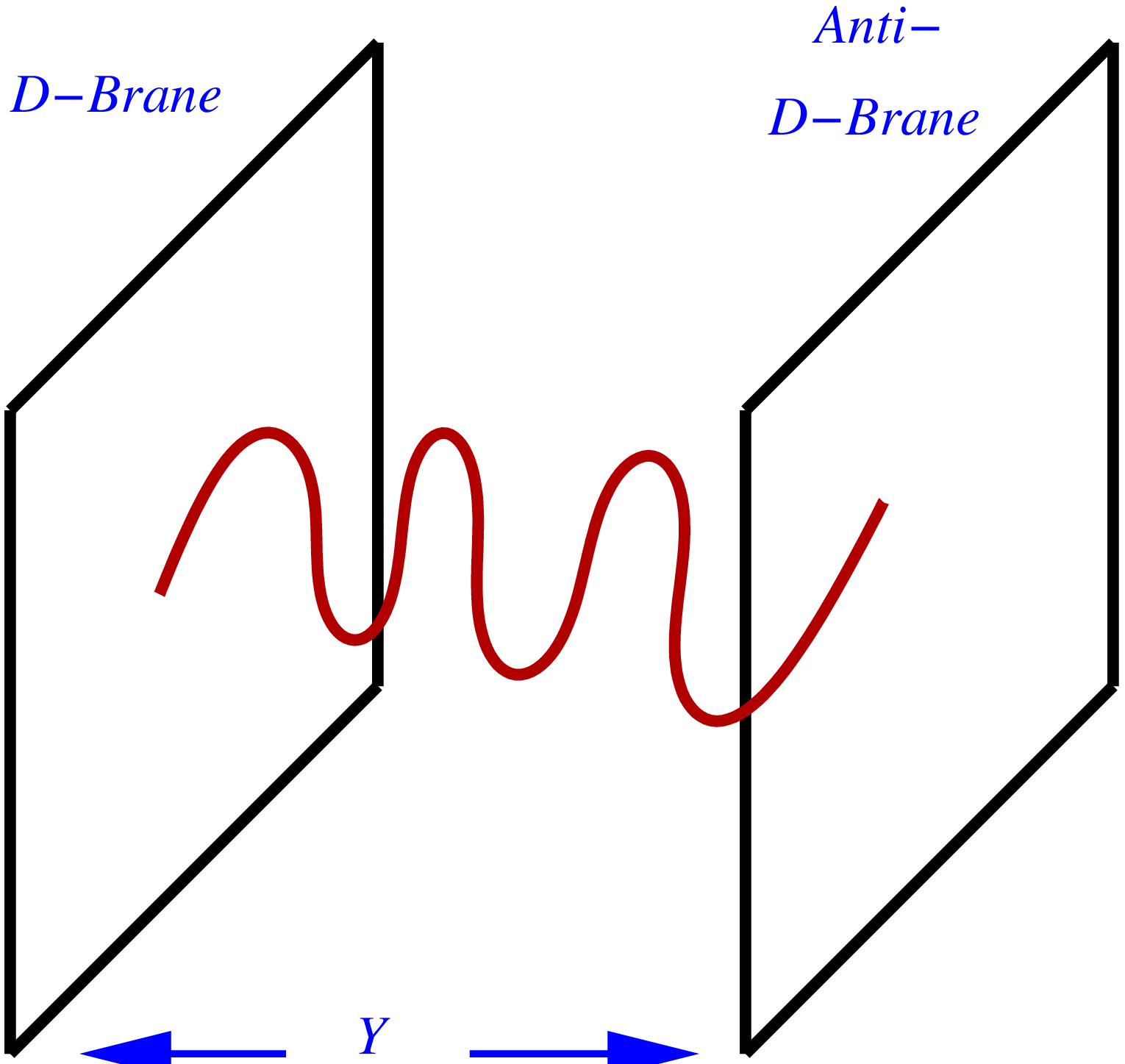,width=6cm}{An open string state becomes tachyonic at a critical inter brane separation.}
{\label{figure13}}

 We can imagine a configuration of a gas of branes and antibranes. Most of the
times they will interact and annihilate each other without giving rise to 
inflation but upon collision they will keep generating $p-2$ branes, these 
ones will 
generate $p-4$ branes and so on, implying a cascade of daughter branes out of
 the original ones. In this system at some point a pair of branes will be at
 a separation close to the equilibrium point and that will give rise to
 inflation, dominating the expansion of the universe and rendering the issue
 of initial conditions for inflation more natural. Notice that as usual, once
 inflation is generated it dominates. Furthermore we may imagine the brane gas
 to originate from just one pair of a D$9$/$\bar{\rm D}9$ branes, acting as
 parent branes generating the cascade of daughter brane/antibrane systems.
It remains to ask for the origin of the D$9$/$\bar{\rm D}9$ pair to start with,
probably as some sort of quantum fluctuation, although this is not clear.

The cascade scenario also allows for some speculations about the 
dimensionality of spacetime \cite{quei}. Starting in type IIB strings we know that 
branes of odd dimensionality ($9,7,5,3,1$) appear, therefore we can have 
D$9$ brane/antibrane annihilating immediately, also D$7$ branes 
annihilate their
 antibranes very easily because of their large dimensionality, as well as D$5$
 branes. However D$3$ branes will have a harder time to meet their antibranes 
because of the difference in dimensionality. Remembering the argument of
Brandenberger-Vafa for the dimensionality of spacetime argued that 
the world-sheets of two strings can meet in 4-dimensions but not in larger ones.
This can be generalised to $p$ branes in $D$ dimensions for which the critical
 dimension is: 
\eq
D_{critical}= 2p+2.
\eeq
Therefore we may say that  D branes with 
$p=3$ can meet in dimensions smaller or equal than $8$ 
but miss each other in higher dimensions, 
whereas $p=5$ branes meet in $D<12$. This makes a rough argument why 
D3-brane worlds may survive annihilation in 10 dimensions and be preferred over higher 
dimensional  ones. 
We may actually imagine a scenario where branes of all types are initially
 present and all dimensions are compact and small. The large dimension branes 
annihilate instantly, leading to a population of branes that include 3-branes
 and lower. The windings of these branes keep any dimensions from growing. 
Then the BV
mechanism starts to act, making four dimensions large and six small, with no 
windings about the large spatial directions. After this we have the particular
 collision which causes inflation of the large dimensions.
It would be very interesting to quantify this statement.
For a further discussion and calculations on this regard see \cite{annemahbub,
tye1}.

\EPSFIGURE[r]{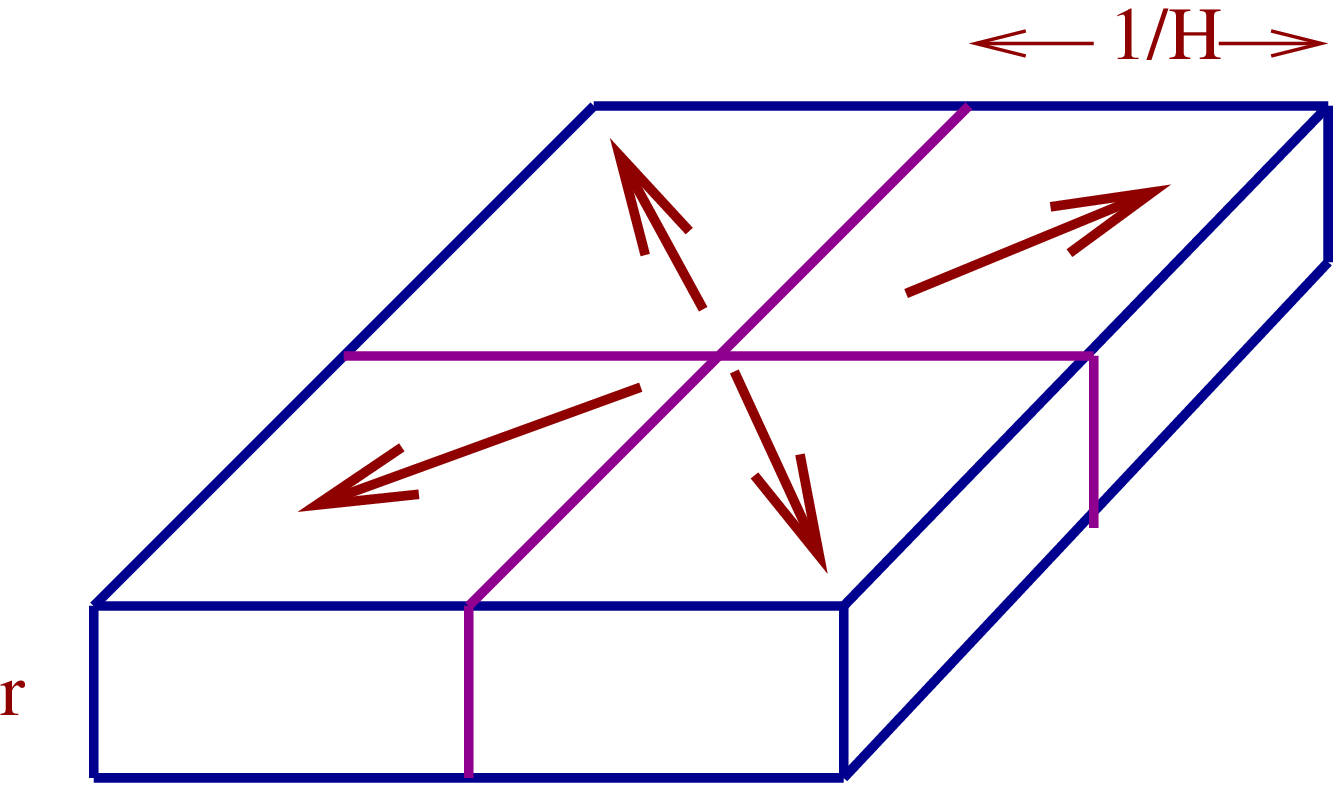,width=6cm}{An illustration on how the Kibble mechanism is not in action in the extra dimensions since the inverse Hubble parameter is much bigger than the size of the extra dimension. Topological defects do appear in the noncompact dimensions.}
{\label{figure14}}

We may still have to worry about D branes of dimensions smaller than 3.
We know that domain walls ($p=2$) and monopoles ($p=0$) can be the source of
 serious cosmological problems, since they  over-close the universe and 
therefore should not survive after inflation. Cosmic strings on the other
 hand are not
ruled out (they have been ruled out as the main source of the density 
perturbations, but they could still exist and contribute at a minor scale
 \cite{tye2}). Fortunately this scenario does not give rise to monopoles nor 
domain walls.
First of all, before inflation these defects may appear as some $p$ branes 
wrapping different numbers of cycles in the extra dimensions, but they are 
diluted away by inflation. After inflation they may be produced from the
 standard Kibble mechanism. But they are not 
because of the following argument.
 Starting from type 
IIB strings we
 know that the dimensionality of the $p$ branes has to be odd, which 
seems to eliminate
 those possibilities from the start. However we have to be more careful
and a complete analysis requires thinking about the mechanism that produces
 topological defects in cosmology, namely the Kibble mechanism

 In FRW the formation of topological
 defects appears since regions separated at distances larger than the
 particle horizon size $H^{-1}$ are uncorrelated and therefore we expect one
 topological defect per Hubble radius. We know that in these models $H\approx
M_s^2/M_{Planck}$, 
therefore $H^{-1} \gg r_\perp$ and there is correlation in the
 compact directions, implying no topological defects in those dimensions. We may then 
say that the Kibble mechanism is  not at work in
 the extra dimensions and therefore that the topological defects that appear
 after inflation are the D 
$(p-2k)$ branes wrapped around the same cycles as the
 original $p$ branes that produced them (see figure~14). In this way, if 
inflation is generated by the collision of two $p$ branes wrapped on an 
$n=p-3$ cycle,
the topological defects of dimension $p-2k$ will also wrapp the $p-3$ cycle 
appearing as a $3-2k$ dimensional defect in the four large dimensions, 
excluding then monopoles and domain walls. Notice that this argument holds 
for $p$ being even and odd.

The interesting point, as far as observations are concerned
\cite{tye1,tye2}, is that
since cosmic
strings will have a tension $\mu\sim M_s$, they may contribute to the
CMB anistropies. Current sensitivity rules out models for which $G\mu
\geq 10^{-6}$. In this case  $G\mu\sim 10^{-9}$ is still consistent
and could be eventually tested in future experiments.

Finally, if  inflation is caused by the collision of 3-branes we may ask 
where will our universe be. One possible answer to this is that the collision 
is between stacks of branes and anti-branes with different number of branes on
 each stack (remember that by being BPS we may have a stack of D-branes which 
do not interact with each other). Therefore we may have, say, 10 branes 
colliding with 4 antibranes leaving then 6 branes after the collision where 
the standard model can live.

This scenario is certainly very interesting. It provides the first example of 
a string-derived potential that gives rise to inflation and has also a stringy 
mechanism to end inflation by the appearance of the open string tachyon at a 
critical distance. Therefore hybrid inflation is realised in a stringy way.
It then shares all the good experimental success that inflation has at present,
in terms of the spectrum of CMB fluctuations. 
It provides many other interesting features like the 
apparent critical dimensionality
 of 
3-branes and the natural suppression of monopoles and domain walls after 
inflation. It was built however following several assumptions. First, it is 
assumed that an effective 4D FRW background is valid (implying that we have
 to be in a regime where the effective field theory description  of string 
theory is valid). Second, the branes were assumed to be parallel and velocity
 effects were neglected \footnote{Velocity effects were considered by Tye and Shiu
in \cite{dvali}. Recent proposals for using the repulsive velocity effects 
in branes for cosmology
are \cite{ramzi}.}. The major assumption, however, is considering the
 moduli 
($r_\perp$ in this case) and dilaton to have been already fixed by some unknown
stringy effect. This is a very strong assumption that prevents from claiming 
that this is a full derivation of inflation from string theory. 
A more dramatic way to see this problem is that since the configuration 
brane/antibrane breaks supersymmetry, there will naturally be NS tadpoles
 which generate a potential for the moduli. This potential is at the level of 
the disk (tree level) whereas the interaction described by the cylinder 
diagram was one-loop in 
open string terms. The assumption of having fixed
 the moduli refers to 
having found a mechanism that compensates the NS tadpole terms \cite{tadpoles}\ 
and induces a minimum for the moduli potential. This is not impossible,  
but needs to be addressed and tree level terms are in principle dominant.
Moreover the reheating mechanism is not completely understood, in particular, as we will see later, 
the tachyon's relaxation to its minimum is not standard in field theory.
Furthermore, the scenario was presented without reference to realistic
D-brane models. It is known that the way to get a chiral spectrum in D-branes
corresponds to branes at singularities and intersecting branes.
 We will move now to these topics and mention how some of the problems mentioned above can 
be relaxed (although not completely solved).

\subsection{Intersecting Branes, Orientifold Models and Inflation}

The idea of the previous subsection can be extended to more general
 string 
constructions. Probably the simplest to consider is 
the intersection of branes at  different angles. It is known that branes 
intersecting at nontrivial angles have chiral fermions in their spectrum, 
corresponding to open strings with endpoints on each of the branes. 
Again, as in the brane/antibrane case, we can compute the amplitude of 
the interaction between two branes at angles, which is given by:

\eq
\mathcal A = 2 \int\frac{dt}{2t}\,{\rm Tr}\, e^{-tH}\,, 
\eeq
where $H$ is the open string Hamiltonian. For two D$p$-branes
making  $n$ angles in ten dimensions  this
amplitude can be computed to give
\cite{jab,poli,g-b}:

\eq\
\mathcal A =
V_p\int_0^\infty\frac{dt}{t}\exp^{-\frac{t\,Y^2}{2\pi^2\alpha'}}
(8\pi^2\alpha't)^{-\frac{p-3}{2}}
\left(-iL\eta(i\,t)^{-3}(8\pi\alpha't)^{-\frac{1}{2}}\right)^{4-n} 
  (Z_{NS}-Z_{R}) \,,
\eeq
with  $V_p L^{4-n}$ the volume of the common dimensions to both branes
and
\eqa
Z_{NS}&=&(\Theta_3 (0\mid it))^{4-n}
 \prod_{j=1}^n\frac{\Theta_3(i \Delta\theta_j t\mid it)}
  {\Theta_1(i \Delta\theta_j t \mid it)} - 
(\Theta_4 (0\mid it))^{4-n}
\prod_{j=1}^n\frac{\Theta_4(i \Delta\theta_jt \mid it)}
  {\Theta_1(i \Delta\theta_j t \mid it)}\,, \nonumber\\
Z_{R} &=&(\Theta_2 (0\mid it))^{4-n}
\prod_{j=1}^n\frac{\Theta_2 (i\Delta\theta_j t \mid it)}{
     \Theta_1(i \Delta\theta_jt \mid it)}\,,\label{zeta}
\eeqa
being the contributions coming from the $NS$ and $R$ sectors. Also in 
(\ref{zeta}) $\Theta_i$ are the usual Jacobi functions and $\eta$ is 
the  Dedekind function and  $Y^2=\sum_k Y_k^2 $ with $Y_k$ the distance 
between the branes in the $k$th direction. This expression generalises the one we wrote before for parallel branes $\Delta\theta_i=0$ and
brane-antibranes for which one angle $\Delta\theta=\pi$.

In order to obtain the effective interaction potential at distances larger
 than the string scale $Y\gg l_s=M_s^{-1}$ we take the limit of
 $t\rightarrow 0$ and find that (for the compact extra dimensions being 
tori of radius $r$)\footnote{I am following closely the discussion of
 Gomez-Reino and Zavala in \cite{g-b}.}: 
\eqa
V_{int}(Y,\Delta\theta_j)&=& -\frac{\left(2\pi r\right)^{p-5}F(\Delta\theta_j)}
{2^{p-2}(2\pi^2\alpha')^{p-3}}\, \mbox{\LARGE{$\Gamma$}}
\left(\frac{7-p-n}{2}\right)\,Y^{(p+n-7)}\,\qquad p+n\neq 7\nonumber\\
&=& \frac{F(\Delta\theta_j)}
{(4\pi^2\alpha')^{p-3}}
\, \ln\frac{Y}{\Lambda_c}\, \qquad p+n=7.
\eeqa
Where the function $F$ contains  the dependence on the
relative angles between the branes, and is extracted from the small
$t$ limit of (\ref{zeta}). The exact form of this function is given by
\eq
F(\Delta\theta_j)=\frac{(4-n)+\sum_{j+1}^n\cos2\Delta\theta_j
-4\prod_{j=1}^n \cos\Delta\theta_j}{2\prod_{j=1}^n 
\sin\Delta\theta_j}\,.\label{ef}
\eeq

The total potential will then be the sum of this interaction potential plus 
the part coming from the brane tensions, similar to the brane-antibrane case.
If we consider the extra six dimensions as products of three two-tori 
(see figure~15), the $i$th ($i=1,2$)  brane will wrap around each of the two cycles of
 the Ith 
tori $(n_I^{(i)}, m_I^{(i)})$ times. The wrapping numbers $n_I^{(i)}$
and $m_I^{(i)}$  determine the angles
between the branes. Also, the energy density of
 the two branes is given by 
\eq
E_0\ = \ E_1+E_2\ = \ T_p\ \left(A_1+A_2\right)
\eeq
where $A_i$ is the volume generated by the $ith$ brane: 
\eq
A_i\ = \ \left(2\pi r\right)^{p-3}
\sqrt{\left( (n_1^{(i)})^2+ (m_1^{(i)})^2\right)\ \left( (n_2^{(i)})^2+ (m_2^{(i)})^2\right)}
\eeq
Therefore the total potential is
\eq
V\ = \ E_0 + V_{int}.
\eeq

\EPSFIGURE[r]{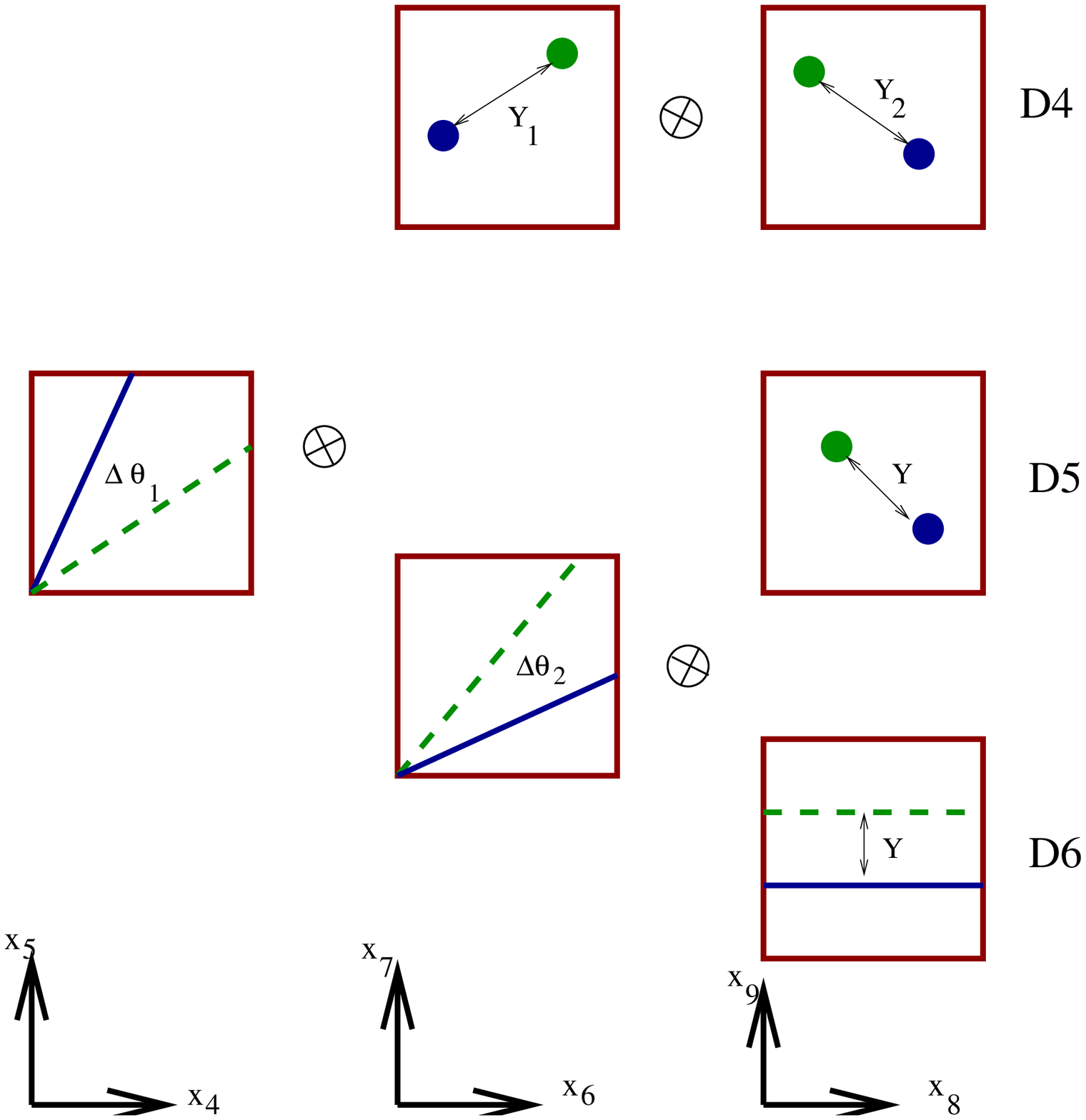,width=8cm}{Possible
 configurations of intersecting D4, D5 and D6
branes when the extra dimensions are products of three two-tori.}
{\label{figure15}}

This  is a typical potential suitable for inflation with a constant piece plus an attractive interaction.
We can  see that, for small values of the angular parameter $F$,
these potentials are inflationary with inflaton field $Y$. From here on the
 analysis is similar to the brane/antibrane system and we refer to the 
literature for the explicit numbers obtained by requiring the right 
number of e-foldings and the consistency with the COBE normalisation to fix the string scale ($M_s\approx 10^{13}- 10^{15} $ GeV).
 In reference \cite{tye1},   the intersecting models were compared with the brane/antibrane models regarding the amount of fine tuning. In both cases there is some  fine tuning, in the brane/antibrane system inflation is obtained only
in special configurations (like being close to the antipodal points of a circle) whereas in branes at angles the fine tuning requires a very small angular
separation (of order $10^{-3}$ or so). Reference \cite{tye1} argues for a less severe fine tuning in the case of intersecting branes.

The other difference is the way of ending inflation. In this system also 
there is a tachyon appearing at a critical separation realising again the 
hybrid inflation model. However the end result of tachyon condensation may
 differ. In some cases the two branes decay into a single brane and in other
 cases they recombine to produce a different configuration of intersecting
 branes but this time being a supersymmetric configuration. The decay
 product is always the 
configuration with the same charges but minimum energy. The second 
possibility is interesting because if the final configuration is still of 
intersecting branes, it has chiral fermions on the intersections and may
 allow a realistic
model at the end of inflation. The reheating temperature can also be computed
 in terms of the difference in energy between the initial and the final 
configuration. Again topological defects will be produced in the process.
 After inflation no domain walls nor monopoles will survive but  cosmic 
strings could, contributing to the CMB anisotropies at a rate that could be a 
few percent without contradicting experiment. This requires that $G\mu\lesssim
10^{-6}$ which for the ranges obtained for  $M_s$ (remember $\mu\sim M_s^2$)
it is safe but relatively close
 to the limit as to be observable in the near future once Planck and
MAP data are analysed.
 Notice that, since the string scale here tends to be a bit higher than
in the brane-antibrane case, 
the brane at angles scenario is closer to be tested experimentally.

\EPSFIGURE[r]{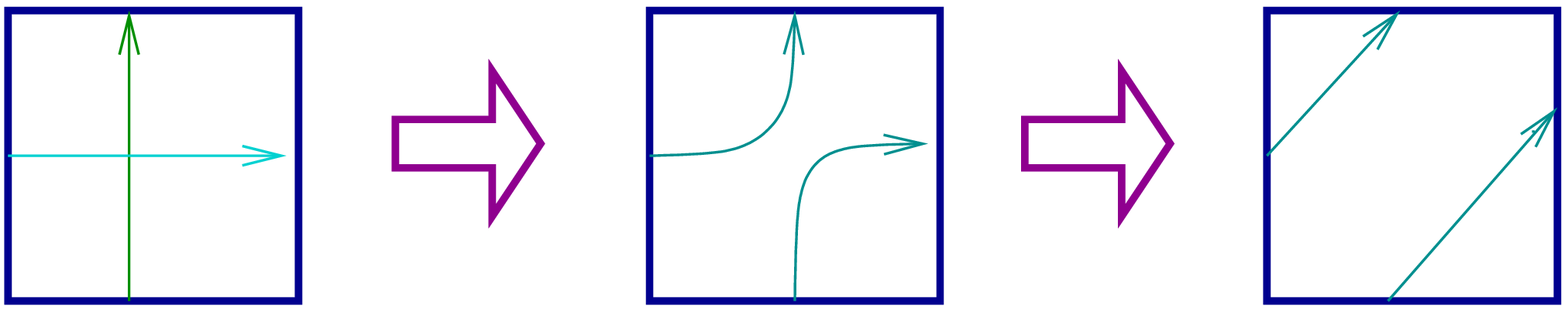,width=8cm}{The decay of two intersecting
branes to a single one wrapping around 
the torus.}
{\label{figure16}}

We must recall that the main problems of the brane/antibrane system are shared
by the intersecting branes, namely the assumption of fixed dilaton and moduli
 by some unknown string effect as well as the details of reheating. For a 
recent discussion of reheating see
\cite{cline}.

\EPSFIGURE[r]{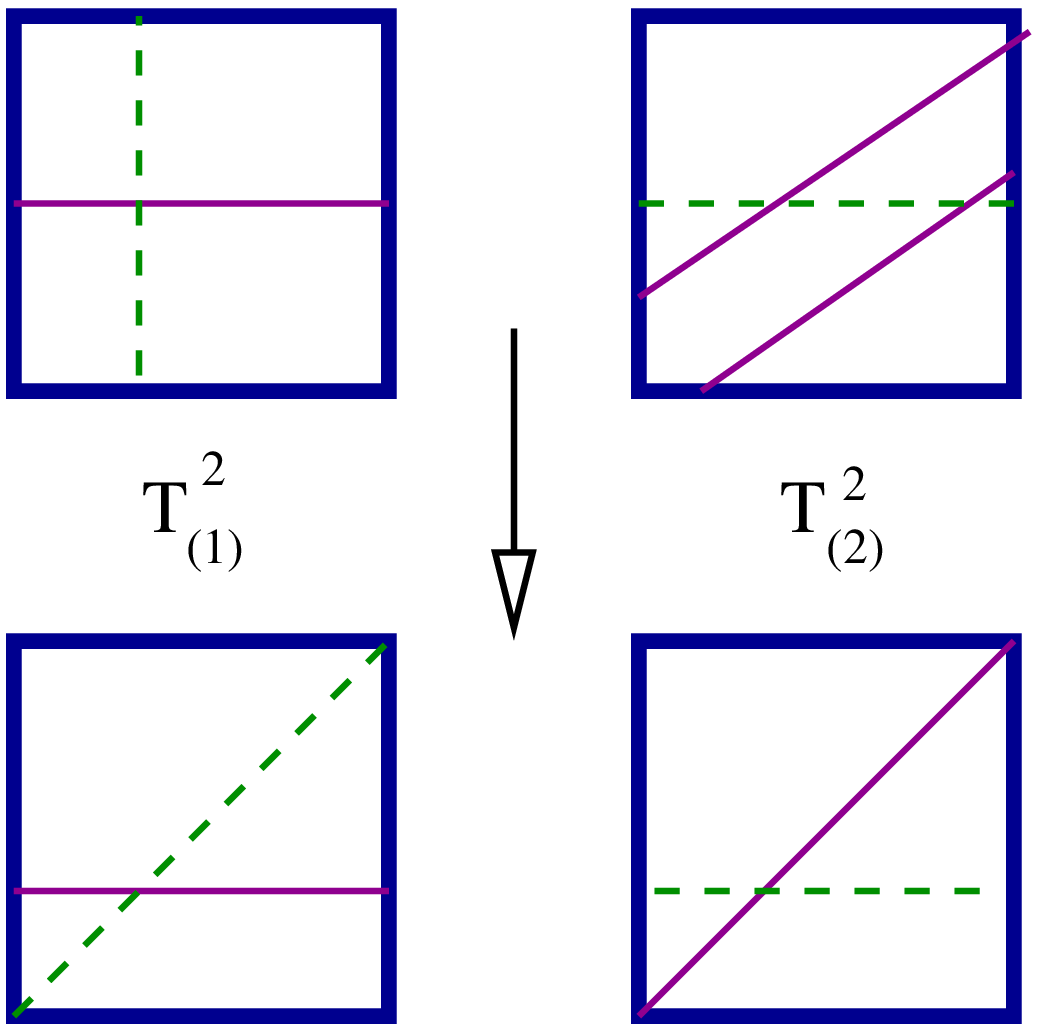,width=6cm}{Recombination of two intersecting D branes to a supersymmetric state with also two intersecting D branes.}
{\label{figure17}}

Finally, a related, very interesting, scenario based on the attraction of
D3 and D7  branes with magnetic fluxes, was proposed in 
\cite{kali}. Again, inflation was obtained with inflaton field being
the separation between the branes. The tachyon condensation mechanism
was further studied and found to lead to the D3/D7 supersymmetric
bound state.
This further illustrates the universality of the hybrid inflation realisation
of
D-brane models.

Let us finish this subsection by briefly considering the other extension of 
these models that addresses at least partially the issue of the moduli. The 
construction corresponds to the other way of constructing chiral models from 
D-branes, namely branes at singularities. In this scenario we typically have 
the extra dimensions being an orbifold or orientifold and have fractional
 branes and orientifold planes. Fractional branes correspond to branes which
 are
attached to the fixed points of the orbifolds and cannot move from there 
(otherwise there is a problem with tadpole cancellations in the twisted
 sector and the model is inconsistent).

Once the branes are trapped at the singular fixed points of the orbifold it 
 allows the existence of chiral fermions in its world volume and there exists
 realistic string models in which the standard model lives in one of these
 branes.
An example of such a model is pictured in  figure~18. It corresponds to the 
${\bf Z}_3$ orbifold that has 27 fixed points where we can put D3 branes to 
include the standard model. Consistency conditions coming from the 
cancellation of twisted R-R tadpoles require the appearance also of D7 branes.
Untwisted tadpoles also imply that we have to have anti D7 branes as well 
as extra D3 branes in different fixed points. Therefore, for the six extra 
dimensions being a product of three two-tori 
 we have three sets of parallel D7 branes and antibranes and several stacks
 of D3 branes located at several of the 27 fixed points, one of them 
includes the standard model. $T$-duality on all the dimensions map the D3
 branes to D9 branes and D7 branes to D5 branes, a configuration that is 
usually simpler to deal with.

%\EPSFIGURE[r]{z2.eps,width=4cm}{The ${\bf Z}_2$ orbifold with one brane at one of the four fixed points,
% one antibrane at a second point and the other two points empty.}
%{\label{figure19}}

Having branes and antibranes trapped at fixed points
 allows for an extension of the brane/antibrane system discussed before in a
 way that the
attraction between the brane and its antibrane corresponds to a potential 
for the size of the extra dimension (the distance $Y$ is no longer a modulus
 because the branes cannot move from the fixed point). Therefore this allows
 to a natural reduction of the moduli, $Y$ is just frozen $Y\sim r$,
 and 
the candidate for inflaton field is the modulus corresponding to the 
size of the extra dimensions \cite{queii},  alleviating the issue of 
assuming it fixed.

Naively we can see that the interaction potential is proportional 
  to $1/r^{d_\perp -2}$ so
\eq
V(r)\ = A-\frac{B}{r^{d_\perp -2}},
\eeq
we have to also know the kinetic term for $r$ to work with the canonically
 normalised field. It is well known that the kinetic term takes the form 
$\partial 
r\partial r/r^2$ and so the canonically normalised field is $\Phi=\log r$,
 and the potential seems to be of the form $V=A-Be^{-a\Phi}$ which is very 
flat and would easily lead to inflation. However we have to be careful with 
this naive
analysis because of two reasons. One is that 
 this will provide the potential in a Brans-Dicke frame and not the Einstein
 frame because the Einstein Lagrangian will have a power of $r$ multiplying 
the scalar curvature. Therefore to 
analyse inflation 
we have to go to the Einstein frame and then the potential above gets an
 overall factor
of $r^{-6}$ becoming a `repulsive' rather than attractive potential in the 
sense that it has a runaway behaviour to the decompactification limit
$r\rightarrow \infty$. Furthermore, in claiming that $A$ and $B$ were constants
we were assuming that the ten-dimensional dilaton was fixed. However it is 
well known that in the effective 4D theory the dilaton combines with the radius
 $r$ to make proper fields like the moduli $S$ and $T$. More precisely, for a
 configuration of D9 and D5 branes in a compactification which is the product
 of 
three two-tori, the relevant fields are:
\eq
s\equiv {\rm Re} S= e^{-\phi}M_s^6 r_1^2 r_2^2 r_3^2,
 \qquad t_i\equiv {\rm Re} T_i= e^{-\phi}M_s^2 r_i^2.
\eeq

The potential in the Einstein frame then takes the form:
\eq
V\ = \ M_{Planck}^4\left[ \frac{A}{t_1 t_2 t_3}+ \frac{B}{st_2
t_3}\left(C-\frac{D}{at_2+bt_3}\right) +
 \cdots \right]\ ,
\eeq
where $A,B,C,D,a,b$ are well defined constants.
We can easily see that this is a runaway potential and does not give
rise to inflation. The only way to get
 inflation is assuming that all of the moduli have been fixed by some
higher energy effect except from one of
 them that we can focus. This actually happens in some models with
fluxes in which the dilaton can be fixed but 
not the overall $T$ field and it is then a plausible assumption.
Once this is assumed there are several options at what the inflaton
field which then has to be redefined in 
order to have canonical kinetic terms.
The remnant potential is of the form:
\eq
V(X)\ =\ K_1 + K_2\ \exp\left(\frac{-\sqrt{2} X}{M_{Planck}}\right) + \cdots\ ,
\eeq
which depending of the relative signs of the constants $K_{1,2}$ it gives rise to inflation, without fine tuning. Again the numerical details can be found in the original literature.

\EPSFIGURE[r]{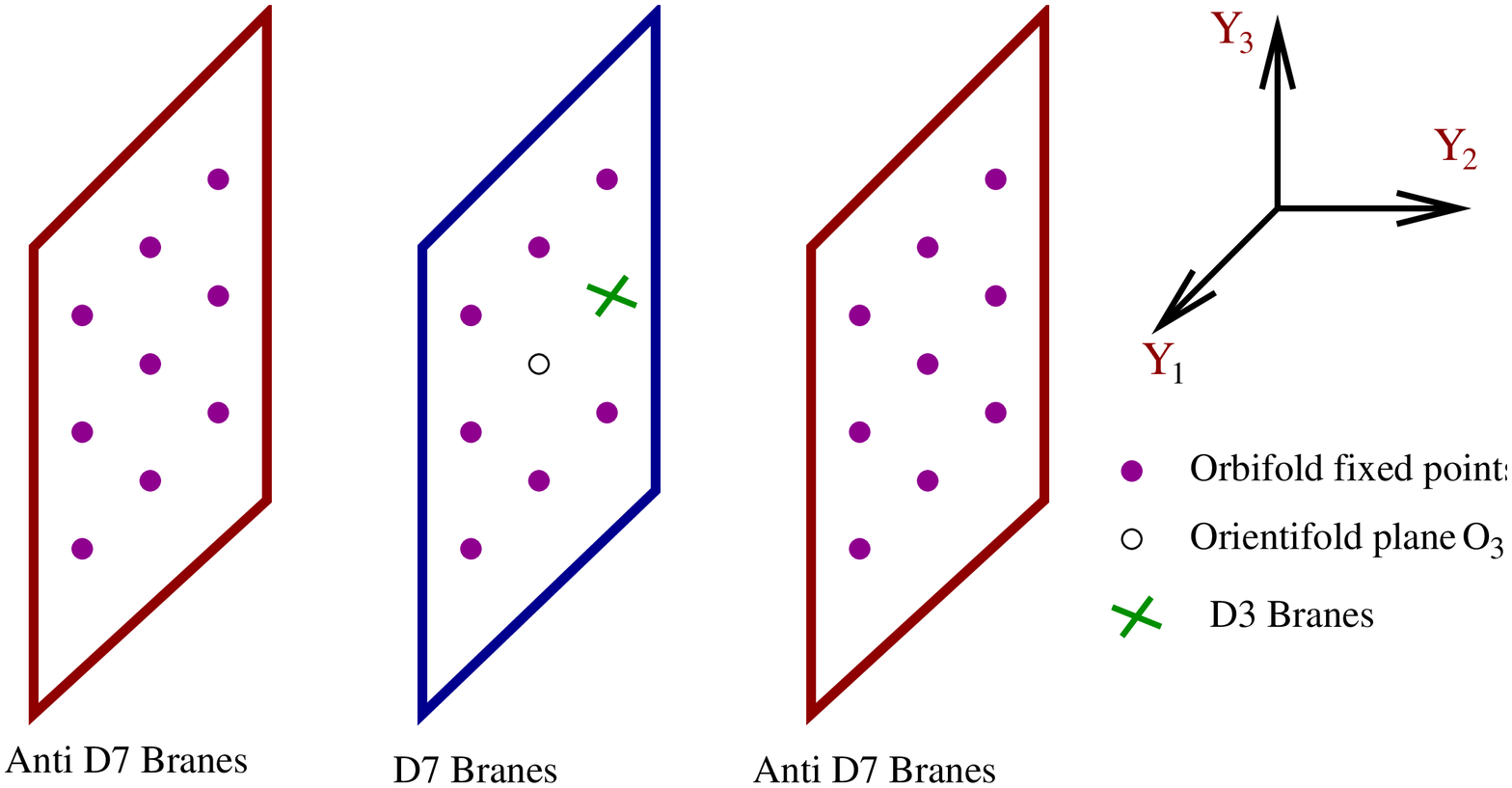,width=6cm}{A configuration of a realistic
  model of branes and antibranes at singularities.
  The branes are trapped at fixed points. The three parallelogramas
  represent D7 and anti D7 branes whereas
 D3 branes containing the Standard Model are trapped in one of the 27 fixed points.}
{\label{figure18}}

A final interesting aspect of this scenario is the ending of inflation. 
Again when the size of the extra dimension is smaller than a critical value
 there 
appears a tachyon in the spectrum that can realize hybrid inflation. The end 
point after inflation is not the vacuum but actually a non BPS D-brane that is
stable for smaller radii (and decays to the brane anti-brane at larger
radii) \cite{norma}.
The reheating temperature  is then the difference between the tensions.
It is worth remarking that this decay process is only known in detail for 
${\bf Z}_2$ orbifolds for which the structure of non BPS branes 
and regions of stability 
has been understood \cite{norma}. For other orbifolds this has not been understood yet,
in particular for the ${\bf Z}_3$ orbifold mentioned above. An extension of this scenario for intersecting
 brane models has been considered in \cite{lusti}\ with similar conclusions. 
Similar considerations were discussed previously by Tye {\it et al} in
\cite{branecosmology}.

We may say that this scenario has several advantages. It naturally freezes 
the modulus related to the brane separation and makes the radius 
(or dilaton) the inflaton, reducing in a dynamical way the number of moduli.
 Inflation, due to the exponential dependence in the potential is obtained in
 a natural way without fine tuning. Still preserving the nice features
of the tachyon field realising hybrid inflation. It has also some
 disadvantages, for instance it is clearly more complicated than the
 original scenario. 
The main weak point is the 
ad-hoc assumption of some moduli fixed by other string effects leaving the 
potential depending on one of them, the inflaton. We mentioned that there are
models in the
literature that can achieve this partial fixing of the moduli, but at the
moment there is no model in the literature that leaves unfixed the modulus
 that gives rise to inflation. Without this assumption the
potential is runaway and surprisingly, although in the Brans-Dicke frame 
is attractive, in the Einstein frame is repulsive even though describes the
 brane/antibrane interaction.

 \subsection{The Rolling Tachyon}

We have mentioned that the open string tachyon can play an important role in 
the D-brane inflation scenarios by providing the graceful exit of inflation 
and realising hybrid inflation. The effective action for the tachyon has 
been subject to intense study during the past several years. It has been probably the most important 
result that has emerged of string field theory and it provides, together with the potentials for 
D-brane interactions, one of  the few
concrete potentials
 derived from string theory. It is then worth investigating the possible implications 
of this potential in detail. 

In the context of brane/antibrane or intersecting brane inflation, it is important to understand the reheating mechanism that the tachyon is responsible for.
But more generally, it is interesting  to isolate the tachyon by itself and ask what kind of cosmological implications it may have.

The results of different formalisms within string theory have provided
an explicit expression for the tachyon effective Lagrangian which depending on the string theory it may take different  forms. For the bosonic string, up to 
two derivatives:
\eq
{\cal L}_b\ = -\sqrt{-g} e^{-T}\left((1+k_bT)\partial_\mu T\partial^\mu T+
(1+T)\right)\ .
\eeq
With $k_b$ a constant usually taken to be $k_b=0$.
For the supersymmetric NSR string, the boundary conformal field theory 
and other related formalisms have provided the expression:
\eq
{\cal L}_s\ = -\ \sqrt{-g} e^{-T^2}\left( (1+k_s T^2)\partial_\mu T\partial^\mu T
+1\right)\ .
\eeq
With $k_s$ again usually taken to be $k_s=0$ \footnote{It is usually argued that these constants can be set to zero
by means of a field redefinition.}. And $T$
 stands here for the modulus of the tachyon field that is complex in this case.

Both Lagrangians provide interesting potentials for the tachyon field
which runaway to $T\rightarrow \infty$.
 It can easily be seen, working with  the canonically normalised field, 
that the tachyonic mass is of order $-M_s^2$ whereas in the minimum
the second derivatives give us a mass$^2$ 
of order $1/k_b, 1/k_s$ respectively. 
That is, for those constants taken to zero the physical 
tachyon has an infinite mass. Otherwise the physical field would have
a finite mass ( a double well potential in the supersymmetric case).

More generally, the string calculations suggest that to all orders in derivative expansion these actions can take a Born-Infeld form.
\eq
{\cal L}\ = -\ V(T)\ \sqrt{1- g^{\mu\nu} \partial_\mu T \partial_\nu T}\ ,
\eeq
where $V(T)$ can take different forms depending on the type of string theory, 
namely bosonic or supersymmetric. It is this form of the Lagrangian that has been studied recently.

First, Sen studied the rolling of the tachyon to its asymptotic minimum 
$T\rightarrow \infty$ and concluded that even though the vacuum should correspond to the closed string vacuum and the unstable D-brane system 
(such as brane/antibrane pairs or non BPS D-branes) has decayed. The energy density is still localised. Furthermore he was able to prove that the resulting gas corresponded to a pressureless gas. This is easy to see from the
effective action above for which the stress energy tensor give
for a time dependent tachyon:
\eq
\rho\ =\ \frac{V(t)}{\sqrt{1-\dot T^2}}\ , \qquad p\ =\ -V(T)\sqrt{1-\dot T^2}
\ .
\eeq
 For constant energy density the pressure goes like $p=-V^2/\rho$ and at the
 minimum in $T\rightarrow \infty$ we know that $V\rightarrow 0$ and so 
$p\rightarrow 0$. The equation of state is $p= w\rho$ with $w=-(1-\dot T^2)$ and therefore
$-1\leq w\leq 0$.

Having a time dependent tachyon field we should actually have  considered a time dependent metric such as FRW. 
In \cite{garytac} this was done obtaining the 
Friedmann's equations for this Lagrangian coupled to 4D gravity:
\eqa
H^2 & = & \frac{8\pi G}{3} \frac{V(T)}{\sqrt{1-\dot T^2}} - \frac{k}{a^2}\ , 
\nonumber \\
\frac{\ddot a}{a} & = &  \frac{8\pi G}{3} \frac{V(T)}{\sqrt{1-\dot T^2}}
\left(1-\frac{3}{2} \dot T^2\right)\ .
\eeqa
Without the need to solve these equations it can be seen easily that the
energy density decreases with time, while $T$ increases, relaxing towards the
 asymptotic minimum of the potential. In the meantime the universe expands,
accelerating first ($|\dot T|<2/3$) and decelerating after ($|\dot T|>2/3$).
Depending on the value of the curvature
 $k=0,1,-1$, the scale factor $a(t)$ goes to a constant for $k=0$, 
to a Milne universe $a(t)\rightarrow t $ for $k=-1$ and re-collapses for 
$k=1$.

It is also natural to ask if this tachyonic potential can give rise to
 inflation by itself. In \cite{tacos} it was  proposed that the fact that
 the tachyon potential has topological defects in terms of lower dimensional
 D branes, they may be  a source of topological inflation (inflation generated
 by the existence of a domain wall providing a cosmological constant due to
 its tension).
More generally, in \cite{quei}, it was looked if either of the above
 listed potentials would give rise to slow roll conditions with a negative
 result. The main reason is the absence of small parameters in the potential 
that can be tuned to give enough slow rolling.
Similar conclusions were obtained by Kofmann and Linde in \cite{garytac} when analysing the 
full action (including the higher derivative terms). In this case the density 
perturbations were also off scale. See the last reference of \cite{garytac} for a recent proposal to 
obtain tachyonic inflation adding a brane world.

Still, the more successful cosmological role for the tachyon, is providing the 
mechanism to end inflation in the hybrid inflation realisation in string
 theory. In this sense  the decay of the tachyon has to provide the 
source for reheating. 
 Two recent discussions in this direction have arrived at
 positive results, in the sense that this tachyonic reheating can work, 
although in a way different from standard reheating \cite{cline}.
Also,  there is much to be learned about the 
dynamics of the brane/antibrane annihilation process \cite{gsy}.
Certainly more work is needed in this direction before having a 
successfull scenario.

\subsection{S-Branes, dS/CFT and Negative Tension Branes}

\EPSFIGURE[r]{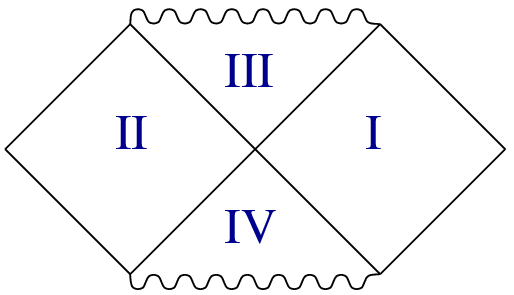,width=8cm}{Penrose diagram for the
 Schwarzschild black hole. There are four regions separated by horizons. 
Regions I and II are static and asymptotically flat. Regions III and IV are 
time dependent with a spacelike singularity.}
{\label{figure19}}

Recently a new class of objects were introduced in field and string theory 
named spacelike or S-branes. An S-brane is a topological defect for which 
all of its longitudinal dimensions are spacelike, and therefore it  exists
 only for a moment of time. There are several reasons to introduce these 
objects.  The simplest example  in field theory corresponds to 
a (tachyon) potential of the form:
\eq
V(\phi)\ = \ \left(\phi^2 - a^2\right)^2\ , 
\eeq
with minima at $\phi_{\pm}=\pm a$. In 4D this has the standard domain wall 
topological defect extrapolating between the regions where the field is in 
the
$\phi_+$ and $\phi_-$ vacua. This domain wall would be a  2-brane.
For a time dependent configuration in which we start at the maximum 
of the potential  $\phi(x,t=0)=0$
but with nonzero velocity $\dot\phi(x,t=0)=v$, we know that the field
will roll towards $\phi_+$ if $v$ is positive, until it eventually 
arrives at the minimum. A time reversal situation would have the minimum 
starting in $\phi_-$ and going to $\phi=0$ and therefore we can say that the
 field evolves from $\phi_-$ at $t=-\infty$ to $\phi_+$ at $t=\infty$ looking
 as a kink in time filling all spatial dimensions, so this would correspond to
 an
S2 brane (an S$p$ brane has $p+1$ spatial dimensions to follow with the
 tradition of standard $p$ branes notation). In practice this kind of process
 requires some fine tuned exchange of energy to the field to climb the barrier.
The process of rolling tachyon is then 
 a concrete realisation of S branes in string theory. 

The main motivation for the  introduction of the S-branes was  the conjectured
dS/CFT correspondence.
 The de Sitter (dS)  space has become more interesting due to the indications
 that the universe seems to be approaching a 
de Sitter geometry in the future.
The correspondence
 was proposed in \cite{strom} following some parallels with the
well established AdS/CFT correspondence,
 given the close connection between dS and AdS spaces.
(See however \cite{dls}).
The point is that a boundary at infinity of, say,  dS$_4$ corresponds to a
 Euclidean $R_3$
space for which the symmetry group of de Sitter space, $SO(4,1)$ acts as 
the conformal group
of the Euclidean $R_3$, suggesting that a conformal field theory on this 
boundary is dual to the
full 4D gravity theory in de Sitter space.
 One of the interesting outcomes of
 this 
conjecture is that the renormalisation group parameter can be identified with
time, in much the same way it was identified with the extra spatial 
coordinate in the AdS/CFT case. A simple way to see this possibility is 
by writing the dS$_4$ metric in FRW coordinates ($k=0$):
\eq
ds^2\ = \ -dt^2\ + \ e^{Ht}\ d\overrightarrow{x}^2,
\eeq
with $\overrightarrow{x}$ the spatial coordinates 
 and $H$ the Hubble parameter. The interesting observation is that this
 metric is invariant under
$t\rightarrow t+\lambda$, $\overrightarrow{x}\rightarrow e^{-\lambda H}\overrightarrow{x}$ which 
generates time evolution in the 4D bulk and scale transformations in the
 Euclidean boundary.
Late times (large values of $\lambda$) correspond to small distances (UV
 regime) whereas earlier 
times to IR regime.
Generic expressions for the scale factor $a(t)$ will not have this symmetry 
but if we assume that
$H(t)$ goes to a constant in the infinite past and infinite future we can 
see the time evolution
between two fixed points under the renormalisation group, which could
 eventually be identified
with early universe inflation and current acceleration. The monotonic 
evolution in time fits well  with the 
expected c-theorem of field theories, shown to hold at least in 2D. 
The RG flow would corresponmd to the direction from future to past. 
This is a very tantalising proposal but unlike the AdS/CFT correspondence 
there is no much support yet for the
dS/CFT one. S-branes are
 an attempt to bring this correspondence closer to the AdS/CFT one, with the
 S-branes playing the role of the D-branes in the
boundary (the Euclidean $R_3$ in the example above).

Using the analogy with $p$ branes, we expect that the S branes could
also  be 
found as explicit solutions of Einstein's equations
 (coupled to dilaton and antisymmetric tensor fields). In the same way 
that $p$ brane solutions
are black hole-like, we then expect that S brane solutions are 
time-dependent backgrounds of the theory, and therefore, they may have a
 cosmological 
interpretation. This is actually the case. Recently, solutions with these
 properties have been found, some of them were previously known. Rather than 
describing the general solutions of the Einstein-dilaton-antisymmetric tensor
system, I will choose to describe one simple example and extract its physical 
properties. The reader is referred to the literature for the general
cases \cite{forste,andre,pope,gqtz,gutperlestrom,gutperle,rob}.

The example we will concentrate on is the simple case of just 4D Einstein's equations in vacuum.
Let us first recall the Schwarzschild black hole solution.
 We know the general static solution with spherical symmetry is just the
 Schwarzschild  black hole solution, which in modern terminology is a black
 0-brane.
The solution is usually written as:
\eq
ds^2_I\ = \ -\left[1-\frac{2M}{r}\right]\ dt^2\ 
+\ \left[1-\frac{2M}{r}\right]^{-1}\ dr^2 \ + \ r^2 \left(\sin^2\theta\
 \ d\phi^2\ +\ d\theta^2\right).
\eeq 
This metric is only valid in the region $r>2M$ at $r=2M$ there is a horizon 
which changes the relative signs of the metric and then 
 for $r<2M$ the role of $t$ and $r$ are exchanged and the metric becomes:
\eq
ds^2_{II}\ = \ -\left[\frac{2M}{t} - 1\right]^{-1}\ dt^2\ + \ 
\left[\frac{2M}{t}-1\right]\ dr^2\ + \ t^2\left(\sin^2\theta d\phi^2\ +
 \ d\theta^2\right).
\eeq
This is then a time dependent region that ends in the singularity $t=0$
 (usually called $r=0$). Actually it is well known that there are two copies
 of each of these regions to have the complete causal structure of this
 spacetime.
This is properly obtained by going to Kruskal coordinates. As in the FRW
 case we can write a Penrose diagram describing the structure of this 
spacetime.

In  the Penrose diagram, Fig.~19 we illustrate the two copies of the asymptotically flat static
 regions separated by $45$ degrees lines corresponding to the horizons.
 The spacelike 
singularity described by the wiggled line is in the time dependent
 regions.
This $(p=0)$ brane  can be used to look for a S0 brane. In this case we need 
the symmetries not to have the  spherical symmetry
 $SO(3)$ of the black hole,  but actually the  hyperbolic symmetry
 $SO(2,1)$ as suggested by the fact that S branes are kinks in time.
 We can then see that 
we need instead of a sphere ($k=1$) a hyperbolic space ($k=-1$). This is 
very simple to obtain since both are related by an analytic continuation.

Therefore we can take the Schwarzschild solution and perform the
 following transformation $t\rightarrow ir, r\rightarrow it,
 \theta\rightarrow i\theta, \phi\rightarrow i\phi$ together 
with $M\rightarrow iP$. Both metrics above become:
\eq
d\hat s_I^2\ = \ -\left[1-\frac{2P}{t}\right]^{-1}\ dt^2\ + 
\ \left[1-\frac{2P}{t}\right] \ dr^2\ + \ t^2\ \left(\sinh^2\theta\ \ 
d\phi^2\ + \ d\theta^2\right), 
\eeq
whose surface of constant $r$ and $t$ is the 
hyperbolic plane ${\cal H}_2$ rather than the two-sphere, as expected. 
In addition to the symmetries of the hyperbolic space
 it has a spacelike Killing vector $\xi=\partial_r$ but is time dependent, 
again as expected for a S0 brane. The apparent singularity at $t=2P$ is
 again a horizon. For $t<2P$ the metric is:
\eq
d\hat s_{II}^2\ = \ -\left[1-\frac{2P}{r}\right]\ dt^2\ 
+ \ \left[1-\frac{2P}{r}\right]^{-1} \ dr^2\ + \ r^2\ \left(\sinh^2\theta\ \
d\phi^2\ + \ d\theta^2\right), 
\eeq
which is now  static with the timelike singularity at $r=0$.

\EPSFIGURE[r]{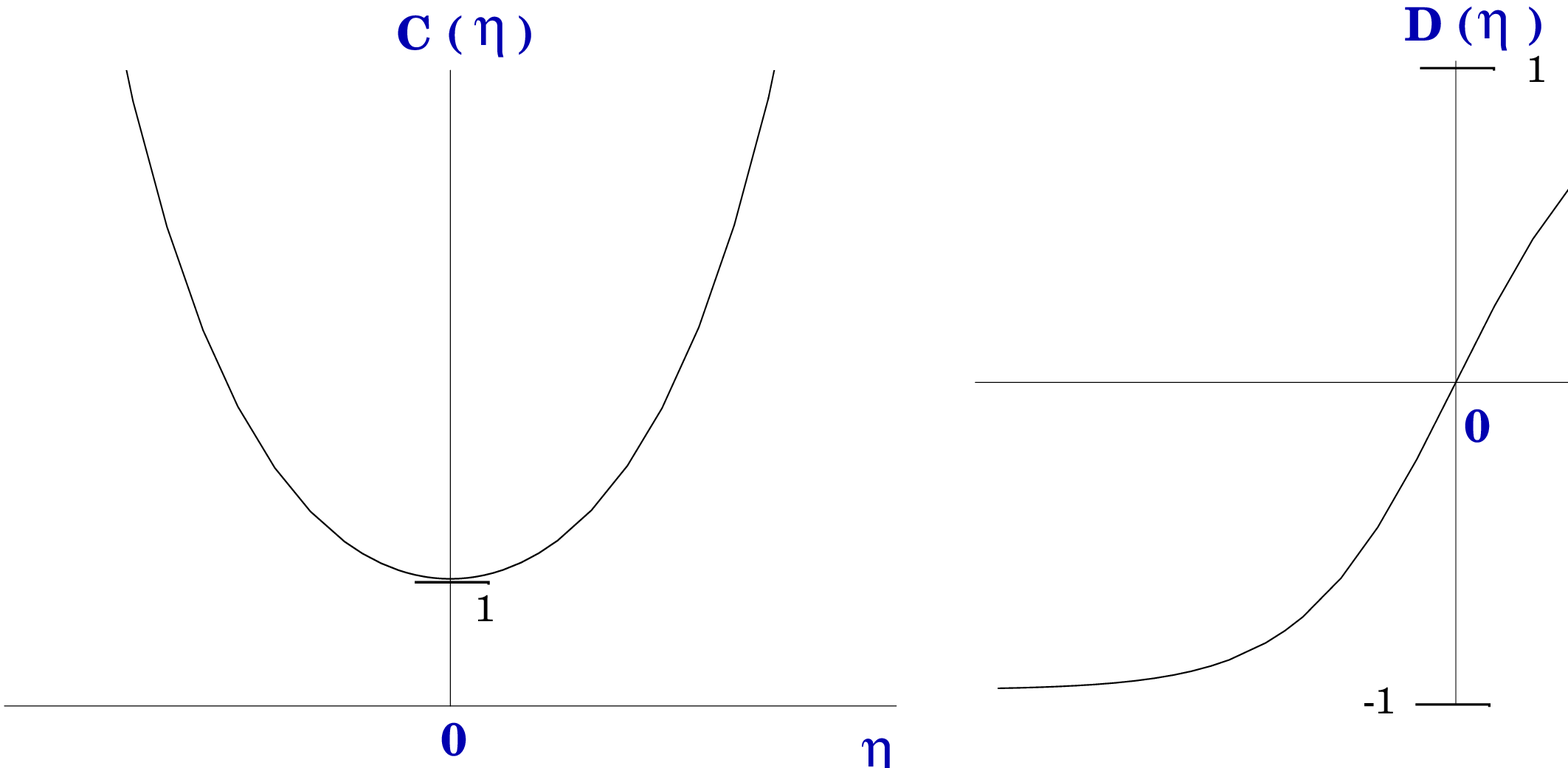,width=8cm}
{\sl The bounce solution for the metric and the kink behaviour of the
`extra dimension' $r$. The kink corredponds to the location of the
horizon that is identified with the S-brane.}

The corresponding Penrose diagram looks very interesting. It is just a $90$
 degrees rotation of the one for the Schwarzschild solution. 
Now the asymptotically flat regions ($I,III$) are time dependent. One can be
 thought to correspond to 
 past cosmology and the other as future cosmology, looking as
an appropriate metric for a pre big-bang scenario. Except that here there
 is no big-bang. Since extrapolation to the past for an observer in region
 $I$ brings him to the horizon,
 which is identified with the position (in time) where the
 S0 brane is located \cite{gutperlestrom}.
 The singularity is in the static region 
that connects the two time dependent regions.
It is easy to see that the metric is asymptotically flat, in the time 
dependent regions, and the near horizon geometry is a 2D Milne universe
($ds^2=-dt^2+t^2 dr^2$) times the hyperbolic surface.

%\cleardoublepage
\EPSFIGURE[r]{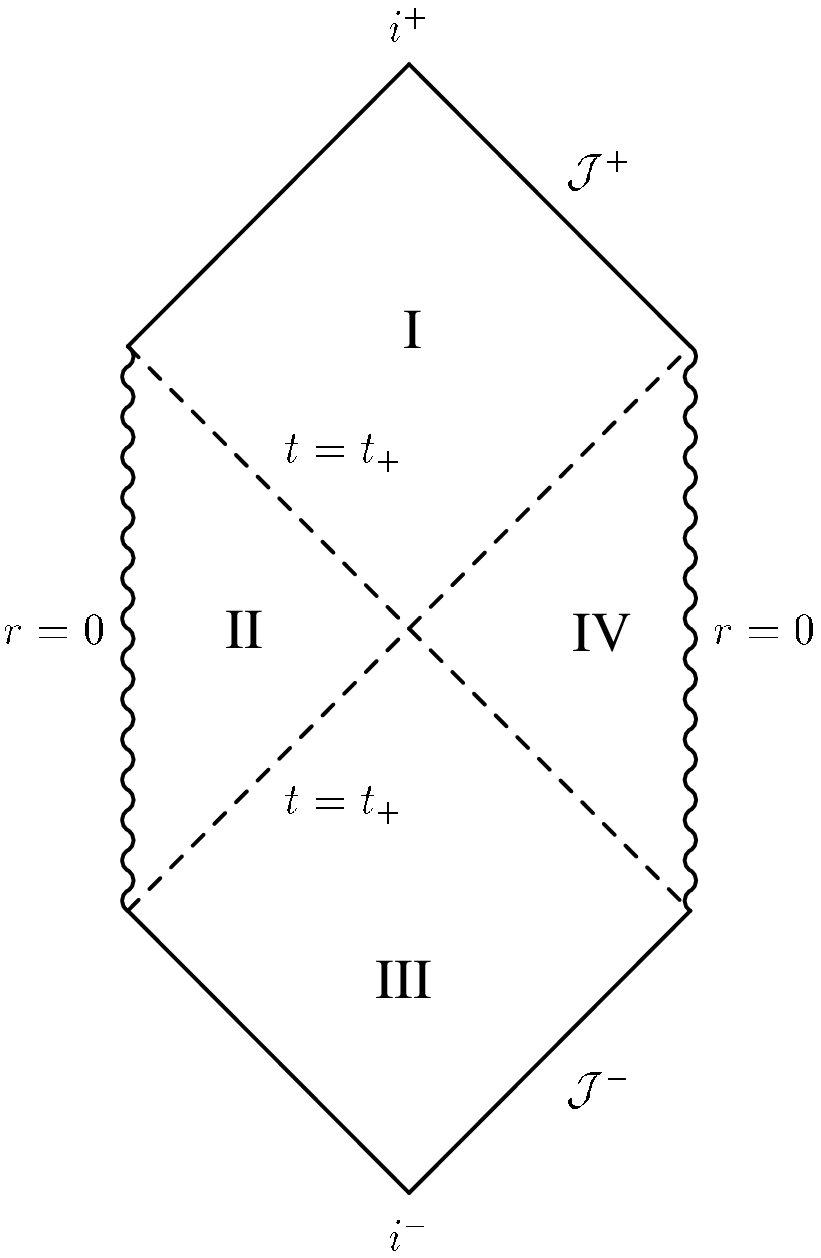,width=10cm}
{\sl Penrose diagram
for the $k=0,-1$ brane solution. This diagram is very similar to the
Schwarzschild black hole (rotated by $\pi/2$),
but now region $I$ ($III$) is not
static, but time-dependent with a Cauchy horizon (at $t=t_+\equiv P$)
 and region
$II$ ($IV$) is static.\label{fig20}}
%\clearpage

To see better the S brane interpretation we can go to a frame for which the
 metric in the cosmological regions takes the form:
\eq
ds^2\ = \ C^2(\eta )\left[ -d\eta^2 + d\Sigma_{k=-1} \right] \ +\ D^2(\eta)
dr^2 ,
\eeq
where $d\Sigma_{k=-1} $ is the metric for the hyperbolic space in $2$ 
dimensions.
 The conformal time is defined by 
\eq
C(\eta)\ = \ t(\eta)\ = \ P \cosh^{2}\left[\frac{\eta}{2}\ \right],
\eeq
and so $\eta$ lies within the range $-\infty<\eta<\infty$. The scale factor 
for $r$ becomes:
\eq
D(\eta)\ = \ \tanh\left[\frac{\eta}{2}\ \right] \ .
\eeq
 These 
expressions exhibit the bouncing structure of the $3$  dimensional space and 
the timelike kink 
structure of the radial dimension. The position of the kink is precisely at 
the horizon, fitting very nicely the S brane interpretation. The bounce 
would seem to  indicate that if we concentrate only on the time dependent part
of the metric it looks like a bouncing cosmology with a contracting universe
 in the past, passing smoothly to an (exponentially) expanding universe in
 the future.

We know of course that the geometry includes also the singular static regions
and therefore the transition from the past to the future cosmology has to
 pass through this region. Actually, remembering that the geometry is a
 rotated black hole
we can borrow a nice interpretation from the Schwarzschild
black hole, namely the wormhole or
 Einstein-Rosen bridge that connects the two asymptotically flat regions
 of the 
black hole. In our case, this  is a timelike wormhole connecting 
the past and future cosmological regions. See the figures~22,23 that illustrate the 
wormhole as the bridge between the two cosmological regions and then 
producing the bouncing.

\cleardoublepage
\EPSFIGURE[r]{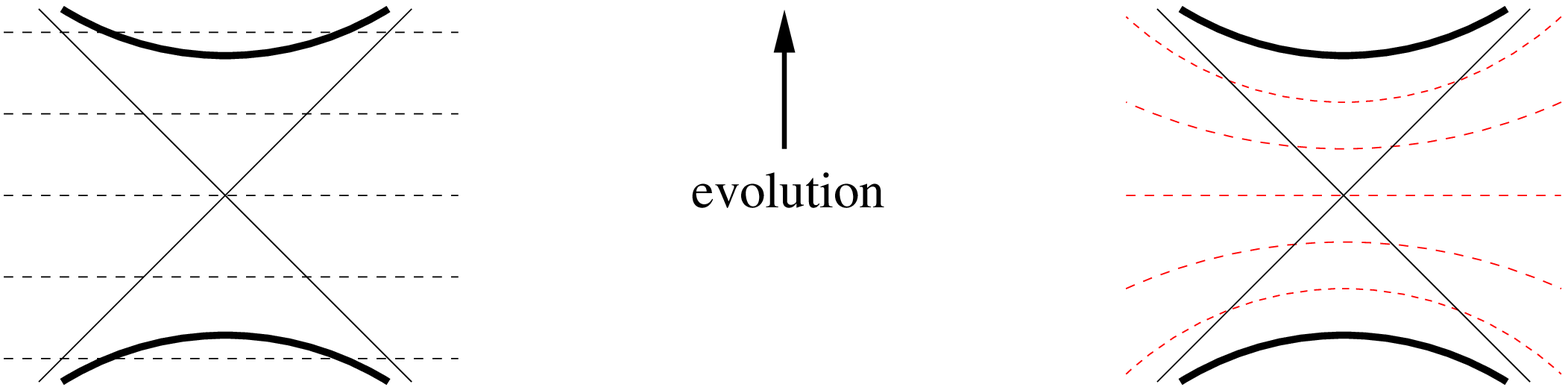,width=12cm} {Different foliations for the timelike
 wormhole
 connecting past and future cosmologies.}
{\label{figure21}}

\EPSFIGURE[r]{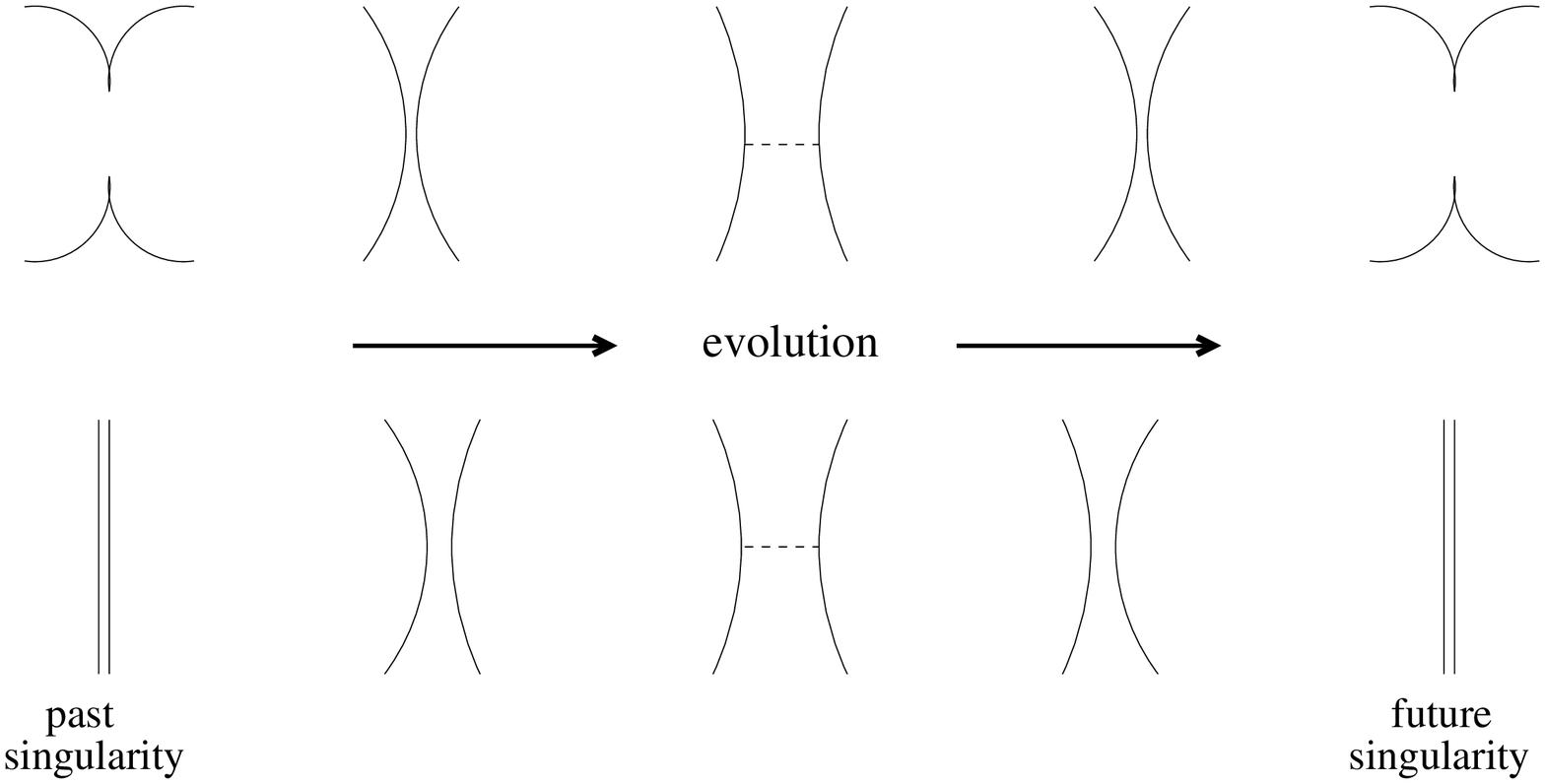,width=12cm}{The time like wormhole.}
{\label{figure22}}
\clearpage

As mentioned before this example is just a particular case of a general
 class of supergravity solutions representing S $q$-branes
in $d$ dimensions, found starting 
with 
the Lagrangian for the Einstein, dilaton, antisymmetric tensor $F_{q+2}= 
dA_{q+1}$
\eq
{\cal L} \ = \ \sqrt{-g}\left( R-\frac{1}{2} 
g^{\mu\nu}\partial_\mu\varphi\partial_\nu\varphi-\frac{1}{2(q+2)!}
 F_{q+2}^2\right)
\eeq
There are solutions similar to the ones just discussed for 
hypersurfaces with $k=-1$ and also $k=0$ with identical Penrose diagram.

%\cleardoublepage
\EPSFIGURE[r]{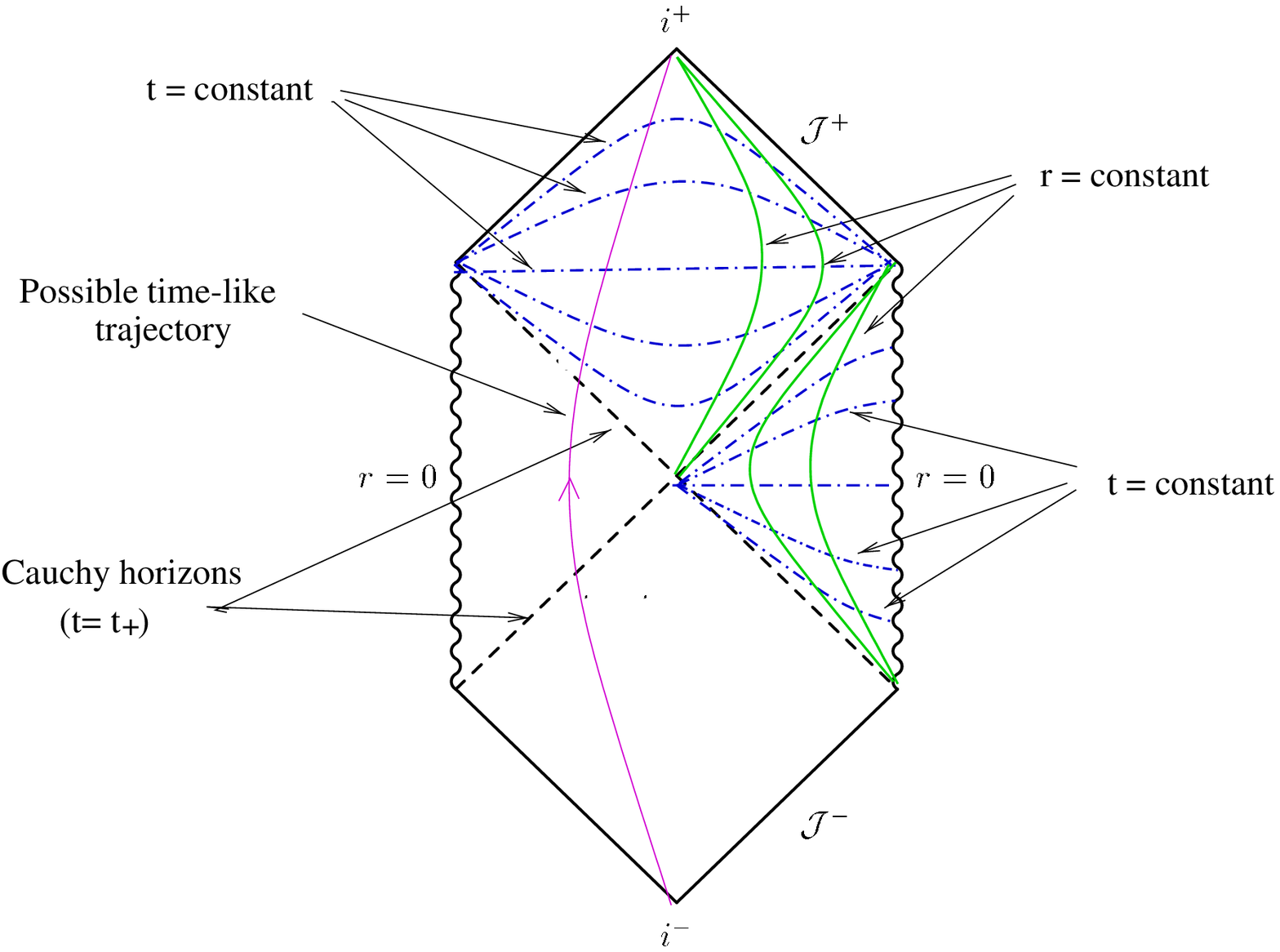,width=8cm}{Penrose diagram illustrating the 
geodesics trajectories and the 
surfaces of constant time. The singularities are repulsive to timelike
 geodesics.}
{\label{figure24}}
%\clearpage

In this case we can assign a charge with respect to the field $F_{q+2}$. 
However this charge, defined as an integral in the cosmological region will 
not be conserved in time (it will be `conserved' in space, {\it i.e.}
 along a 
surface of constant $t$). A similar argument could be used to define a
 mass for this object. This is somehow unsatisfactory since there does 
not seem to be a really 
conserved quantity that  identifies this geometry, 
contrary to the mass and charge of the black holes. Actually, this is 
only the case if we ignore 
the static regions.
 However,
having the static region provides us with a way to actually identify 
correctly this geometry. It turns out that the singularities are the
 physical objects (as 
it should be expected) to which mass (or tension) and charge can be
 assigned unambiguously.
It is found that the two singularities correspond to negative mass 
objects with opposite charge \cite{bqrtz,cck}. 
Furthermore, the similarity with black
hole geometry 
indicates that there will be particle production 
 and then we 
 can also compute a generalised Hawking temperature, which could have 
interesting cosmological interpretation. Furthermore the entropy can
be computed. It was found in \cite{bqrtz}\ 
that the entropy  density is proportional to the determinant of the
metric in the hyperbolic or flat space,
 generalising then the famous $1/4$ area 
expression valid for $k=1$, for which the entropy is finite. For $k=-1,0$ the 
area of the horizon is infinite but the relation holds locally.

Finally in this kind of geometries we have to worry about
 stability. Even though the solution seems to be stable  under small
 perturbations, the horizons may suffer some instability similar to
 the Reissner-Nordstrom black hole, in which the interior horizons are
 unstable. A
 naive
 calculation indicates that the past horizons seem to be unstable
 under some scalar field  perturbations \cite{bqrtz}.
 A more complete analysis  is needed to confirm this is
actually  the
 case.

Other generalisations of S$p$-branes  exist. For $p\neq 0$, requiring
 the full
$SO(n,1)\times ISO(p+1)$ symmetry,  naturally generalises the
 D-brane solutions of $k=1$ to $k=-1$.
These solutions yield generically to different
 Penrose's diagrams than the one discussed here, with
the location of the S brane corresponding to a singularity instead of
 a horizon,
\cite{gutperle,rob}. 
However there are some general classes 
that have the same global structure as the S$0$-brane just described.
The generalisations for which the symmetry is 
$SO(1,1)\times O_k(n)\times ISO(p)$ where $O_k(n)$ refers to
 $SO(n-1,1)$ for $k=-1$ and $ISO(n)$ for $k=0$, do have the Penrose
 diagram of Fig.~20 \cite{bqrtz},   for which the singularity is $p$
 dimensional and are the analogue for $k=0,-1$ of the black $p$-brane
 solutions of Horowitz and Strominger ($k=1$) for which the symmetry
 $O_k(n)$ is $SO(n)$.

\subsection{Ekpyrotic/Cyclic Scenarios}

So far we have mostly discussed cosmology associated with the physics of D-branes appearing in type IIA, IIB
closed string theories 
and type I open strings. 
Let us now discuss brane cosmology in the other way of getting realistic
 string models, namely the Horava-Witten scenario. This scenario corresponds
 to M-theory compactified on an interval $S^1/{\bf Z}_2$ for which the two 
10D 
end points have an $E_8$ gauge theory providing the strong coupling 
realisation of the heterotic string. Further 
compactification on a six-dimensional Calabi-Yau manifold then leave two 
4D worlds at the ends of the interval in the 5D bulk. Again quasi realistic
 models can be obtained from this approach using mostly the topological
 properties of Calabi-Yau manifolds. It turns out that besides the end
 of the interval 
world (which we will refer to boundary  branes)
 there are also in the compactifications 5 branes that are not restricted to
live at the fixed points and can actually move through the bulk.
 These are called bulk branes. We have then configurations very similar
 to D-branes at orbifolds, although there are no D-branes
 in this construction.

\EPSFIGURE[r]{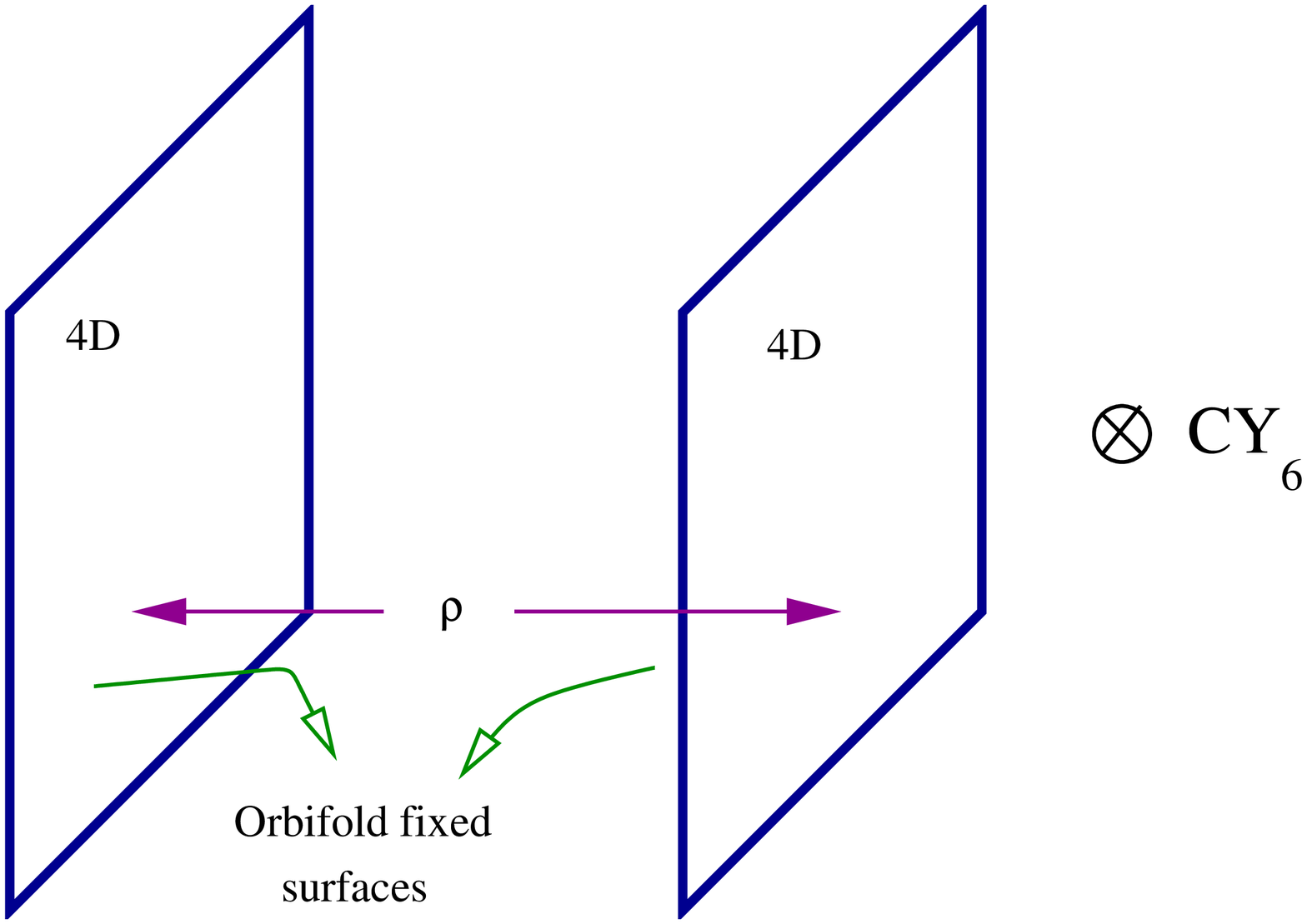,width=8cm}{The Horava Witten scenario. 
Two surfaces each at the end of 
the interval provide chiral matter and possibly interesting cosmology.}
{\label{figure25}}

A very interesting proposal was 
made in \cite{ekpyrosis} regarding  the collision of branes, this time
not to obtain inflation but an alternative to inflation. The original idea
was to assume that a bulk brane going from one boundary of the interval
 to the other end, would collide with the second boundary 
brane and 
produce the big-bang. The bulk brane would be almost BPS by which it was 
meant that it is essentially parallel to the boundary branes, moves slowly
 from one end to the other of the interval and small quantum fluctuations 
induce some ripples on this brane which when colliding with the visible 
brane would produce the density fluctuations measured in the CMB. There is 
no need of an inflation potential for this. A potential of the type 
$-e^{-\alpha Y}$ was proposed (although not derived) describing  the attraction of the branes. 
The 5D metric is taken with a warp factor that implies that the
motion is
 from smaller to larger curvature across the interval. Therefore the scale 
factor depends on the position of the brane in the interval. 

%\EPSFIGURE[r]{ekpyrotic1.eps,width=8cm}{The first ekpyrotic
% scenario with one bulk brane moving and colliding
% later with one boundary brane.}
%{\label{figure26}}%

Several criticisms have been made to this proposal. First, regarding
the standard problems solved by inflation. The horizon and flatness
problems require the  branes to be very parallel before collision
which may require fine tunning of initial conditions. Relics such as
monopoles will not be present if the collision temperature is low
enough, but this has to be quantified. There is no general natural
dilution as in inflation, making the solutions of these problems more
difficult in general.  The issue of fine tuning
 the initial conditions has been 
very much debated \cite{pyrot}.
However, the main difficulty of this scenario  is the following
\cite{kosst}:
it so happens that in a 4D description,
 $\dot a<0$ before the collision and  is expected that $\dot a>0$ after the collision,
 which means passing from contraction to expansion, without crossing a 
singularity. This is a problem because it violates the null energy condition.
 Let us review this argument briefly.
Starting with gravity coupled to a scalar field:
\eq
{\cal L}\ = \ \sqrt{-g}\left( R-\frac{1}{2} 
\partial_\mu\phi\partial^\mu\phi-V(\phi)\right)
\eeq
we know that the energy density and pressure are given by:
\eq
\rho\ = \ \frac{1}{2}\dot\phi^2 + V\ \qquad p\ = \ \frac{1}{2}\dot\phi^2 - V 
\eeq
Einstein's equations then give:
\eq
\dot H\ = \ -\frac{1}{4}\left(\rho+p\right)\ =\ 
-\frac{1}{4}\dot\phi^2\ \leq \ 0.
\eeq
Therefore $H$ is  monotonically decreasing and 
we cannot go from contraction ($H<0$) to expansion ($H>0$).

A way to avoid this problem is to simply get rid of the bulk brane and 
consider the collision between the two boundary branes. In this case, there is 
a singularity at the moment of collision, since the size of the fifth dimension reduces to zero, 
 that could allow the transition from 
contraction to expansion. This is the second version of the ekpyrotic scenario.
The singularity happens to be only in the extra dimension because the scale 
factors of the branes remain finite during the process. After the collision 
the 
two branes separate again and the scale factor increases. 
\cite{kosst}.

\EPSFIGURE[r]{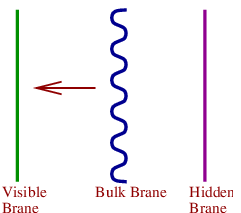,width=6cm}{The first ekpyrotic scenario. 
The motion of the bulk brane with small ripples caused by quantum fluctuations, 
could be the origin of density perturbations after colliding with the other
 boundary brane.}
{\label{figure27}}

%\EPSFIGURE[r]{ekpypot.eps,width=8cm}{
%A possible potential for the realisation of the first 
%ekpyrotic scenario.}
%{\label{figure28}}%

In a 4D effective action, 
this process can be understood in terms of the discussion we had 
in the pre big-bang section. Neglecting the scalar potential and identifying 
the separation between the branes with the string dilaton (as it happens in 
the Horava-Witten scenario) we can use equations (\ref{ven6})-(\ref{ven7})
for the case $d=4$. Out of the four possibilities provided by the choices of
 sign we can choose:
\eq
a(t)\ = \ |t|^{1/2}\qquad \varphi\ = \ \varphi_0\pm \sqrt{3}\log|t|.
\eeq

%\EPSFIGURE[r]{ekpyrotic2.eps,width=6cm}{Second ekpyrotic
% scenario with the boundary branes
% moving and colliding.}
% {\label{figure29}}

The behaviour of the scale factor $a(t)$ is clearly from contraction
at negative $t$  to
 expansion at $t>0$. But this still leaves the choice of sign for the 
dilaton open. Since the string coupling is proportional to 
$e^{-\phi}$, the $-$ sign  choice (taken in the pre big-bang scenario) 
corresponds to strong string coupling whereas the $+$ choice 
(chosen in the ekpyrotic scenario) implies weak coupling at $t=0$. Therefore 
they conjectured that the transition is smooth at the singular point and 
the branes start to separate again. We show in  figure~27 a cartoon version
 of the process writing the sizes of the branes different just to remind that 
the warp factor changes with the separation of the boundary branes.

\EPSFIGURE[r]{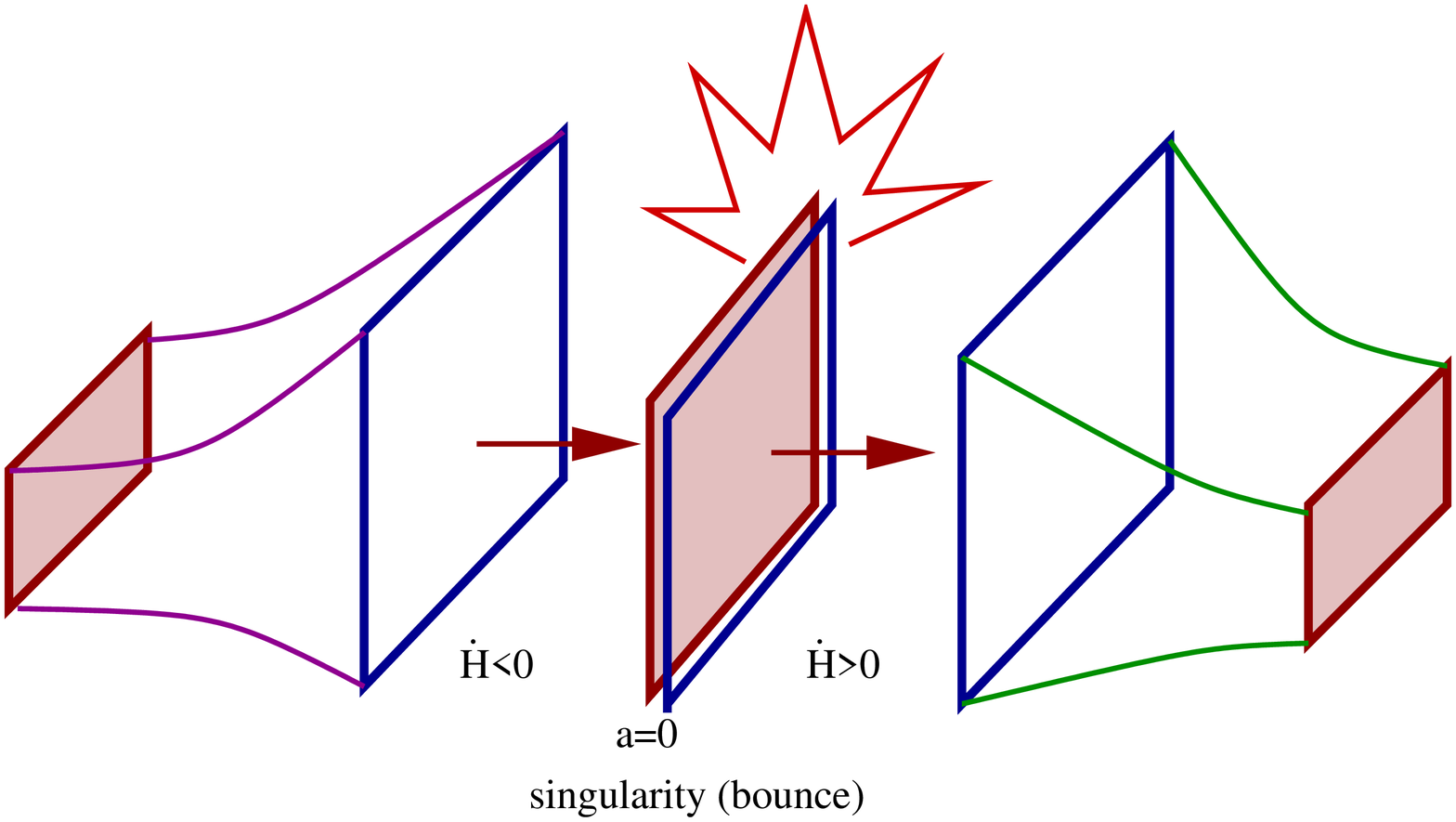,width=8cm}{A cartoon of the collision between
 branes and the passing through of the branes with the different scale
 factors in each case.}
{\label{figure30}}

Finally this leads us to the third version of this scenario that corresponds
 to the cyclic universe
\cite{cyclic}. In this case the two branes could keep separating and
 passing through each other an infinite number of 
times as long as the interacting 
potential has a very particular form. For instance for a potential like in 
 figure~28. We may describe the history as follows. Let us start with the right
 hand side that would correspond to today. The potential is slightly positive 
and flat reflecting the fact that the universe accelerates today 
(a mild inflation or quintessence).
 Since the slope of the potential is slightly negative the scalar field
will start rolling towards smaller values: the branes approach each other.
At some point the field will cross the $V=0$ point and its energy density
 will be kinetic. The potential becomes negative very fast
 and at some point the
 energy density $\rho= V+\frac{1}{2}\dot\varphi^2$ touches zero, implying 
that the universe starts contracting. Since the kinetic energy is also 
large the 
field easily passes through the minimum, towards the flat region at infinite
$\varphi$ (zero string coupling) where the branes collide
and bounce back, with enough energy as to cross again the steep minimum and
go to the right hand side of the 
potential, where it will get to the radiation dominated era, then repeat the
 whole cycle again.

\EPSFIGURE[r]{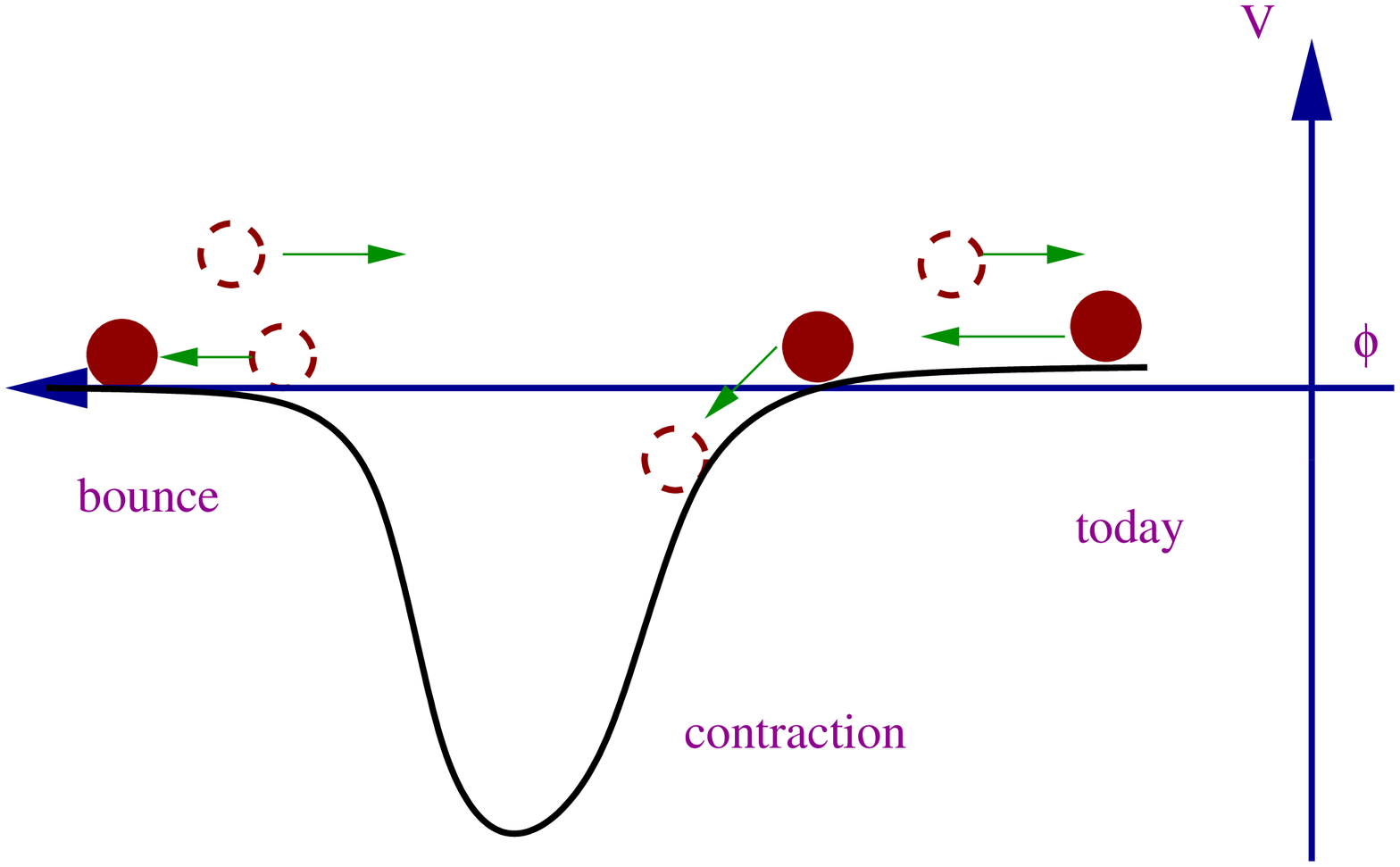,width=8cm}{An illustration of the potential and
 trajectory of the field in the cyclic universe.}
{\label{figure31}}

These scenarios claim to  approach the questions solved by inflation. 
For  instance, the horizon problem does not exist if there is a bounce, since there will be
 clear causal contact between different points. In the cyclic version,
the late period of mild inflation plays a similar role as the
original inflationary scenario by dissolving some wanted objects, 
like magnetic monopoles, and emptying the universe for the next cycle, solving the
flatness problem.
 Also the spectrum of perturbations has been claimed to be 
consistent with observations although there has been a debate on this issue, 
which I am not qualified to judge \cite{brandenbergerfin}.
All parts seem to agree on the fact that the methods used so far are not 
conclusive one way or another. A full 5D treatment should be performed
and then face the singularity 
at the collision. 

There are very interesting aspects on these proposals, 
especially regarding  the revival of the cyclic universe.
Remember that this was proposed in the 1930's \cite{cyclicold}\
but it was immediately 
realised that the entropy increases on each cycle meaning that the length of
 the cycles also increase and extrapolating back in time we hit again 
an  initial singularity. Therefore making the model semi eternal. 
Similar to eternal inflation which also requires a beginning.
The entropy problem is solved as follows. It is true that the total entropy 
 increases with the cycles but the entropy of matter is always the same 
at the end of each cycle. This is due to the accelerated expansion at present
 which will dilute matter until bringing the universe essentially
 empty (one particle per Hubble radius), before restarting the cycle.
Even though this idea has been found in the context of the ekpyrotic scenario,
it is clearly independent of it and may have far reaching implications as
 well as different realisations.

Another interesting point of this scenario is that it connects the early
 universe and 
late universe in a coherent way. The current acceleration is used as a
 virtue to prepare
 the universe to the next cycle.

A weak point about these scenarios is the dynamics of the scalar field. 
Even though the scenarios are motivated in terms of string theory, the 
kind of potentials that work are relatively contrived and have not been 
derived from theory.
 This is definitely an urgent question to approach before these models 
can be considered genuine M-theory models. In this sense these scenarios are 
at
 present in the same stage as D-brane inflation was in 1998 where the scalar 
potential was only guessed, instead of explicitly calculated as in the 
brane/antibrane and intersecting brane models. Finding a potential with the 
proposed properties is certainly an interesting challenge.
 
The problems of the D-brane models
 also apply to this scenario. In particular the assumption of having 
fixed  
the moduli of the Calabi-Yau manifold is not justified.
  Although the main problem to deal with is the singularity 
giving rise to the bounce, which is a very strong assumption.
Observationally, the important points to address 
refer to the spectrum of density perturbations since this is what could 
rule out the model. A criticism of 
this scenario and its comparison with inflation has been presented 
in \cite{lindeekp}.

\subsection{Time Dependent Orbifolds}

The major  assumption of the ekpyrotic and cyclic universe scenario
is the smooth passing through the singularity. This has motivated much recent 
effort in trying to describe 
field and string theory
in such singular spaces. Given that the singularity can be associated to an
 orbifold singularity
 and the fact that  orbifolds have proven to be backgrounds in which string
 theory is well behaved, despite 
the  singularities, it is then natural to investigate the behaviour of string
 theory in orbifolds
which are time-dependent. The simplest case illustrated in \cite{hs,kosst} is
 the following.
Take the spacetime to be the product of a  2D space and a $D-2$ one,  with 
the 2D metric
\eq
ds^2\ = \ -dt^2 \ + \ t^2\ dx^2.
\eeq

\EPSFIGURE[r]{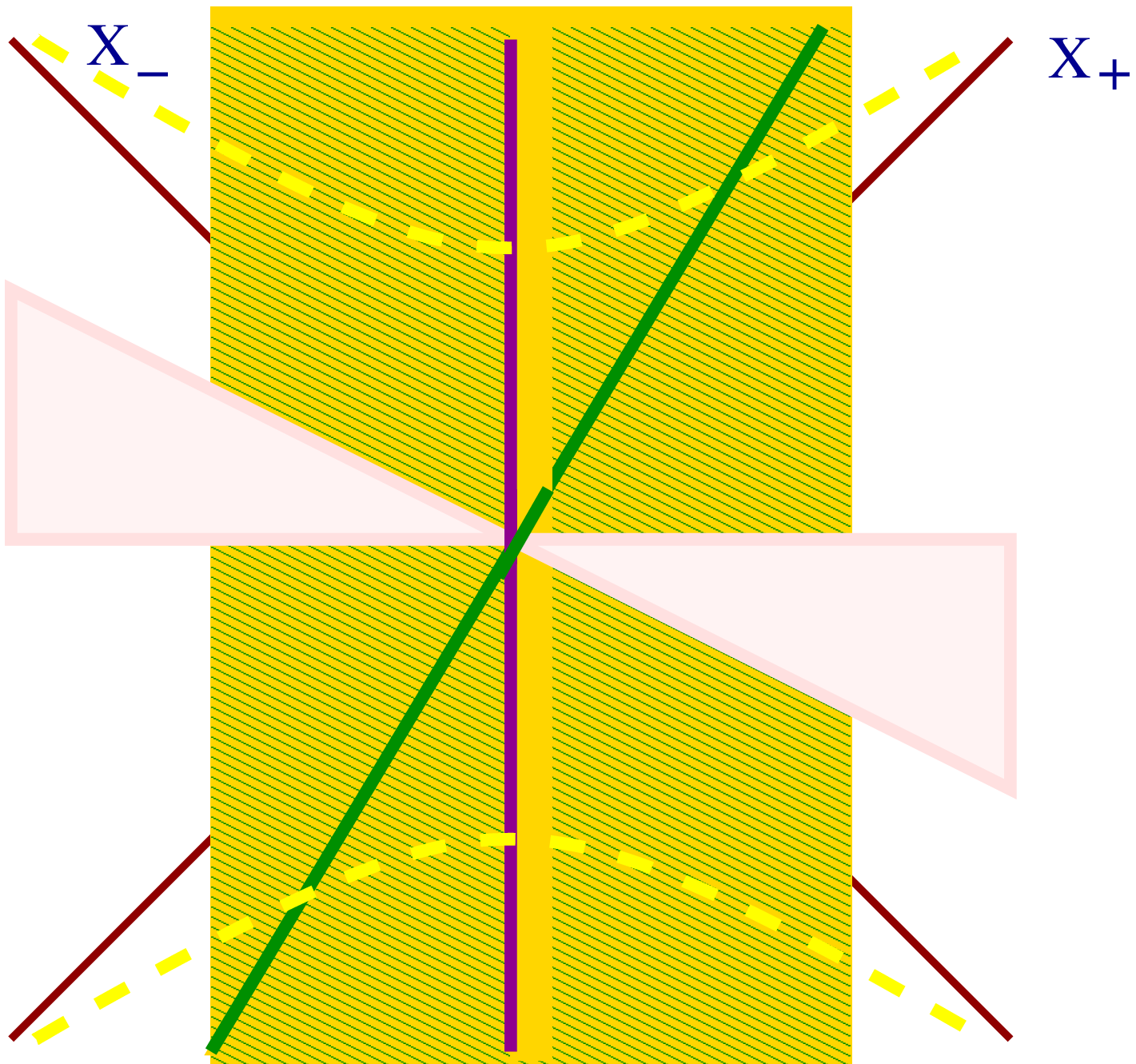,width=8cm}{An orbifold of the 2D Milne universe.
The horizontal wedges correspond to static regions with closed timelike curves.
The vertical wedges correspond to past and future cosmological regions
joined by the (non Haussdorf)
 singularity. If the radial direction is an interval, the 
heavy lines mark the trajectory of boundary branes, colliding and
moving apart.
The discontinuous lines are curves of constant time.}
{\label{figure33}}

With $x$ a periodic coordinate $x\sim x+\alpha$.  Before the identification 
this is just a realisation of flat 
2D Minkowski space (similar to the polar coordinates representation of the 
flat metric in 2D) which 
corresponds to the Milne universe. We can define the Kruskal coordinates:
\eq
X_{\pm}\ = \  t e^{\pm x}
\eeq
for which the metric looks as $ds^2= - dX_+ dX_-$. The identification 
$x\sim x+\alpha$ corresponds to multiplying
$X_\pm$ by a factor, which is a boost. This means we are orbifolding by a 
discrete
element of the Poincare group in 2D.  This defines an orbifold
with $X_\pm=0$ as a fixed point, which identifies the singularity. This space
was studied in \cite{hs, kosst}. 
This space has  closed time like curves in the left and right sections of 
 figure~29,
and the singularity would be such that the space 
is not separable (non Hausdorff). We may  still choose only the 
top and bottom `wedges', see the figure, reflecting the singularity at the 
origin connecting
past and future cosmologies. If the space defined by $x$  is an interval
 $(S_1/{\bf Z}_2$ as in Horava-Witten)
we can see the trajectories of the boundary branes (thick green and purple 
lines) which join and split again.
It was conjectured in \cite{kosst} that the branes  could actually pass the
 singularity smoothly.

This simple space as well as other variations have been recently studied
 in the context of trying to formulate a 
consistent string theory in time dependent orbifold backgrounds and ask how 
sensible to the singularity string
theory is. Several interesting results have been found, including the formulation 
of string amplitudes in this
 backgrounds and the construction of 
explicit time dependent backgrounds with at least one 
supersymmetry (something rare in time
 dependent backgrounds since the non-existence of timelike Killing vectors is 
usually an obstacle for
 supersymmetry). 

Furthermore, combining  
boosts with shifts, in reference 
\cite{costa}
 cosmological backgrounds with precisely the same Penrose diagram as the one 
discussed above in the
 S-branes section, were discovered (see figure~20): with  past and future cosmology 
separated by 
the static region with the timelike singularity. Although 
this geometry has closed time-like 
curves in the static region which were
 absent in the S-brane solutions.
 The causal structure of the Penrose diagram
 motivated the interpretation of the singularities as 
orientifold planes, causing the universe to expand
and a general discussion of the cosmological implications was done
in 
\cite{cck,costa} (see also \cite{bqrtz}). Several interesting 
cosmological issues were discussed
including the avoidance of the horizon problem, given that it is a horizon
 rather than the big bang singularity,
the starting point of the future cosmological region. It is interesting to 
see if these geometries have
more explicit relation with the solutions of \cite{gqtz,gutperle,rob}. Also if the apparent 
instability of the horizon also happens in this case.

 In the study of \cite{TDbackgrounds}, 
even though some interactions seem to pass smoothly through 
the singularity,
 some divergences were also found \footnote{See for instance the
article 
of Liu, Moore and 
Seiberg in \cite{TDbackgrounds}. We need to keep in mind that
this analysis was done in 10D string theory and not in 11D as in the
Horava-Witten scenario of the previous subsection.}.
 Moreover, a general argument 
was found in \cite{hp} in which
the presence of a single particle in this kind of spaces would induce a 
black hole to develop immersing the 
whole space into a large  black hole and having to face the standard 
singularity 
problem. The argument goes as follows.
Taking an orbifold like the one defined above, with 
\eq
(X_+,X_-)\rightarrow 
(e^{n\alpha}X_+, e^{-n\alpha}X_-)
\eeq
for any $n$ and a constant $\alpha$, 
and all transverse coordinates invariant. Two massless particles in the
 orbifold with impact parameter $b$
 smaller than the Schwarzschild radius would produce a black hole if: 
\eq
G\sqrt{s}\ > \ b^{D-3}.
\eeq
With $G$ Newton's constant and $s$ the squared canter of mass energy of the
 particles in total $D$ dimensions.
In the original space they
 need to be very close to each  but in the orbifold we have to include all the 
image particles under the orbifold twist, since the twist is a Lorentz boost,
 each of the image
 particles has boosted energy (the momenta transform as  the coordinates
 under the orbifold)
whereas the impact parameter $b$ does not change with $n$. Therefore for 
large enough $n$ the condition above
 is satisfied and the particles produce a black hole. Actually the
 Schwarzschild radius becomes
 $R_s\sim G e^{n\alpha}$ and for large enough $n$ $R_s$ can be as large as 
the whole space.
 Having the whole universe inside a black hole and then having a normal 
future big crunch 
singularity. Therefore it seems that time dependent orbifolds do not help
 into the 
problem of cosmological singularities and the regions of large curvature have
 to be dealt with in string theory. A possible exemption to this
problem are the   null branes introduced in \cite{null}.

\section{Final Remarks}

Probably the main ideas that were developed in our field in 
the 1980's are string theory
and inflation. 
Twenty years after, both continue very lively but their connection, if any,
is not yet understood.
Many ideas have been discussed recently in string cosmology: from pre 
big-bang cosmology to mirage cosmology, 
D-brane inflation and rolling tachyon condensation, S-branes, dS/CFT 
correspondence, ekpyrotic/cyclic scenarios, time-dependent orbifolds, 
etc. The important questions related with initial conditions, the treatment 
of singularities, definition of observables and consistency of de Sitter
 space with string theory and the holographic pinciple, 
remain open and may remain open for some
 time \cite{desitter,depression}. It is
still useful to come up with a phenomenological attitude to these issues
 and hope to make little progress
with time, before a possible breakthrough  comes up. This at least
can be a
 guide 
to what the relevant questions to be answered are. In some sense it is
similar to the
search for standard-like models from string theory, that has been very
fruitful over the years, identifying new 
phenomenological avenues in string and field theories, as well as 
increasing our understanding of
 string theory itself.
 Eventually a 
common property of several string cosmology scenarios could be identified that 
would be  close to a prediction from the theory.

We have seen several interesting proposals for a consistent string cosmology.
The S-brane ideas, dS/CFT correspondence and the rolling tachyon
are certainly interesting avenues to explore, as well as 
higher dimensional cosmologies with a richer global structure than the 
standard big bang, as those described in section 4.6. But at the moment they 
are not developed enough as to have some phenomenological impact.
On the other hand, there are three classes of scenarios that have been
put
 forward
that do have some phenomenological implications, namely: the pre
big-bang
 scenario, the
different versions of D-brane inflation and the ekpyrotic/cyclic scenarios.
Both pre big-bang and ekpyrotic/cyclic scenarios are usually presented as 
alternatives to inflation and end up in disadvantage given that
 inflation
is a scenario that has been evolving for more than 20 years and has many
 possible realisations, most of them not necessarily connected with a 
fundamental theory, such as strings. Whereas these alternative scenarios
have some constraints from their original top-down formulation that 
make them less flexible. This is similar to field theory against string theory
model building, which is more rigid. This is the reason why it has been very
difficult
 to obtain 
fully phenomenologically realistic
string models.

 Probably a  fairer comparison 
is between the different concrete string scenarios that have been
proposed so
 far.
 In this sense, each scenario
 has its virtues and problems.
From the top-down approach the D-brane inflation scenarios 
with subsequent rolling tachyon condensation have the advantage 
of being concrete and derivable from direct string perturbation theory
calculations,
as well as sharing the successes of standard inflationary models.
  It is still an interesting challenge to
find string derived potentials with properties 
as those proposed in the pre big-bang and ekpyrotic
 scenarios.
However, from the bottom-up view these two scenarios offer interesting
alternatives to
 inflation that could be theoretically and experimentally tested in
 the
 not too far  future
(for  string theory standards), in
 particular regarding the spectrum of density perturbations and  
gravitational waves. Both ekpyrotic and pre big-bang scenarios
have the possibility 
of being tested, especially in their 
predictions for tensor perturbations, which would differ from inflationary 
models.
For D-brane inflation models, 
the fact that they reproduce inflation is very good because they share 
the successfull predictions of inflation.
This also makes them difficult to be tested independently of inflation.
Fortunately, the generic presence of cosmic strings after inflation 
 may have important implications that could eventually put these proposals to test.

 We do not have to forget that, 
 as emphasised above,
 all these three classes of scenarios  have assumptions. Their common weak
point is  regarding the 
problem of moduli fixing. Before addressing this issue (and others
 that vary
 from 
scenario to scenario) none of the proposals can be called a truly 
top-down approach towards cosmology. Probably the considerations of 
non-vanishing
fluxes of antisymmetric tensor fields \cite{flux}\  could help in 
finding models
with, at least, the relevant moduli fixed, that could be used as
 starting
 point for a 
successfull cosmology. The understanding of the singularity and
 possible 
bouncing
in the pre big-bang and ekpyrotic/cyclic models is their fundamental challenge.
 It is then fair to say that even though there are now several 
interesting frameworks
 in the literature, it is still an open question to
derive a  realistic cosmological scenario from string theory,
 inflationary or non. In any case these concrete attempts, even if
 they prove not realistic, can serve as examples on what  the 
typical  problems to address are, in order to get a successfull
 cosmology from string theory, and could help to eventually identify
 observable model-independent
implications of string cosmology.

On this observational 
 aspect, there has been recent interest on a model independent 
treatment of `trans-Planckian' physics imprints in the CMB. The point 
is to realize that the inflation scale $H$ (usually taken in the range 
$10^{13}-10^{14}$ GeV)   is  smaller than the fundamental scale $M$
(that could be the string scale $M_s$). Therefore the massive states, heavier
 than $M$ (string modes) can be integrated out and an effective field theory
 description of the density perturbation can be done in an expansion in the 
small parameter
$r=\left(\frac{H}{M}\right)^2$. If effects of order $r$ are observable in the
 CMB they would then provide information about the fundamental theory.
If the string scale is of the order of the Planck mass then $r\sim 10^{-11}$
 which may be too small to be observed (still better than the present
 ratio of energies in collider
 experiments at $1$ TeV which gives 
$\left(\frac{10^3{\rm GeV}}{M_{Planck}}\right)^2 \sim 10^{-32}$).
 However if the string scale is smaller,
like in some brane world models,
their imprints may be observable. There may be special cases for which
the leading term in the expansion is of the order $r^{1/2}$ that could have
 more detectable signals although the generic case of
${\cal O}(r)$ is based only on 
very general assumptions such as local effective field theory description.
 This is an 
important
subject given the potential of testing string theoretical effects
in a more or less model independent way. 

The field of string/brane cosmology may no longer be in its infancy,
it may be passing a turbulent adolescence time. Let us hope it will 
arrive  at a successful and productive mature life.

\acknowledgments{
%\vskip 2.0 in
%\centerline{\bf Acknowledgements}
%\bigskip
I would like to thank Bernard de Wit and 
Stefan Vandoren for inviting me to lecture at this school
 and for having infinite patience with my delays to
 finish this written version. I 
acknowledge my collaborators on the original
 topics discussed here: C.P. Burgess,
C. Grojean,  
M. Majumdar, G. Rajesh, S.-J. Rey, G. Tasinato, I. Zavala and  R.-J. Zhang.
I also thank L. Cornalba, M. Costa, G. Gibbons, M. Gomez-Reino,
 C. Kounnas, A. Linde and H. Tye
 for interesting conversations; N.Turok 
for explaining patiently his work; R. Crittenden for providing the 
CMB figure; A. Sen, A. Tseytlin for interesting  correspondence; C.P. Burgess, 
 A. Font, R. Rabad\'an, N. Turok, H. Tye,  G. Veneziano, T. Wiseman and I. Zavala for discussions and
comments on 
the manuscript, and the
 CERN Theory Division for hospitality while this article was written. 
This work is partially supported by PPARC.}

%\newpage

\end{document}